\documentclass[10pt,preprint2,usenatbib]{aastex}

\newcommand{\mum}{\ifmmode{\rm \mu m}\else{$\mu$m}\fi}

\newcommand{\Msun}{\ensuremath{{\rm M}_{\odot}}}      
    
\newcommand{\chisq}{\ifmmode{\chi^{2} }\else{$\chi^2$}\fi}
\newcommand{\rchisq}{\ifmmode{\chi^{2} }\else{$\chi^2_\nu$}\fi}

\newcommand{\SSp}{SAGE-Spec~}
\newcommand{\mbol}{$M_{\mathrm{bol}}$}
\newcommand{\kband}{$K_{s}$}
\newcommand{\jhk}{\emph{JHK}$_{s}$}

\usepackage{amssymb}
\usepackage{graphicx}
\usepackage{threeparttable}
\usepackage{multirow}
\usepackage{url}

\usepackage{multicol}

\shorttitle{SAGE-Spec: Point source classification III.}  
\shortauthors{O.~C.~Jones et al.}

\begin{document}

\title{The SAGE-Spec {\em Spitzer} Legacy program:\\ The life-cycle of dust and gas in the Large Magellanic Cloud. Point source classification III.}


\author{Olivia~C.~Jones\altaffilmark{1},
        Paul~M.~Woods\altaffilmark{2},
        F.~Kemper\altaffilmark{3},
        K.~E.~Kraemer\altaffilmark{4},
    	G.~C.~Sloan\altaffilmark{1,8,12},
    	S.~Srinivasan\altaffilmark{3},     
    	J.~M.~Oliveira\altaffilmark{5},
    	J.~Th.~van~Loon\altaffilmark{5},
    	Martha~L.~Boyer\altaffilmark{1},
        Benjamin A.~Sargent\altaffilmark{1,13},
        I.~McDonald\altaffilmark{6},
        Margaret~Meixner\altaffilmark{1},   
        A.~A.~Zijlstra\altaffilmark{6,11},        
	    Paul~M.~E.~Ruffle\altaffilmark{3,6,*},    
	    E.~Lagadec\altaffilmark{7}, 
        Tyler~Pauly\altaffilmark{8},
        Marta~Sewi{\l}o\altaffilmark{9},
        G.~C.~Clayton\altaffilmark{10},
    	K.~Volk\altaffilmark{1}
 }
\altaffiltext{1}{Space Telescope Science Institute, 3700 San Martin Drive, Baltimore, MD, 21218, USA}
\altaffiltext{2}{Astrophysics Research Centre, Dept. of Mathematics \& Physics, Queen's University Belfast, Belfast BT7 1NN, UK}
\altaffiltext{3}{Academia Sinica, Institute of Astronomy and Astrophysics, Taipei 10617, Taiwan }
\altaffiltext{4}{Institute for Scientific Research, Boston College, 140 Commonwealth Avenue, Chestnut Hill, MA 02467, USA}
\altaffiltext{5}{School of Chemical and Physical Sciences, Lennard-Jones Laboratories, Keele University, Staffordshire ST5 5BG, UK}
\altaffiltext{6}{Jodrell Bank Centre for Astrophysics, Alan Turing Building, School of Physics and Astronomy, The University of Manchester, Oxford Road, Manchester, M13 9PL, UK}
\altaffiltext{7}{Observatoire de la C\^ote d'Azur, CNRS, Lagrange, France}
\altaffiltext{8}{Cornell Center for Astrophysics and Planetary Science, Cornell University, Ithaca, NY 14853-6801, USA}
\altaffiltext{9}{NASA Goddard Space Flight Center, 8800 Greenbelt Rd, Greenbelt, MD 20771, USA}
\altaffiltext{10}{Department of Physics \& Astronomy, Louisiana State University, Baton Rouge, LA 70803, USA}
\altaffiltext{11}{Department of Physics, University of Hong Kong, Pokfulam Road, Hong Kong}
\altaffiltext{12}{Department of Physics and Astronomy, University of North Carolina Chapel Hill, Chapel Hill, NC 27599-3255, USA}
\altaffiltext{13}{Center for Imaging Science and Laboratory for Multiwavelength Astrophysics, Rochester Institute of Technology, 54 Lomb Memorial Drive, Rochester, NY 14623, USA}
\altaffiltext{*}{Paul M. E. Ruffle passed away on 21 November 2013.}

\slugcomment{Accepted for publication in MNRAS.}


\begin{abstract}
  \noindent The Infrared Spectrograph (IRS) on the {\em Spitzer Space Telescope} observed nearly 800 point sources in the Large Magellanic Cloud (LMC), taking over 1\,000 spectra. 197 of these targets were observed as part of the \SSp \emph{{\em Spitzer}} Legacy program; the remainder are from a variety of different calibration, guaranteed time and open time projects. We classify these point sources into types according to their infrared spectral features, continuum and spectral energy distribution shape, bolometric luminosity, cluster membership, and variability information, using a decision-tree classification method. We then refine the classification using supplementary information from the astrophysical literature.  We find 
  that our IRS sample is comprised substantially of YSO and H\,{\sc ii} regions, post-Main Sequence low-mass stars: (post-)AGB stars and planetary nebulae and massive stars including several rare evolutionary types. Two supernova remnants, a nova and several background galaxies were also observed. 
  We use these classifications to improve our understanding of the stellar populations in the Large Magellanic Cloud, study the composition and characteristics of dust species in a variety of LMC objects, and to verify the photometric classification methods used by mid-IR surveys. We discover that some widely-used catalogues of objects contain considerable contamination and others are missing sources in our sample. 
\end{abstract}

\keywords{galaxies: individual (LMC) --- infrared: galaxies --- infrared: stars --- Magellanic Clouds --- surveys --- techniques: spectroscopic.}

\section{Introduction}
\label{sec:intro}

There are many ways to probe the properties of a galaxy. One of the most important is to characterise its stellar content. Particularly revealing is to identify (stellar) sources that are bright in the infrared: these sources are generally dust-rich, and form an important step in the cycle of baryonic matter throughout a galaxy, for which dust is an important tracer. IR-bright stellar sources cover a wide range of stellar masses, at both ends of stellar evolution.

The Large Magellanic Cloud (LMC) is a nearby galaxy (50\,kpc away; \citealt{Pietrzynski2013}) that has been the subject of, or included in, a plethora of past and on-going surveys at all wavelengths. As such, a wealth of information exists about its stellar sources, which can be compiled and understood in terms of different stellar populations. Within the LMC, stars can be assumed to be all at approximately the same distance from the Earth and to have a narrow range of metallicity, [Fe/H] $\sim -0.3$ dex \citep{Westerlund1997}.

The Surveying the Agents of Galaxy Evolution (SAGE) project \citep{Meixner2006} performed a uniform, unbiased photometric survey of the LMC with the {\emph{Spitzer Space Telescope}} 
\citep{Werner2004}, using all bands of the Infrared Array Camera \citep[IRAC; 3.6, 4.5, 5.8, and 8.0 $\mu$m;][]{Fazio2004} and the Multi-Band Imaging Photometer for Spitzer \citep[MIPS; 24, 70, and 160 $\mu$m;][]{Rieke2004}. This complements the optical Magellanic Clouds Photometric Survey \citep[MCPS;][]{Zaritsky2004} and the near-infrared surveys: Two Micron All Sky Survey \citep[2MASS;][]{Skrutskie2006} and the VISTA Survey of the Magellanic Clouds \citep[VMC;][]{Cioni2011}. From the SAGE imaging, a point source catalogue has been constructed, containing 8.5 million sources with detections in at least one of the \emph{Spitzer} photometric bands \citep{Meixner2006}. The \emph{Herschel Space Observatory} \citep{Pilbratt2010} with the Herschel Inventory of the Agents of Galaxy Evolution survey (HERITAGE; \citealt{Meixner2013}) has also targeted the Magellanic Clouds, probing the far-infrared (FIR) and submillimetre (sub-mm) regime, detecting the emission from the coldest dust grains.

Some of the initial spectroscopic surveys in the LMC with the Infrared Spectrograph (IRS; \citealt{Houck2004}) on  Spitzer were conducted by \citet{Buchanan2006, Buchanan2009} who targeted 60 IR bright sources in the LMC with an 8 $\mu$m {\em MSX} detection \citep{Egan2001}.
A larger spectroscopic follow-up to the SAGE photometric survey, \SSp \citep{Kemper2010}, charts the life cycle of gas and dust in galaxies by observing a variety of circumstellar and interstellar environments. These observations of 197 sources covered the full luminosity and colour range found in the SAGE photometric survey. 
\citet[hereafter Paper I]{Woods2011} developed a comprehensive classification scheme for point sources based on the mid-IR spectra, and applied this to the initial 197 LMC point sources in the \SSp sample. The second paper, which applied this scheme to sources from the Small Magellanic Cloud (SMC), was published by \citet[hereafter Paper II]{Ruffle2015}. In the current paper (Paper III) we extend the spectroscopic inventory to all IRS staring-mode observations within the SAGE footprint.
These classifications improve our understanding of the stellar populations of the Magellanic Clouds and ultimately, allow us to relate mid-IR photometry to the spectral characteristics of different types of objects within the LMC. These results provide a global view of the dust content at different stages of the evolutionary cycle in the LMC and thereby trace the life cycle of baryonic matter that drives galactic evolution.

In Sect.~\ref{sec:data} we describe the entire IRS sample of over 1\,000 spectra and the data reduction process we used to process them. In Sect.~\ref{sect:classmethods} we describe the classification method using the different spectral features in individual sources, and in Sect.~\ref{sec:classifications} we present the results of the spectral classification and discuss the different object classes contained in the catalog. We compare these results to classifications that are commonly used in the literature in Sect.~\ref{sec:photCompare}. 
Finally, we end with a discussion and conclusions in Sect.~\ref{sec:conclusion}.

\section{The {\em Spitzer} data and data reduction}
\label{sec:data}

\begin{table*}\footnotesize
  \caption{\emph{{\em Spitzer}} IRS Staring mode data and their sources \label{tab:ArchiveProjects}}
\begin{tabular}{lclll}
~\\
  \hline
  \hline
Cycle & PID & program PI & Primary target & Spectra count\tablenotemark{a} \\
\hline
Calsfx& M -- 28  & Various & Various calibrators & 14 hi, 43 lo \\
1$^\mathrm{(GTO)}$ & 18    & Houck/Uchida        & H\,{\sc ii} regions & 10 hi \\
1$^\mathrm{(GTO)}$ & 63    & Houck/Brandl        & H\,{\sc ii} regions & 19 hi, 17 lo  \\
1$^\mathrm{(GTO)}$ & 103   & Houck/Bernard-Salas & Planetary nebulae &18 hi, 18 lo \\
1$^\mathrm{(GTO)}$ & 124, 129 & Gehrz/Polomski      & YSOs/LBVs & 6 slhi, 2 lo \\
1$^\mathrm{(GTO)}$ & 200   & Houck/Sloan         & Evolved stars & 29 lo \\
1$^\mathrm{(GTO)}$ & 249   & Indebetouw    & Star-forming regions & 6 lo \\
1$^\mathrm{(GTO)}$ & 263   & Woodward      & Novae & 1 lo \\
1$^\mathrm{(GTO)}$ & 464, 472   & Cohen    & Stars &14 lo \\
1$^\mathrm{(GTO)}$ & 485   & Ardila        & Stars & 6 lo \\
1   & 1094  & Kemper        & O-rich evolved stars & 9 hi, 2 slhi, 9 lo  \\
1   & 2333  & Woodward      & Novae & 4 slhi \\
1   & 3426  & Kastner       & Evolved stars & 60 lo \\
1   & 3470  & Bouwman       & Herbig AeBe stars & 1 hi \\
1   & 3483  & Rho           & Supernova remnants & 1 lo \\
1   & 3505  & Wood          & Evolved stars & 31 lo \\
1   & 3578  & Misselt       & Interstellar medium & 13 lo \\
1   & 3583  & Onaka         & O-rich evolved stars & 12 lo \\
1   & 3591  & Kemper        & O-rich evolved stars & 53 lo \\
2   & 20080 & Polomski      & SN 1987A & 1 lo \\
2   & 20443 & Stanghellini  & Planetary nebulae & 20 lo \\
2   & 20752 & Reynolds      & Supernova remnants & 1 hi \\
3   & 30067 & Dwek          & SN 1987A & 4 lo \\
3   & 30077 & Evans         & Evolved stars & 1 lo \\
3   & 30180 & Fazio/Patten     & Unknown object & 1 lo \\
3   & 30332 & Houck/Sloan         & O-rich evolved stars & 1 lo \\
3   & 30345 & Houck/Sloan         & O-rich evolved stars & 4 hi \\
3   & 30372 & Tappe         & Supernova remnants & 7 lo \\
3   & 30380 & Clayton       & R Coronae Borealis stars & 1 lo \\
3   & 30544 & Morris        & Wolf-Rayet stars, supergiants & 1 lo \\
3   & 30788 & Sahai         & Post-AGB stars & 24 lo \\
3   & 30869 & Kastner       & B[e] stars & 1 slhi, 8 lo \\
4   & 40031 & Fazio/Reach     & Interstellar medium & 2 lo \\
4   & 40149 & Dwek          & SN 1987A & 4 lo \\
4   & 40159 & Kemper/Tielens & Various (\SSp) & 197 lo \\
4   & 40604 & Reynolds      & Supernova remnants & 2 lo \\
4   & 40650 & Looney        & Massive YSOs & 8 hi, 331 slhi \\
5   & 50092 & Gielen        & Post-AGB stars &10 lo \\
5   & 50147 & Sloan         & Evolved stars & 6 hi \\
5   & 50167 & Clayton       & Red supergiants & 7 lo \\
5   & 50338 & Matsuura      & Post-AGB stars &  23 lo, 1 slhi, 11 hi  \\
5   & 50444 & Dwek          & SN 1987A & 3 hi, 2 lo \\
\hline
\tablenotetext{a}{`hi' indicates a spectrum taken in high-resolution  mode (Short-High and/or Long-High), and `lo' in low-resolution mode  (Short-Low and/or Long-Low). `slhi' indicates a combination of data  from the {\em Spitzer} Short-Low, Short-High and Long-High modules.}
\end{tabular}
\end{table*}

We used the spectra from the \SSp database at the NASA/IPAC Infrared Science Archive (IRSA)\footnote{\url{http://irsa.ipac.caltech.edu/cgi-bin/Gator/nph-dd?catalog=ssid2}}. This database includes all staring-mode observations with the  {\it Spitzer} IRS taken in the area of the sky covered by the SAGE survey \citep{Meixner2006}, which fully encompasses the LMC.  Most of the spectra cover the full $\sim$5--37~\mum\ range of the low-resolution modules of the IRS, although some spectra are limited to the $\sim$5--14~\mum\ range and others use the high-resolution modules. Included in this selection of 1080 high and low-resolution spectra were the 197 spectra obtained under the \SSp program (discussed in Paper I). The remainder comprised other science projects led by a variety of different Principal Investigators (PIs; see the list in Table~\ref{tab:ArchiveProjects}), and were obtained with a diverse range of observing parameters.  
Additionally, 61 objects were observed with MIPS in SED mode, as part of the \SSp project. This gives a very low resolution spectrum in the wavelength range 52--93\,$\mu$m. Those observations and objects are described fully by \citet{vanLoon2010}.

\citet{Kemper2010} describe how the IRS spectra were processed and calibrated. We reduced all the spectra using this method.  In outline, the low-resolution spectra were extracted from spectral images using the tapered-column algorithm\footnote{Available in both the SMART \citep[Spectroscopy Modeling Analysis and Reduction Tool;][]{Higdon2004} and SPICE ({\it Spitzer} IRS Custom Extraction) data reduction packages.}.  High-resolution spectra used full-slit extraction.  All spectra were flux-calibrated using assumed truth spectra for the K giants HR 6348, HD 166780, and HD 173511, as described by \cite{Sloan2015}.  The scientific rationale for the observations varied considerably among the programs; in some cases, the point sources we consider here were not even the primary target.  \citet{Woods2011b}  provide more detailed comments in the documentation accompanying the data.


The $\sim$300 spectra from program 40650 (PI: Looney) required special treatment, as they included low-resolution spectra from 5 to 14~\mum\  and high-resolution spectra from 10 to 37~\mum. These segments were extracted with the tapered-column and full-slit algorithms, respectively, as for most of the spectra.  Due to possible background contamination in the high-resolution data, the spectra were normalized to the 5--14~\mum\ segments.  For sources in this program that were not spatially extended, we also used optimally-extracted spectra from the CASSIS database \citep[Combined Atlas of Spitzer IRS Sources;][]{Lebouteiller2015}.  These spectra generally have higher signal/noise ratios than those produced by the other algorithms, because the optimal extractions exclude off-source data dominated by the sky and background noise.  They also have better background removal, which reduces the contributions of nebular emission lines to the final spectra.


In total, 1080 spectra were obtained from  947 locations on the sky (i.e., there were 133 repeated observations, usually at a different resolution). In order to select spectra of point sources only, we matched the position of each extraction from the raw data (\texttt{RA{\_}FOV}, \texttt{DEC{\_}FOV}) to one of the SAGE point source catalogues (IRAC 3.6-8.0 $\mu$m and MIPS 24-160 $\mu$m; see the SAGE documentation\footnote{\url{http://irsa.ipac.caltech.edu/data/SPITZER/SAGE/doc/SAGEDataProductsDescription_Sep09.pdf}}) or the catalogue of \citet{Gruendl2009} which is based on aperture photometry. This recovers sources which are slightly extended in the IRAC bands (but are point-sources at 24 \mum) that were missed by the SAGE point source catalogues. We discarded any spectra without a 3\arcsec\ radius photometric match to an object in one of these catalogues. 


Using the IRAC coordinates, the {\em Spitzer} IRS spectra were matched by position to photometry from point sources in complementary LMC surveys: the \emph{UBVI} photometry comes from the Magellanic Clouds Photometric Survey \citep[MCPS;][]{Zaritsky2004}; \emph{JHK$_s$} photometry comes from both the Two Micron All Sky Survey catalogue \citep[2MASS;][]{Skrutskie2006} and the Infrared Survey Facility (IRSF) survey \citep{Kato2007}, and the PACS and SPIRE data from the HERITAGE Band-Matched catalogue \citep{Seale2014}. In general, positional matches were chosen to be the closest point source in each of these catalogues within 3\arcsec\ of the observed target. Exceptions to this were the MIPS, PACS and SPIRE catalogues, where 5\arcsec, 10\arcsec\ and 20\arcsec\ radii were used, respectively. This is due to the difference in spatial resolution between the surveys. 
This supplementary data enabled us to construct a spectral energy distribution (SED) for each object, to aid in the classification process. The SEDs were checked to ensure a good match between the spectral data and the photometric data. Only in a few cases  
was the next-nearest photometric match a better fit to the spectrum. Using this SED we also estimated a bolometric luminosity, which was useful in source classification.

In order to be as general as possible in our classification, we classify the object shown in each point source spectrum. Where multiple spectra have been taken of one object, this redundancy gives an extra check, since all the classifications should agree, even if variations between spectra could exist, due to variability. We have identified 816 point-source objects with spectra within the footprint of SAGE, although 27 objects were not detected by the IRS (the spectrum is without signal), resulting in a final count of 789 unique point-source objects with IRS spectra. We numbered each of these LMC objects with an identifier (OBJID), from 1--789. Each individual IRS spectrum was also numbered with an identifier (SSID) from 1--197 for spectra discussed in Paper I, and from 4000--4817 for all other spectra. This numbering corresponds to that used in the \SSp database at IPAC. 

Table~\ref{tab:metatable}  describes the columns of the online table which lists all the IRS staring-mode targets in the LMC, the field-of-view (FOV) position, their spectral classification, bolometric magnitude, variability and other relevant information. The matching photometry for the IRS spectra is also given in the online table.

\begin{table*}
  \caption{Numbering, names and description of the columns present in the classification table which is available on-line only.}
\label{tab:metatable}\small
\begin{tabular}{lll}
~\\
  \hline
  \hline
  Column    &          Name    &          Description \\
  \hline
  1	&	 OBJID 	&	 Object identification number of the target \\
2	&	 SSID 	&	 Unique spectral identification number\\
3	&	 NAME 	&	 Name of point source targeted \\ 
4	&	 SAGE\_{}SPEC\_{}CLASS 	&	 Source classification determined in this paper \\
5--6 	&	 RA\_{}SPEC, DEC\_{}SPEC 	&	 Position of the extracted spectrum \\
7--8	&	 AOR, PID 	&	 {\em Spitzer} Astronomical Observation Request \\ 
        &       & and observing program identification numbers\\
9	&	 PI 	&	 Last name of the PI of the {\em Spitzer} program\\
10--17 	&	 MAGU, DMAGU, MAGB, DMAGB, 	&	 Matching MCPS  $U$, $B$, $V$, $I$ magnitudes\\
 & MAGV, DMAGV, MAGI, DMAGI & and errors \\
18--22   &    OGLE3ID, OGLE3MEAN\_{}V, OGLE3MEAN\_{}I, & Source ID and variability\\ 
        & OGLE3AMP\_{I}, OGLE3PERIOD &	information from the OGLE survey\\
23--29   & TMASS\_DESIGNATION, MAGJ, DMAGJ, & Designation of 2MASS(6X) counterpart and its\\
         &  MAGH, DMAGH, MAGK, DMAGK &  $J$, $H$, and \kband\ magnitudes and errors \\
30--36   & IRSF\_DESIGNATION, IRSF\_JM, IRSF\_JME, &	 Designation of IRSF counterpart and its \\
        & IRSF\_HM, IRSF\_HME, IRSF\_KM, IRSF\_KME & $J$, $H$, \kband\ magnitudes and errors \\
37--53  & IRAC\_DES, MAG3\_6, DMAG3\_6, MAG4\_5, & SAGE-LMC IRAC Archive point source \\
        & DMAG4\_5, MAG5\_8, DMAG5\_8, MAG8\_0,  & designation\tablenotemark{1} matching the extracted spectrum, \\
        & DMAG8\_0, FLUX3\_6, DFLUX3\_6, FLUX4\_5,  & and the corresponding IRAC mean \\ 
        & DFLUX4\_5, FLUX5\_8, DFLUX5\_8, FLUX8\_0, & photometry\tablenotemark{2} with errors.\\
        & DFLUX8\_0 \\
54--58  & MIPS24\_DES, MAG24, DMAG24, FLUX24, & SAGE-LMC MIPS24 catalogue designation and\\
        & DFLUX24 & the corresponding mean photometry with errors \\
59--63  & MIPS70\_DES, MAG70, DMAG70, FLUX70,& SAGE-LMC MIPS70 catalogue designation and\\
        & DFLUX70 & the corresponding photometry with errors \\
64--68  & MIPS160\_DES, MAG160, DMAG160, FLUX160,& SAGE-LMC MIPS160 catalogue designation and\\
        & DFLUX160 & the corresponding photometry with errors \\
69--81 	& HERSCHEL\_DES, HERSCHEL\_RA, &	 HERITAGE designation, positions, fluxes,\\
        & HERSCHEL\_DEC, PACS\_F100, PACS\_DF100, & and errors\\
        & PACS\_F160, PACS\_DF160 & \\
        & SPIRE\_F250, SPIRE\_DF250, SPIRE\_F350,& \\
        &  SPIRE\_DF350, SPIRE\_F500, SPIRE\_DF500 & \\
82	    &	 MBOL\_{}PHOT 	& \mbol\ calculated by interpolation of \jhk,\\ 
        &                   & IRAC, and MIPS24 photometry\\
83--84 	&	 MBOL\_{}MCD, TEFF\_{}MCD 	& \mbol\ calculated using the SED fitting code\\
        &                               & from \citet{McDonald2009,McDonald2012}\tablenotemark{3}\\
85	&	 FGFLAG 	&	 Foreground flag, set to 1 if foreground source \\
\hline
\tablenotetext{1}{By default, IRAC information is provided for the Mosaic Archive match. If no such match was available, the Epoch 1 (or, failing that, Epoch 2) information is provided instead.}
\tablenotetext{2}{All flux densities in this table are in Jy.}
\tablenotetext{3}{Only good fits are included.}
\end{tabular}
\end{table*}

\section{The classification method}
\label{sect:classmethods}

We have classified the objects associated with the 1\,080 IRS staring mode spectra in the footprint of the SAGE project, which covers the entirety of the LMC. The primary basis for this classification is the {\em Spitzer} spectrum itself, but we use the additional information provided by the SED, and also an estimate of the bolometric luminosity, and furthermore, we confirm our classification with a literature study when possible.

To add rigor to our classification method, we use a slightly modified version of the classification decision tree, as presented in Papers I and II. One proceeds through the tree, responding to the {\sc yes} or {\sc no} questions in order to reach a terminal classification. Where questions cannot be answered, e.g., when some part of the required data is not available (for instance, longer wavelength data from the IRS Long Low modules), one may be left with a number of potential classifications, which can be constrained with a literature search, or comparison with similar spectra. 

The final classification groups, with a tally of each class of object are listed in Table~\ref{tab:sourcecounts}. An explanation of these groups follows:

{\bf YSO-1}. These objects are embedded young stellar objects (YSOs), which display spectral features due to ices: molecules adsorbed onto the surface of dust grains, that indicate a cold, shielded and relatively dense environment. A detailed description of the ice features found in the IRS spectral range can be found in \citet{Oliveira2011} and references therein; we summarise the main features here. A relatively narrow absorption feature close to 15.2\,$\mu$m is attributed to CO$_2$ ice; the structure and width of the feature is determined by the composition and physical conditions in the ice mantles. The broad and sub-structured feature between 5--7\,$\mu$m is dominated by the contribution of water (the most abundant ice species in YSO environments and the quiescent interstellar medium) at 6\,$\mu$m. Other molecular species that contribute to this broad complex are ammonium (NH$^+_4$, possibly the main contributor to the absorption at 6.8\,$\mu$m), formic acid (HCOOH), formaldehyde (H$_2$CO), and ammonia (NH$_3$). Objects in this class may also show refractory dust features in their spectra.

{\bf YSO-2}. These YSOs are more evolved than {\tt YSO-1}s, either having removed some of their surrounding circumstellar envelopes or heated it up sufficiently to destroy icy mantles. Dust features are present, typically silicate absorption seen as a characteristic broad dip in the spectrum around 10\,$\mu$m. 

{\bf YSO-3}. {\tt YSO-3}s are the most evolved stage of massive YSO, where the envelope may have largely dispersed, and the YSO is beginning to ionize its surrounding region. This evolutionary stage manifests itself in strong PAH emission (at 3.3, 6.2, 7.7, 8.6, 11.2, 12.7 and 16.4\,$\mu$m), but with no other features such as atomic emission lines, usually superposed on a rising dust continuum (in $F_{\nu}$ versus $\lambda$ space) over the whole IRS range.

{\bf H\,{\sc ii} region}. H\,{\sc ii} regions are characterised by atomic emission lines in their spectra, which may also contain PAH emission features. The class of H\,{\sc ii} region includes both compact H\,{\sc ii} regions (dense regions of size $\sim$\,1\,pc surrounding emerging massive stars) and more classical H\,{\sc ii} regions (more diffuse regions of size on the order of several parsecs, ionized by single or multiple massive stars in the near vicinity). Even though a rising continuum, indicative of the presence of a significant amount of dust, could hint at a compact younger source, these two types cannot easily be distinguished based on IRS spectra alone.  For classification purposes, the distinction between sources classed as {\tt YSO-3} and (PAH-rich) H\,{\sc ii} is the presence of atomic emission lines. (H\,{\sc ii} sources are distinguished from PNe by the shape of the continuum or the presence of higher excitation emission lines.) For more diffuse H\,{\sc ii} regions the continuum may be due to free-free emission not dust, making them strong radio emitters.

{\bf H\,{\sc ii}/YSO-3}. As mentioned above the distinction between the {\tt YSO-3} and H\,{\sc ii} region classes is not straightforward. Furthermore, telescope resolution and sensitivity limitations make it very difficult to separate the contributions to the spectrum of a point source and the wider energetic environment, especially in complex regions of high background. Thus sources in this intermediate class share properties with both the {\tt YSO-3}s and ultra-compact H\,{\sc ii} regions classes, and could not be categorically be placed in either class.  

{\bf YSO-4}. This class is composed of sources with a 10\,$\mu$m silicate emission feature usually superposed on a moderately rising continuum over the  IRS range. These objects are expected to be analogues of Galactic Herbig Ae/Be (HAeBe) objects.  

{\bf Star}. The {\sc star} classification contains dustless objects which show approximately blackbody spectra, with SEDs which peak shortwards of 2\,$\mu$m. The class contains both hot massive stars and low-mass dustless (red giant branch) stars. Low-mass stars on the Main Sequence are not bright enough in the IR to be detected by {\em Spitzer}  at the distance of the LMC.  

{\bf O-AGB, O-PAGB, O-PN}. These categories contain evolved stars showing the presence of oxygen-rich dust in their spectra. The {\tt O-AGB} category contains objects which in general have a blackbody-like SED, with superimposed dust or molecular features: either an absorption feature due to SiO close to 8\,$\mu$m ({\tt O-EAGB}; \citealt{Ruffle2015}), or the more common 10 and 20\,$\mu$m features due to silicate dust. Stars which are less luminous than the traditional AGB luminosity limit of $M_\mathrm{bol}=-7.1$ are put into this category, whilst similar but more luminous SEDs are classified as {\tt RSG} (see below). 

{\tt O-PAGB} objects show oxygen-rich dust spectral features, but their SED features two distinct peaks, one due to the stellar emission, and one due to the cooler dust contained in a disk or detached shell. 
{\tt  RV Tau} stars are a sub-category of {\tt O-PAGB} stars that are identified by the distinct variability pattern of their lightcurves and chemical depletions in the photospheres of these objects \citep[see e.g.,][]{Gielen2009}.  Due to the limits of RV Tauri studies, there may be unidentified RV Tauri stars included in our {\tt O-PAGB} category.

{\tt O-PN} objects exhibit a rising continuum up to $\sim$30\,$\mu$m, but then a decreasing continuum. They show atomic emission lines, particularly the [O {\sc iv}] line at 25.913\,$\mu$m, and an absence of PAH features. Their spectra may include 10 and 20\,$\mu$m silicate features.
{\tt O-PN}e are distinguished from H\,{\sc ii} regions by the presence of high-excitation lines (e.g., [Ne {\sc v}], [Ne {\sc vi}]) in the spectrum of the former, but not the latter.

{\bf RSG}. The spectra of red supergiants exhibit oxygen-rich dust features similar to {\tt O-AGB} stars. Due to their large mass ($> 8\,M_{\rm \odot}$) they are generally more luminous than AGB stars \citep{Wood1983} and for the purposes of our classification have either a bolometric luminosity of $M_\mathrm{bol}<-7.1$,  reside in a young cluster  ($< 55$ Myr; \citealt{Schaller1992}), or show no large-amplitude variability.

{\bf C-AGB, C-PAGB, C-PN}. These three categories encompass carbon-rich evolved stars, usually identified by molecular absorption features, most notably C$_2$H$_2$ (acetylene) at 7.5\,$\mu$m and 13.7\,$\mu$m, and dust features at 11.3\,$\mu$m (SiC), 21\,$\mu$m (unknown carrier) and 30\,$\mu$m (MgS). In addition, these sources may exhibit PAH or related molecular emission lines in their spectra, which in planetary nebulae seem to indicate a carbon-rich \citep{Stanghellini2007} or carbon-enhanced  \citep{Guzman2014} nebula.
{\tt C-AGB}s often have a broad and red SED, characterised by acetylene absorption features and/or an SiC feature. %
{\tt C-PAGB}s exhibit a double-peaked SED, with a strong 30\,$\mu$m feature or a 21\,$\mu$m feature \citep{Volk2011}.
{\tt C-PN}e show atomic emission lines and the presence of PAH features; a 30\,$\mu$m feature may also be seen.

{\bf RCrB}. R Coronae Borealis stars are a class of carbon-rich supergiant stars, which are rare in the Galaxy (over 100 are now known), but which may be more common in the LMC (about 20 have been discovered; \citealt[][2017 in prep.]{Clayton2012, Tisserand2013}). These objects have unusual infrared spectra, showing few features apart from a very cool continuum (due to very \emph{hot} dust), with an SED that peaks at wavelengths longer than 2\,$\mu$m \citep[e.g.,][]{Kraemer2005}. Candidates are confirmed through other means, including analysis of lightcurves, which show irregular and fast declines in brightness \citep[up to 9 visible magnitudes;][]{Clayton1996} thought to be caused by episodic ejections of carbon dust which occlude the starlight.

{\bf GAL}. Galaxies show a variety of red-shifted features in their spectra, including atomic emission lines and oxygen-rich dust features. 

{\bf Other}. {\tt OTHER} is a catch-all class which incorporates objects we have been able to classify, but which are rare in the present sample. We find a small number of Luminous Blue Variable (LBV) stars, B[e] stars, Wolf-Rayet stars, supernova remnants, blue supergiants, a yellow supergiant and a nova.  In many cases, these spectra are not satisfactorily classified by the decision tree, and instead are classified with reference to the literature.

The SEDs of LBVs and blue supergiants generally show the short wavelength characteristics of a hot star (hence the `blue'), coupled with a dramatic rise at longer wavelengths, with a minimum at $\sim$10\,$\mu$m. They exhibit erratic variability. 
%
Wolf--Rayet stars present an infrared spectrum which shows the presence of a hot star, as well as atomic emission lines and in some cases, a large amount of warm dust. 
%
B[e] stars are young supergiants characterised by forbidden line emission and an infrared dust excess \citep{Lamers1998,Kastner2010}. They often show oxygen-rich dust features, with signs of crystallinity, and weak emission lines, and in the infrared, may be confused with oxygen-rich giant stars.

{\bf UNK}. Point-source objects which we are not able to classify are placed into the {\tt UNK} (unknown) category. There are many reasons that an object may not be reliably classified, including incomplete data, unusual (and therefore unrecognised) object, data issues. We say a few words about these individual objects in Sect.~\ref{sect:UNK}.

\subsection{Bolometric luminosities}
\label{sec:lum}

Bolometric luminosities ($M_{\rm bol}$) were calculated via a trapezoidal integration of the SED over the range of available photometric points (typically \emph{U}-band to IRAC 8\,$\mu$m or MIPS 24\,$\mu$m) and compared with published values \citep[e.g.,][]{Srinivasan2009, Groenewegen2009, Jones2012, Riebel2012}. For stellar photospheres and oxygen-rich evolved sources with little infrared excess, bolometric magnitudes were also calculated using the SED-fitting code of \cite{McDonald2009, McDonald2012}. This code performs a $\chi^2$-minimisation between the observed SED (corrected for interstellar reddening) and a grid of {\sc bt-settl} stellar atmosphere models \citep{Allard2011} which are scaled in flux to derive a bolometric luminosity.  This provides a better $\chi^2$ fit to the optical and near-IR photometry than fitting a Planck function. For the most enshrouded stars, fitting the SED with `naked' stellar photosphere models leads to a under-estimation of the temperature and luminosity due to circumstellar reddening and hence this method is not preferred for very dusty sources. For the redder sources the values calculated by trapezoidal integration provide a lower limit estimate of $M_{\rm bol}$. 

\begin{figure}
 \centering
  \includegraphics [trim=2.0cm 2cm 0cm 7cm, width=7.5in]{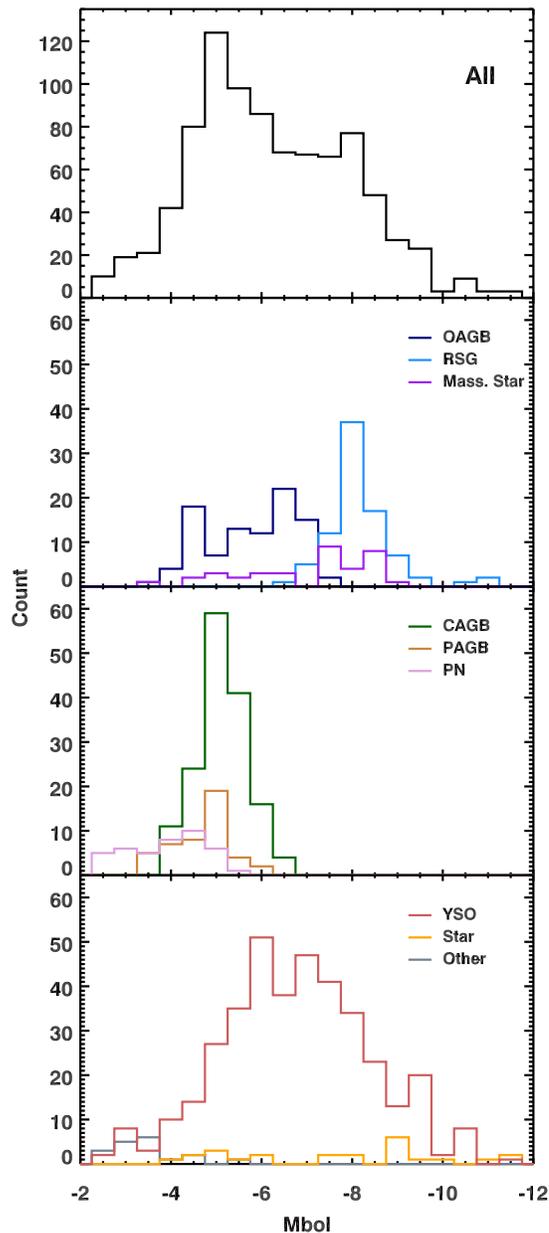}
 \caption[Bolometric luminosities calculated for the LMC spectral sample]{Bolometric magnitudes for the LMC spectral sample.}
 \label{fig:luminositycomparison}
\end{figure}

Figure~\ref{fig:luminositycomparison} shows luminosity functions for AGB stars, RSGs, post-AGB stars, massive stars and YSOs for which IRS spectra are available. The oxygen-rich AGB distribution is double peaked; this is due to low-mass O-AGB stars at the faint end of the distribution, and massive O-AGBs ($M > 5M_{\odot}$) currently undergoing hot bottom burning at the brighter end. A luminosity cut at $M_{\rm bol} =-7.1$ is used to distinguish between the RSG and massive O-AGB stars. Stars brighter than $M_{\rm bol}= -12.0$ are classified as foreground sources. In the LMC, carbon stars form in the mass range 1.5--4.0 $M_{\odot}$, and this effect can be seen in their narrow luminosity distribution, which peaks at the minima ($M_{\rm bol} = -5)$ of the O-AGB stars' distribution. The YSOs in our sample have a broad spread in luminosity. This can be used as a proxy for the YSOs' mass distribution, with the brightest sources corresponding to high-mass YSOs, while the fainter sources are less massive. Background galaxies, PNe, and objects of unknown spectral type are the faintest classes of objects within our sample. The bolometric luminosities for background galaxies have not been corrected for distance; if this were accounted for, they would become by far the most luminous objects in our sample.

\subsection{Foreground sources}
\label{sec:foreground}
To check for foreground sources, we performed a SIMBAD search using a 3\arcsec\ radius for each source in our sample.  We tag as foreground candidates any sources that have a SIMBAD match with a radial velocity outside the range 150--450 km s$^{-1}$ determined by \citet{Neugent2012} for LMC stars, and include sources for which our bolometric luminosities are above $-12.0$ mag (see previous section). There are 19 foreground objects in our sample; these sources have been flagged as foreground in the online table and have been classified by us as stellar photospheres or O-AGB stars.

\subsection{Variability}
\label{sec:var}

To aid in the classification, a literature search was conducted to compile variability information. The IRS sample was matched to the Optical Gravitational Lensing Experiment (OGLE-III) catalogue of variable stars \citep{Soszynski2008} and the catalogues of \citet{Vijh2009} and \citet{Groenewegen2009}. 

\citet{Vijh2009} identified IR variable sources using 2 epochs of data from the SAGE catalogue, including 99 AGB stars, six massive stars and four variable YSOs from the \SSp sample. The sample of \citet{Groenewegen2009} has 130 matches to our catalogue, all of which are AGB stars. They determine pulsation periods for 105 of these sources.

\section{Presentation of the classification}
\label{sec:classifications}


\begin{table}
\tabcolsep=0.11cm
\caption{Spectral classification groups and source counts} 
\label{tab:sourcecounts}
~\\
\begin{tabular}{lllc}
\hline
\hline
Main & Sub- & Object type  & \#  \\
class & class &   &   \\
\hline
{\tt YSO-1}  & & Young stellar object                              &    \phantom{1}53       \\
{\tt YSO-2}  & & Young stellar object                              &    \phantom{1}14       \\
{\tt YSO-3}  & & Young stellar object                              &    \phantom{1}80       \\
{\tt H\,{\sc ii}}/{\tt YSO-3} & & Probable evolved YSO             &    \phantom{1}35       \\
{\tt YSO-4}  & & Herbig AeBe object                               &     \phantom{1}21        \\
{\tt H\,{\sc ii}} & & H\,{\sc ii} region (including compact)      &              134         \\
{\tt STAR}   & & Stellar photosphere                              &     \phantom{1}30       \\
{\tt RSG}    & & Red supergiant                                   &     \phantom{1}71       \\
{\tt C-AGB}  & & C-rich AGB star                                  &               148       \\
{\tt O-AGB}  & & O-rich AGB star                                  &     \phantom{1}77       \\
{\tt C-PAGB} & & C-rich post-AGB star                             &     \phantom{1}19       \\
{\tt O-PAGB} & & O-rich post-AGB star                             &     \phantom{1}23       \\
{\tt C-PN}   & & C-rich planetary nebula                          &     \phantom{1}13       \\
{\tt O-PN}   & & O-rich planetary nebula                          &     \phantom{1}28       \\
{\tt RCrB}   & & R Coronae Borealis star                          &     \phantom{11}6       \\
{\tt GAL}    & & Galaxy                                           &     \phantom{11}8       \\
{\tt UNK}    & & Object of unknown type                           &     \phantom{11}5       \\
{\tt OTHER}  & & Object of known type                             &     \phantom{1}24       \\       
--- & {\tt B[e]}   & B[e] star                                    &       \phantom{11}8       \\
--- & {\tt BSG}    & Blue Supergiant Star                         &       \phantom{11}2       \\
--- & {\tt LBV}    & Luminous Blue Variable                       &       \phantom{11}3       \\
--- & {\tt YSG}    & Yellow Supergiant Star                       &       \phantom{11}1       \\
--- & {\tt WR}     & Wolf-Rayet star                              &       \phantom{11}7       \\
--- & {\tt Nova}   & Nova                                         &       \phantom{11}1       \\
--- & {\tt SNR}    & Supernova remnant                            &       \phantom{11}2       \\
\hline
\end{tabular}
\end{table}


The  1\,080 spectra in our IRS sample are comprised substantially of YSO and H\,{\sc ii} regions (see Table~\ref{tab:sourcecounts}): there are nearly 300 spectra of these types of object. A significant proportion of the remainder is made up of various evolutionary stages of post-Main Sequence low-mass stars: (post-)AGB stars and planetary nebulae are represented by 327 spectra between them. Massive stars contribute largely to the remnants: 71 {\tt RSG}s and a proportion of the {\tt STAR} class. Of course, this sample is heavily biased by the science goals of the projects listed in Table~\ref{tab:ArchiveProjects}. The YSO spectra are predominantly a product of program 40650 (PI: Looney), a catalogue of YSO spectra in the LMC published initially by \citet{Seale2009}. This was a targeted program, drawing heavily from the catalogues of YSO candidates produced by \citet{Gruendl2009}. Various evolved low- and intermediate- mass stars and RSGs appeared in program 3426 (PI: Kastner), which targeted the most luminous 8\,$\mu$m sources in the LMC, presented in a series of papers \citep{Buchanan2006,Kastner2008,Buchanan2009}.  Other programs with a high contingent of AGB stars include PIDs 3505 (PI: Wood) and 3591 (PI: Kemper). Post-AGB spectra come from PIDs 30788 (PI: Sahai) and 50338 \citep[PI: Matsuura,][]{Matsuura2014}, while spectra of planetary nebulae are the result of PIDs 103 (PI: Bernard-Salas) and 20443 (PI: Stanghellini). The 197 \SSp spectra of various targeted objects resulting from PID 40159 (PI: Kemper/Tielens) are presented by \citet{Kemper2010}, in Paper I and in various other publications.


\begin{figure*}
\centering
\includegraphics[trim=0.0cm 0cm 0cm 0cm, width=6.5in]{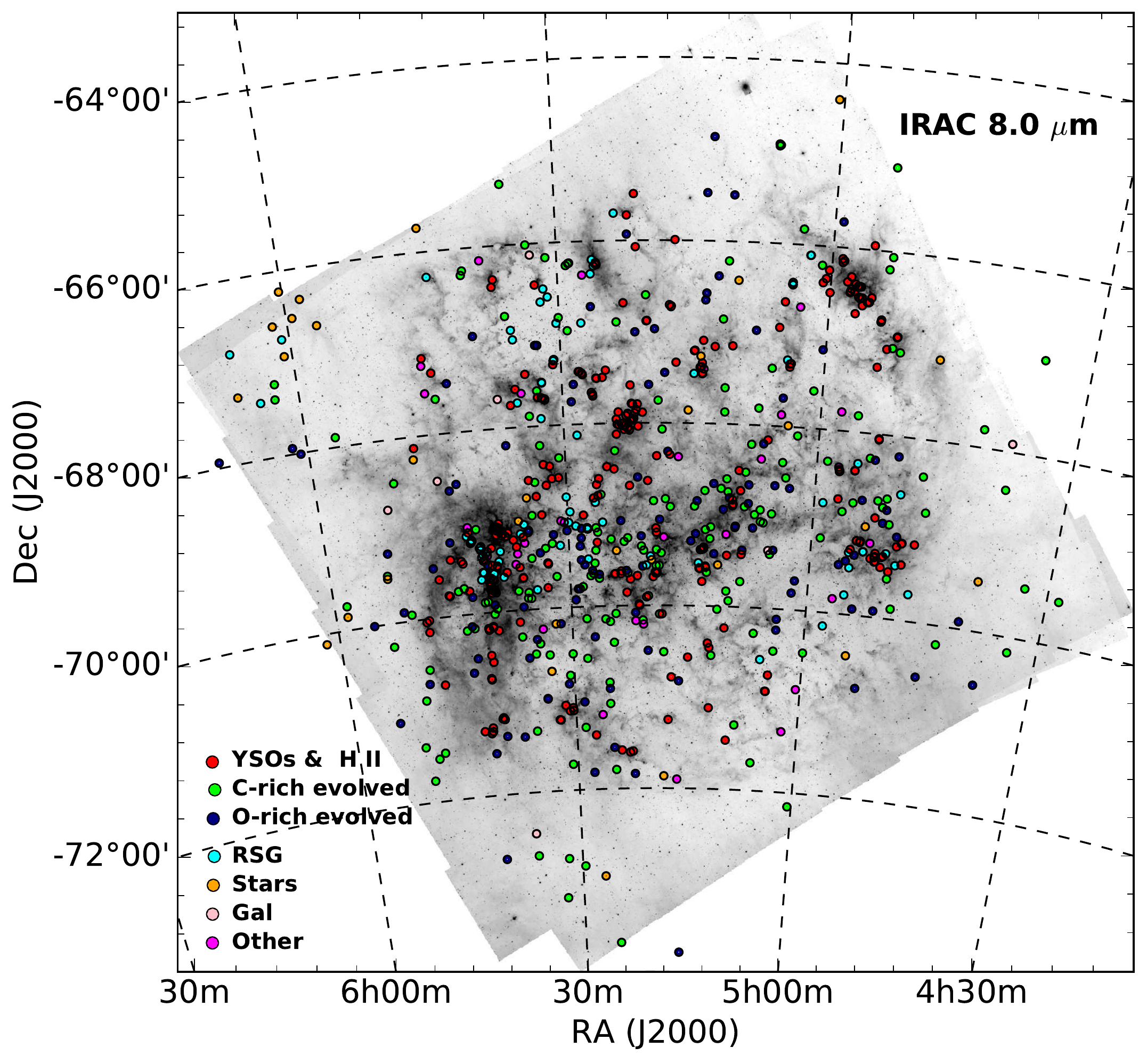}
\caption{The distribution of {\em Spitzer} staring-mode observations plotted over the SAGE 8 $\mu$m sky map. Different groups are given  different colours: YSO ({\tt YSO-1, YSO-2, YSO-3,  YSO-4, H\,{\sc ii}}), red;  C-evolved ({\tt C-AGB, C-PAGB, C-PN}), green; O-evolved ({\tt O-AGB, O-PAGB, O-PN}), navy; {\tt RSG}, cyan;  {\tt STAR}, orange;  {\tt GAL}, pink; and {\tt Other}, magenta. 
Only observations of point sources are shown. \label{fig:lmcdist}}
\end{figure*}

Figure~\ref{fig:lmcdist} shows that sources are distributed widely over the area of sky covering the LMC, with only a slight increase in density in the region of the stellar bar of the LMC. The star-forming regions of 30 Dor, N11 and N79 are well sampled. A loose cluster of stars is located towards the northeast of the LMC, the majority of which are foreground objects.


In the subsections below, we analyse the prevalent shapes and features of the spectra in different classification groups. Some classification groups are large enough such that sub-classes can be identified, based on nuances of the spectra. For each class or sub-class both a weighted mean and median spectra were determined. This accounts for the varying signal-to-noise in the spectra and the non-uniform resolution and wavelength coverage of each spectrum in that class. Each general class of object is discussed in detail below.

\subsection{The YSO sample and H\,{\sc ii} regions}

\begin{figure}
  \includegraphics[trim=3.0cm 7cm 0cm 8cm, width=4.2in]{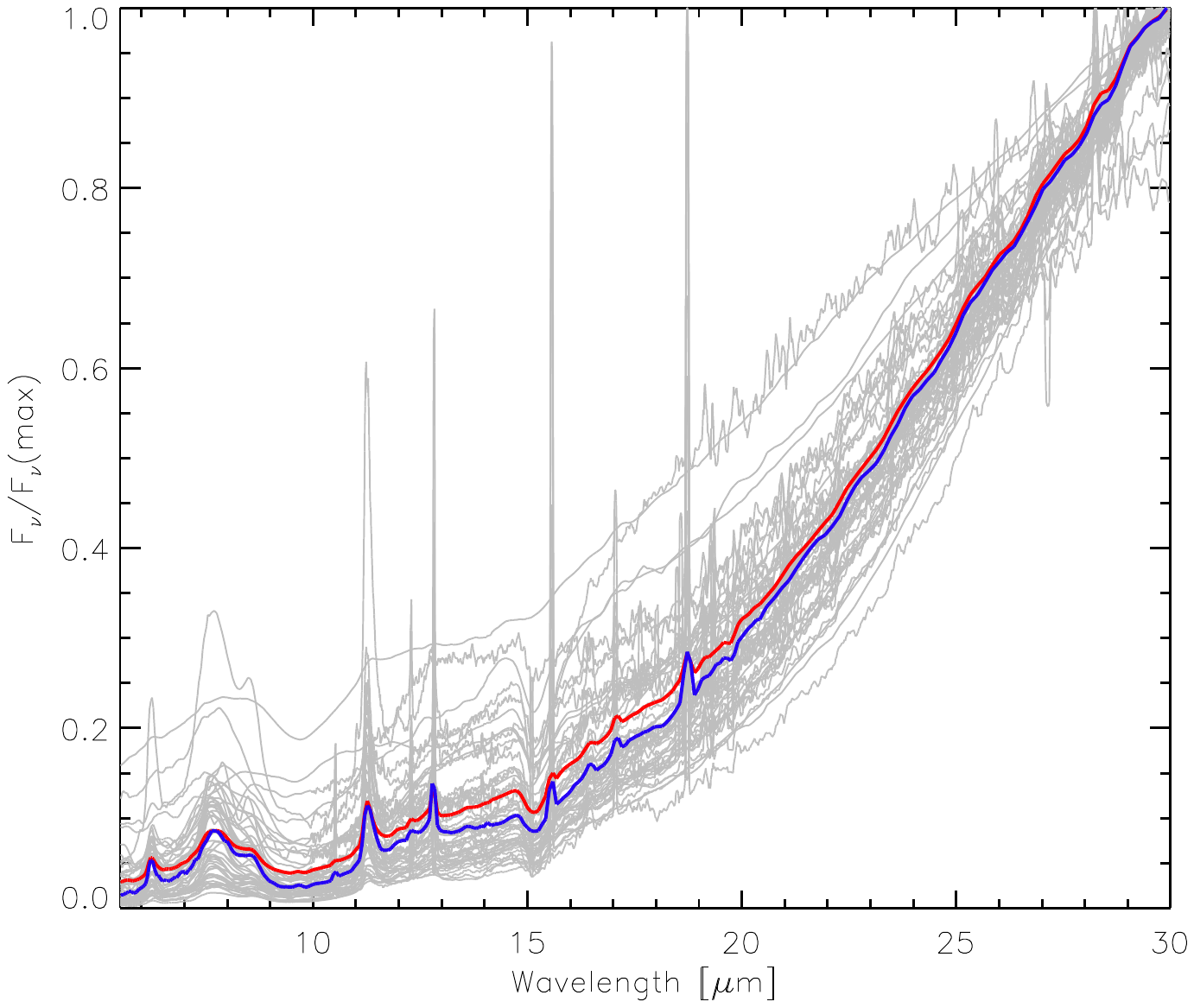}
  \includegraphics[trim=3.0cm 2cm 0cm 8cm, width=4.2in]{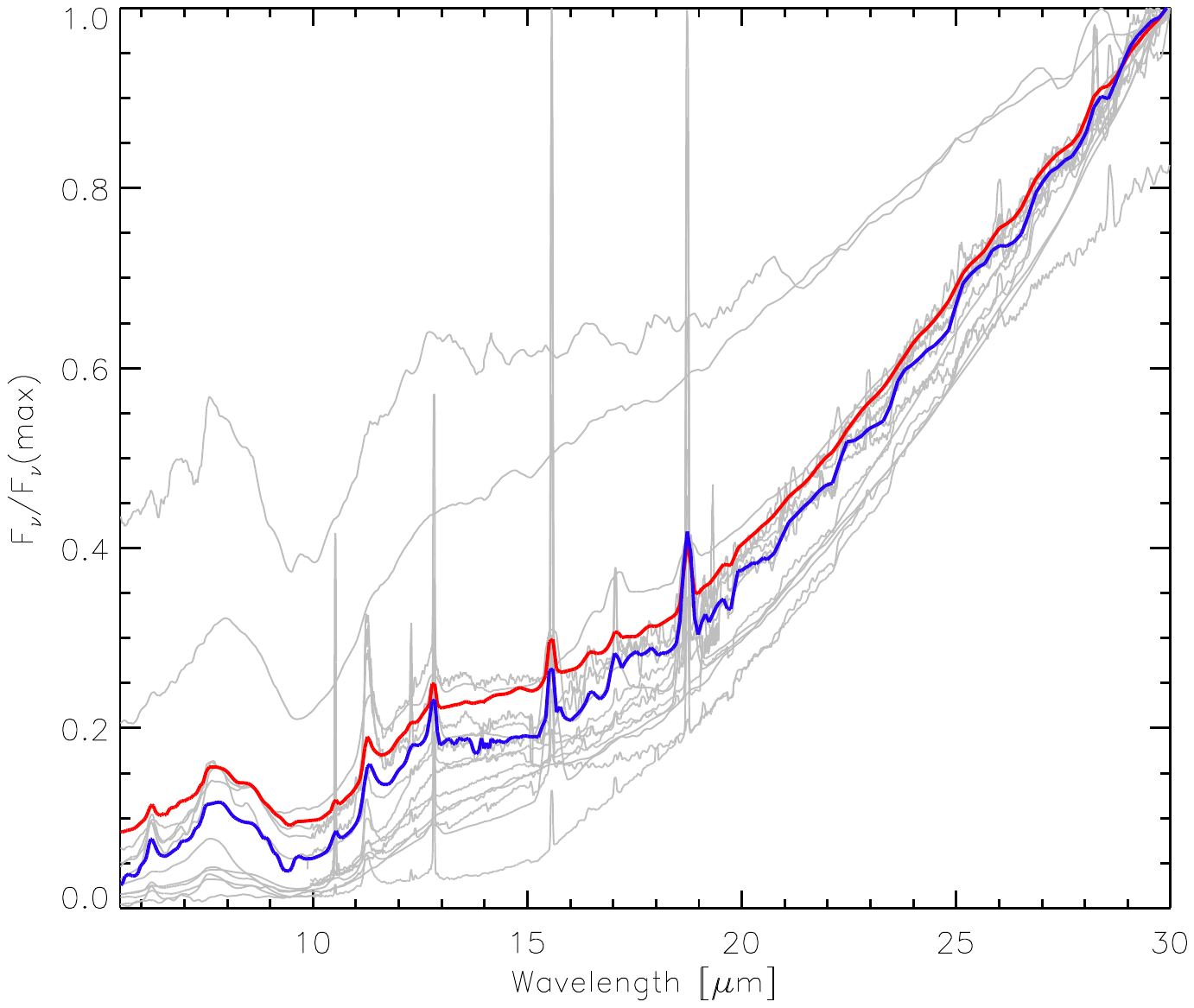}
   \caption{Mean (red line) and median (blue line) spectra for the {\tt YSO-1} (top) and {\tt YSO-2} (bottom) classes. The distinctive 15\,$\mu$m ({\tt YSO-1}) and 10\,$\mu$m ({\tt YSO-2}) absorption features can be seen. The grey lines show the spectral diversity within each class. 
\label{fig:yso12avg}}
\end{figure}

\begin{figure}
  \includegraphics[trim=3.0cm 7cm 0cm 8cm, width=4.2in]{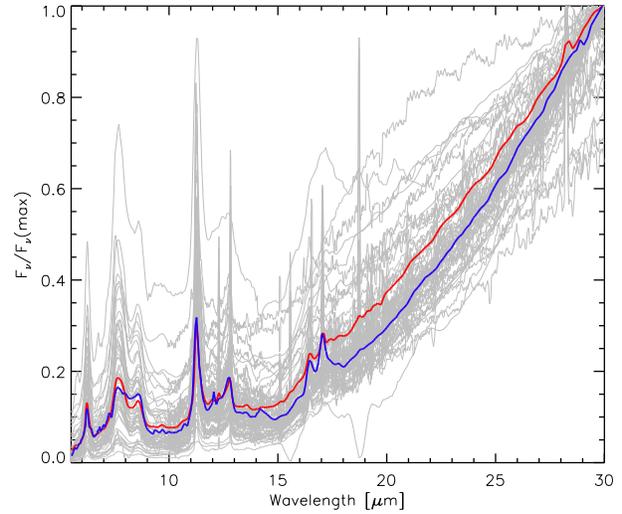}
  \includegraphics[trim=3.0cm 2cm 0cm 8cm, width=4.2in]{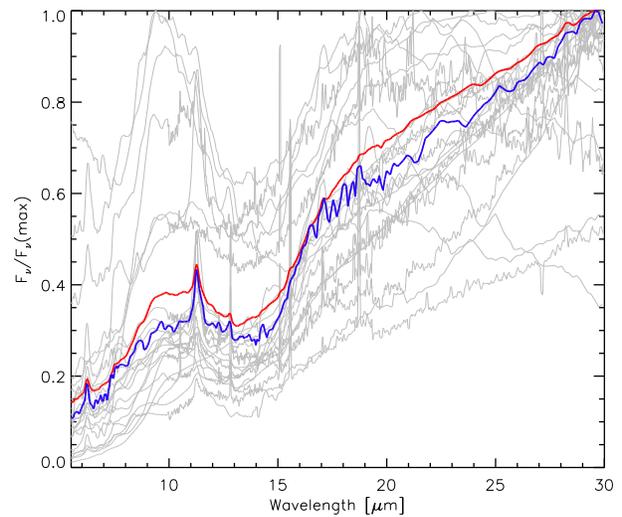} 
\caption{Averaged spectra for the {\tt YSO-3} (top) and {\tt YSO-4} (bottom) classes.  Characteristic UIR/PAH features are seen in the {\tt YSO-3} spectrum, whilst {\tt YSO-4}s show significant 10 $\mu$m features on a rising spectrum at longer wavelengths.\label{fig:yso34avg}}
\end{figure}

\begin{figure}
  \includegraphics[trim=3.0cm 7cm 0cm 8cm, width=4.2in]{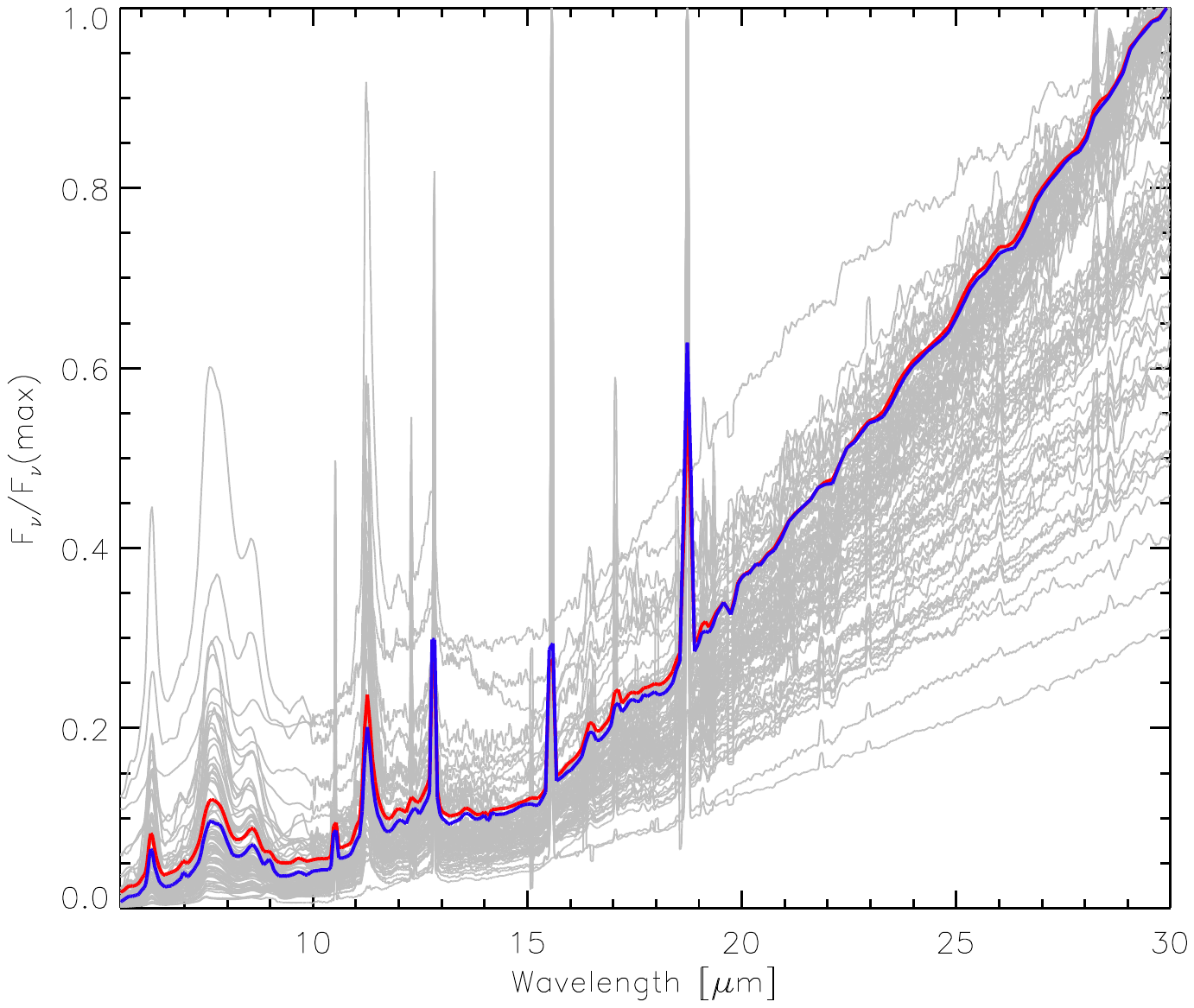}
  \includegraphics[trim=3.0cm 2cm 0cm 8cm, width=4.2in]{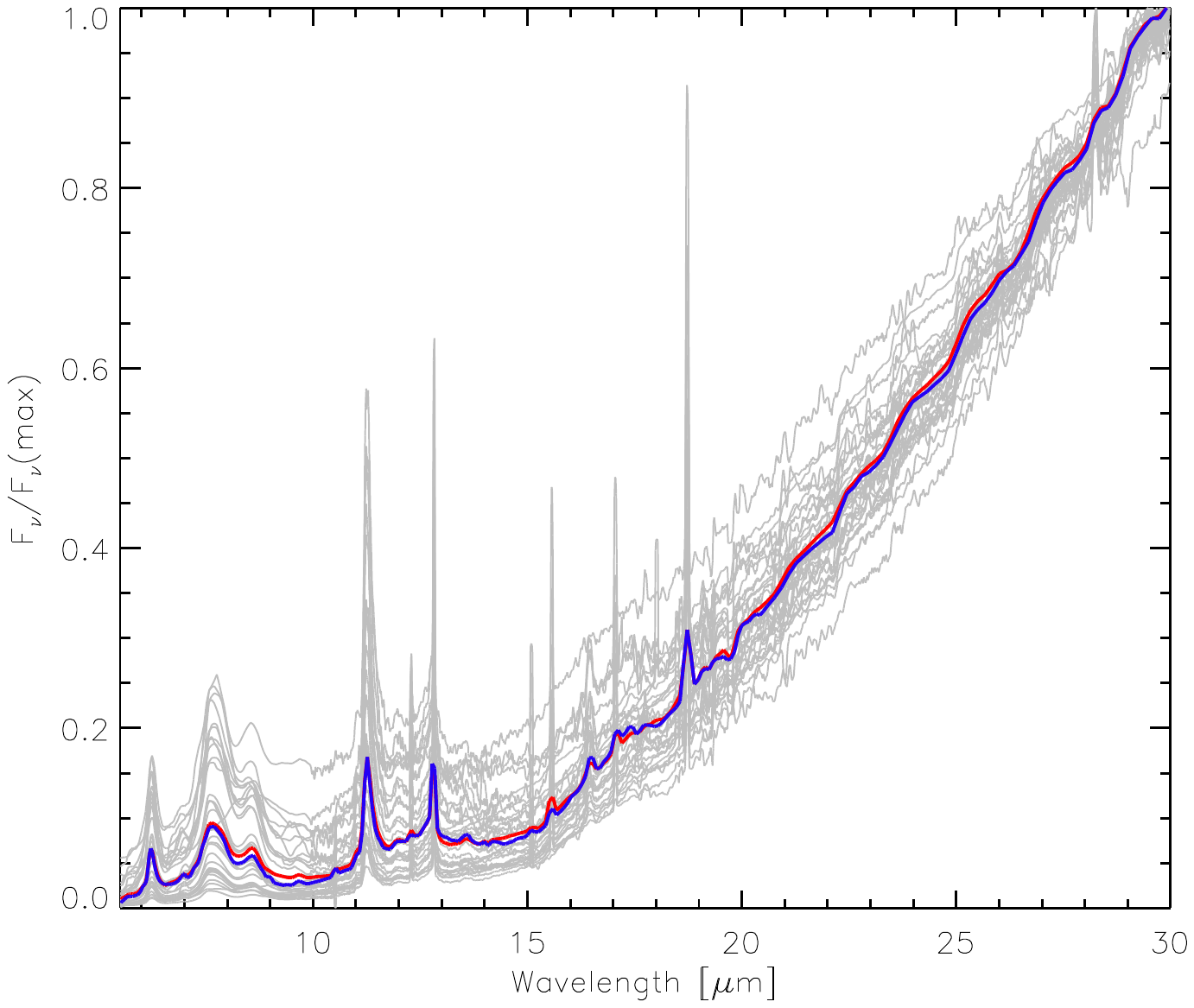}
\caption{Averaged spectra for the {\tt H\,{\sc ii}} (top) and {\tt H\,{\sc ii} }/{\tt YSO-3} (bottom) classes. The spectra have UV-pumped PAHs and fine-structure atomic emission lines. \label{fig:hiiyso3avg}}
\end{figure}

The classification scheme described in Section~\ref{sect:classmethods} can be understood in terms of an evolutionary sequence for massive YSOs, with the {\tt YSO-1} and {\tt YSO-2} classes representing the earlier, more embedded stages. As the young massive objects evolve, they start to interact with and shape their environments, creating ultra-compact or compact H\,{\sc ii} regions; such objects populate the {\tt YSO-3} class. However, several factors make a classification based on IRS spectra alone complicated for YSOs (see also the discussion in \citealt{Oliveira2013}, and \citealt{Seale2009}). The IRS angular resolution varies between 3 and 10\arcsec; at the distance of the LMC, this corresponds to 0.6--2.4\,pc. As shown recently by \citet{Ward2016, Ward2017} for a small sample of YSOs in the SMC, {\it Spitzer}-identified YSOs can be further resolved into multiple sources of different masses (and inferred evolutionary stages) when observed at higher spatial resolution.  Furthermore, in these complex environments often there is a strong environmental contribution to PAH and fine-structure emission that cannot be disentangled, and the IRS spectra sample different spatial scales at different wavelengths. As a result, objects with similar IRS classifications can reveal different spectral properties when studied in more detail.

Another potential complication arises from the intrinsic phenomenological overlap between the {\tt YSO-3} and {\tt H\,{\sc ii}} classes. Since compact H\,{\sc ii} regions are powered by young massive YSOs, the separation between the classes is somewhat artificial. Indeed \citet{Seale2009} include emission line objects in their sample of LMC YSOs (their classes PE and E, see discussion below), and so do \citet{Oliveira2013} for a sample of SMC YSOs. There are a number of sources that we could not unambiguously classify as {\tt YSO-3} or {\tt H\,{\sc ii}} and are thus placed in an intermediate class {\tt YSO-3}/{\tt H\,{\sc ii}}. Therefore both the YSO and {\tt H\,{\sc ii}} classes are discussed in this section.

Galactic HAeBe objects exhibit a great variety of spectral and SED  properties, reflecting effects like disk geometry, evolutionary stage and mineralogy (see e.g.~\citealt{Acke2004} and \citealt{Keller2008}, for studies using {\it ISO} and IRS spectra respectively). Most sources exhibit silicate in emission at 10\,$\mu$m, but occasionally this feature is seen in absorption or it is absent altogether. Many sources exhibit PAH emission but not all. The continuum can either rise over the IRS range or be somewhat flatter. Based on the sources analysed by \citet[][their Fig.1]{Keller2008}, our decision tree would classify sources with silicate emission as {\tt YSO-4}; however sources with a flatter/falling SED could also be placed in the {\tt O-}AGB class. Sources dominated by PAH emission would be classified as {\tt YSO-3}, while sources with essentially a featureless continuum over the IRS range would be  difficult to classify. Only abundant ancillary data would be able to refine these classifications. Therefore, the sources classified as {\tt YSO-4} are Magellanic analogues of Galactic HAeBes, but constitute a somewhat restricted sample.

There are over 300 spectroscopically-confirmed YSOs in the LMC, selected from large photometric samples \citep{Whitney2008,Gruendl2009}. The first massive YSO spectroscopically identified in any extragalactic environment was a serendipitous discovery by \citet{vanLoon2005b} using IRS and VLT-ISAAC spectra. The overwhelming majority of spectroscopic YSOs have been studied using IRS by \citet{Seale2009,Seale2011}, \citet{Oliveira2009,Oliveira2011} and by us in Paper I. A smaller sample using {\it AKARI} spectra was analysed by \citet[and references therein]{Shimonishi2010}.

The objects analysed by \citet{Oliveira2009,Oliveira2011} were classified in Paper I, following the same classification scheme described here. The sample of \citet{Seale2009} includes 294 sources, of which 277 objects were classified as YSOs. Their classification scheme made use of spectral features (PAH and fine-structure emission, and silicate absorption) to classify the objects using principal component analysis (PCA). Accordingly they classified YSOs as follows: S/SE (silicate absorption dominated, without/with fine-structure emission) respectively 12 and 5 sources; P/PE (PAH dominated without/with fine-structure emission) respectively 100 and 142 sources; 2 E sources (emission-line dominated); 11 F sources (weak features only), and 5 U sources (incomplete spectral range that does not allow PCA classification). Their sample also includes 8 type O sources (silicate in emission) that could be HAeBe analogues (not included in the 277-strong YSO sample). We reclassify all 277 sources according to our scheme.

In total, 168 sources are classified as YSOs, respectively 53, 14, 80 and 21 in classes {\tt YSO-1}, {\tt YSO-2}, {\tt YSO-3}, {\tt YSO-4}. Respectively 134 and 35 sources are classified as {\tt H\,{\sc ii}} and {\tt H\,{\sc ii}/YSO-3}. 

The \citet{Seale2009} classification does not make use of the ice features to classify YSOs (the 41 ice-rich sources from this sample are described in more detail by \citealt{Seale2011}). As a result the 53 most embedded and colder {\tt YSO-1} sources ended up in a variety of Seale's classes mentioned above.

Without detailed spectral modeling it can be difficult to identify weak silicate absorption in a spectrum with PAH emission. Most {\tt YSO-2} sources were previously classified by \citet{Seale2009} in the S (or SE) and P (or PE) groups; one additional source was classified in the F group (weak features); another source was left unclassified.

Most {\tt YSO-3},  {\tt H\,{\sc ii}/YSO-3} and {\tt H\,{\sc ii}} sources from Seale's sample belong to either the P (or PE) group, with a few sources belonging to E (emission line) or F (weak features) groups. This reinforces the idea of how difficult it is to classify more evolved YSOs and distinguish them from compact H\,{\sc ii} regions. Any user of the spectral classification discussed in this paper that is interested in YSOs should thus consider {\tt H\,{\sc ii}/YSO-3} and {\tt H\,{\sc ii}} sources as well.

Finally, those {\tt YSO-4} sources originating from Seale's sample were either classified in groups O (silicate emission) or P. It should be noted that group O sources in Seale's classification are not necessarily HAeBe analogues, they may be oxygen-rich AGB stars as discussed elsewhere in this paper.

\subsection{The massive evolved star sample}

\begin{figure}

\includegraphics[trim=3.0cm 7cm 0cm 16cm, width=4.2in]{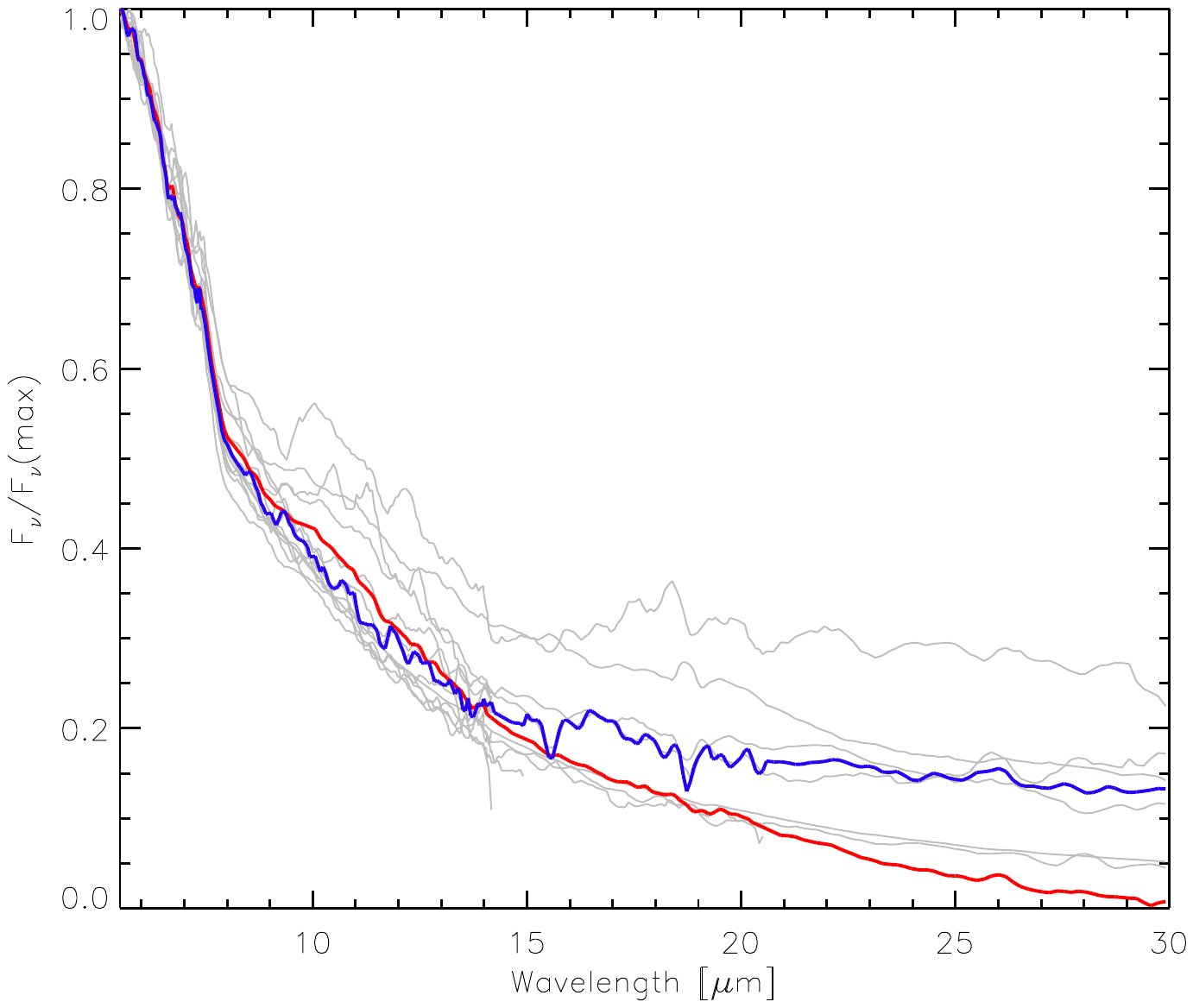}
\includegraphics[trim=3.0cm 7cm 0cm 9cm, width=4.2in]{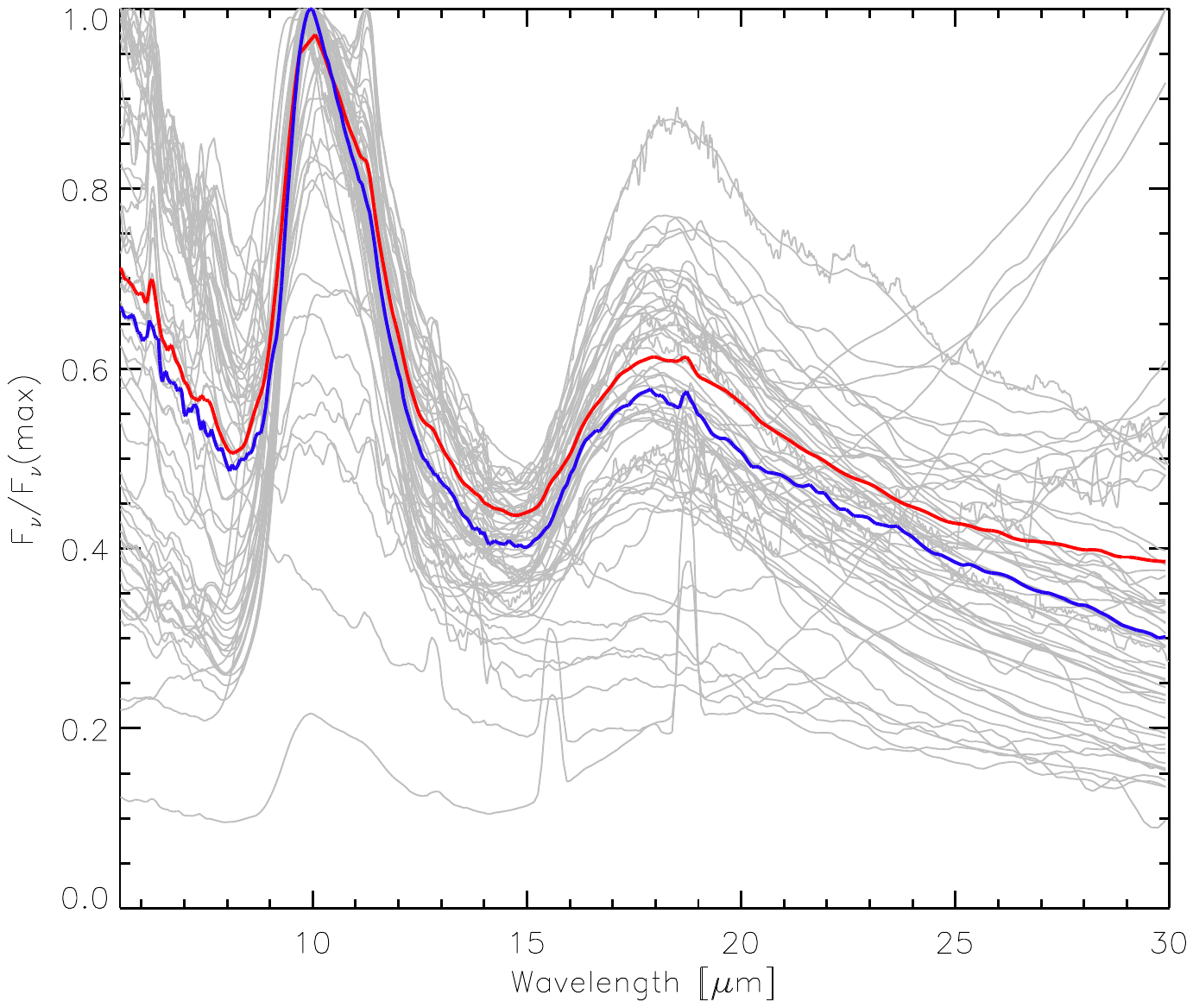}
\includegraphics[trim=3.0cm 3cm 0cm 9cm, width=4.2in]{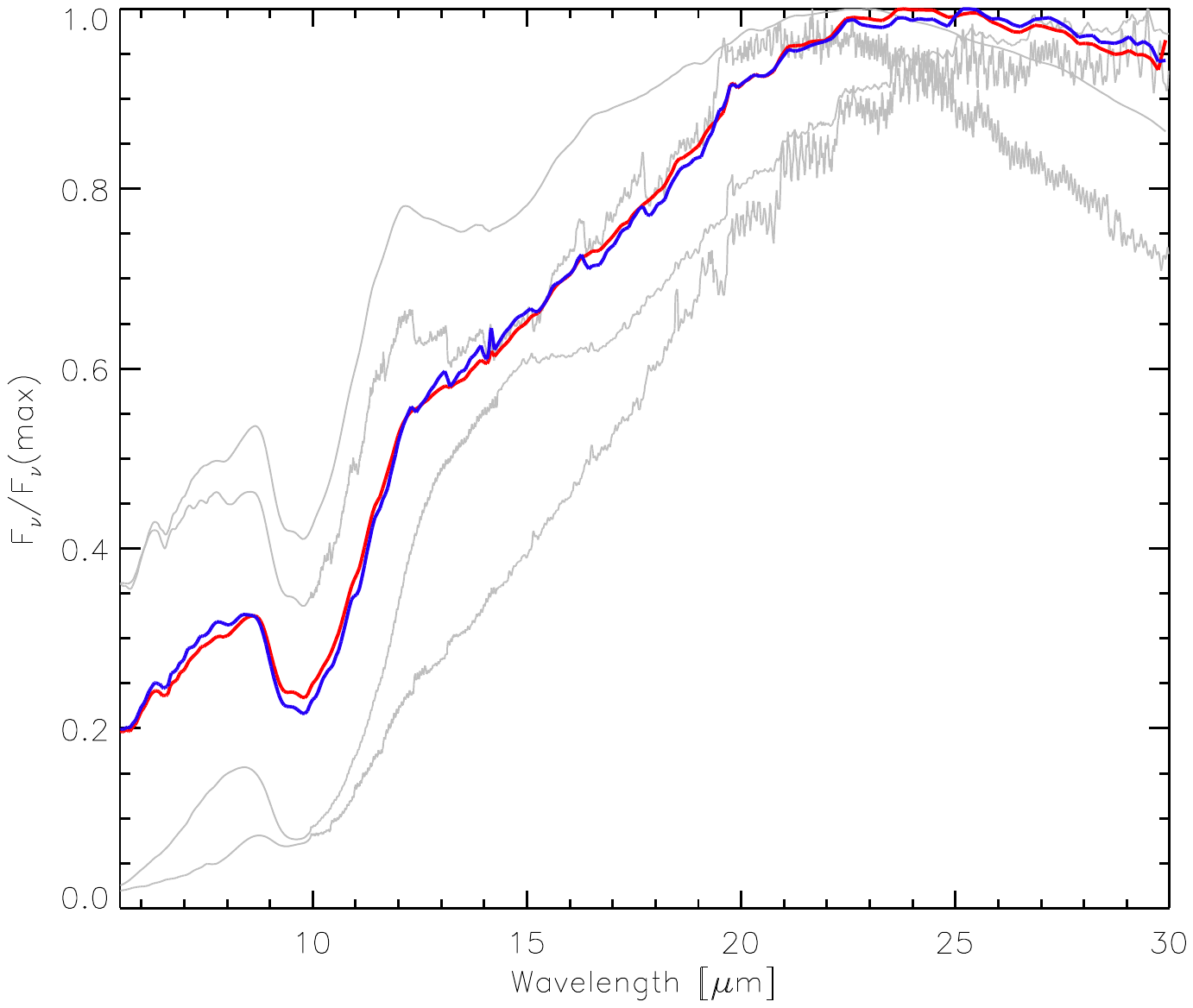}

\caption{Averaged spectra for the {\tt RSG} class. The sample has been sub-divided into three classes: sources with little IR excess that show an inflection at 8 $\mu$m due to SiO, or have a weak 10 $\mu$m emission feature (top); sources with strong 10- and 18 $\mu$m silicate emission (middle); and sources which have silicate features in absorption (bottom). \label{fig:rsgavg}}
\end{figure}

\begin{figure}
  \includegraphics[trim=3.0cm 2cm 0cm 8cm, width=4.2in]{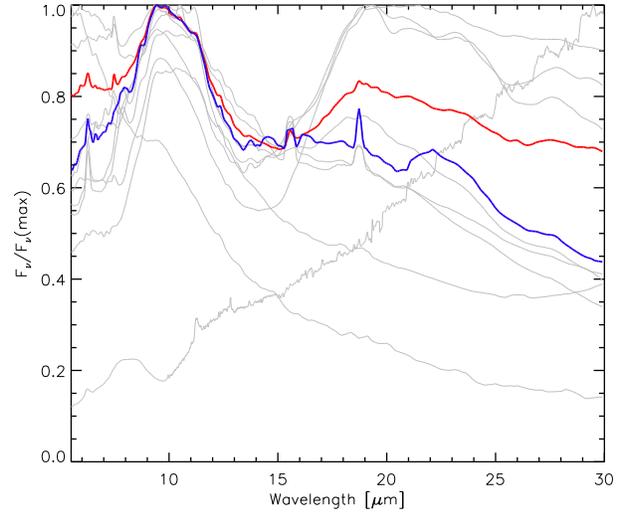}
\caption{Averaged spectra for the {\tt B[e]} class. Emission from  amorphous and crystalline silicates are seen in the spectra. \label{fig:Beavg}}
\end{figure}

\begin{figure}
  \includegraphics[trim=2.0cm 2cm 0cm 4cm, width=6.5in]{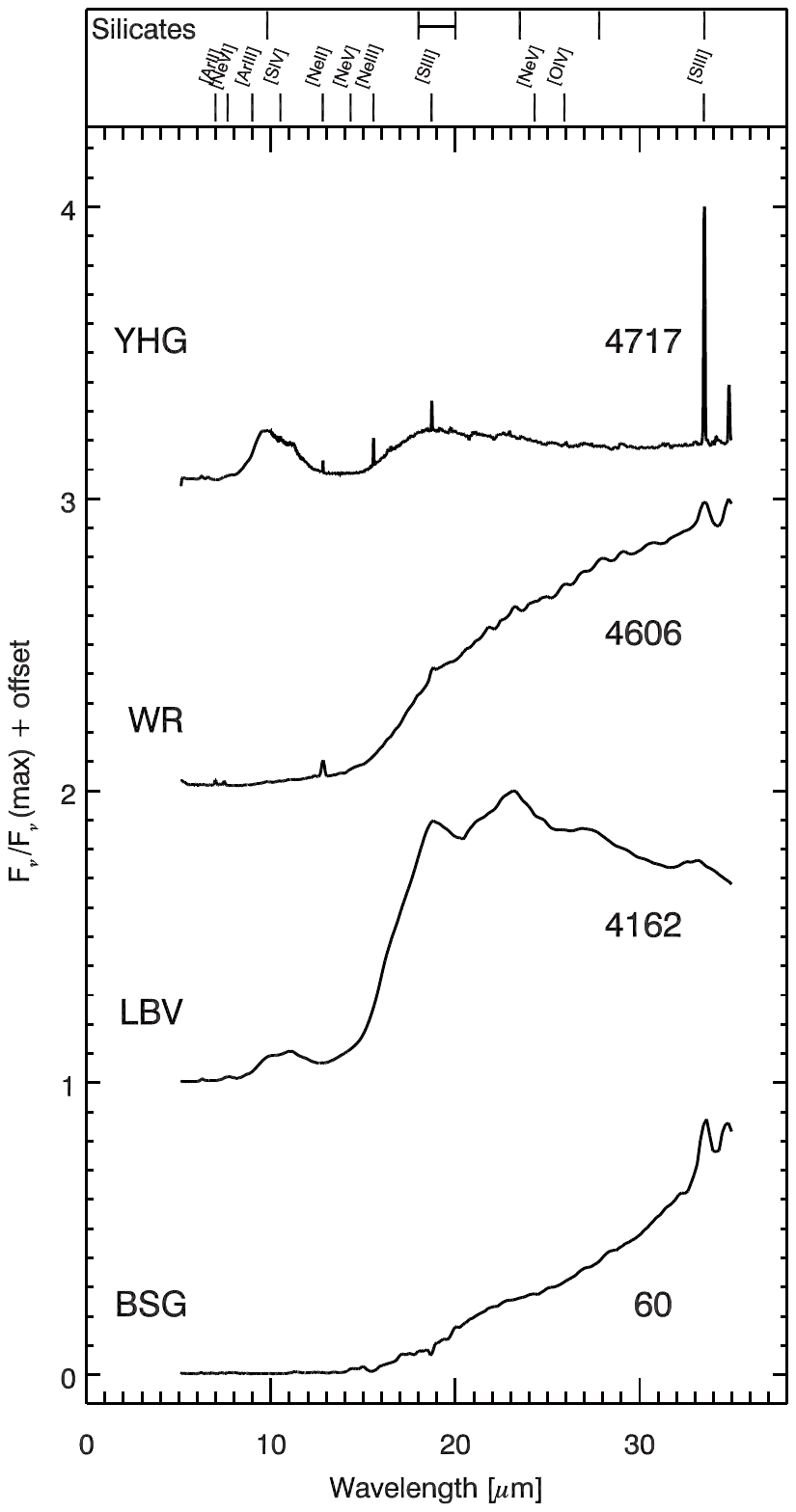}
\caption{Example IRS spectra of objects in the massive star category. Shown here from bottom to top are the spectra of a {\tt BSG}, a {\tt LBV}, a {\tt WR} star and a {\tt YSG}, labelled with their SSID number. The WR and the YSG stars have atomic lines in their spectrum. Silicates are seen in the LBV spectra. The infrared excess in the WR star and BSG objects are caused by dust emission.   \label{fig:bsgavg}}
\end{figure}



The red supergiant ({\tt RSG}) spectra which show strong 10- and 18 $\mu$m emission (Fig.~\ref{fig:rsgavg}) in many cases also show the presence of an 11.3 $\mu$m PAH feature. This feature has previously been observed in the spectra of five Galactic red supergiants \citep{Sylvester1998,Verhoelst2009}, in LMC supergiant Massey SMC 59803 \citep{Sloan2008} and in four red supergiants in the SMC \citep{Kraemer2017}. It arises in a manner akin to the Pleiades effect, with the supergiants embedded in a thin cloud of diffuse material \citep{Sheets2013,Adams2013}. At distances of 50--60 kpc this emission cannot be resolved, but there is sufficient surrounding material so that PAH features are visible. Once PAH molecules are liberated from their matrix (requiring UV irradiation), the free-floating molecules themselves can be excited with optical radiation \citep[c.f.,][]{Sloan2007,Li2002},  although \citet{Verhoelst2009} argued that the PAH molecules are part of the RSG circumstellar environment.  Other spectral features occurring in some of the RSG spectra include crystalline silicates \citep{Jones2012}, most notably at 23\,$\mu$m, and  atomic emission lines. While most spectra are of hot dusty objects, showing strong silicate features on top of an infrared excess, in some cases, a steeply-rising continuum is visible, indicating the presence  of colder detached dust shells, or ambient interstellar dust.

As with the YSOs, some of the {\em Spitzer} spectroscopy presented in this paper has already been published previously, for instance in the studies by \citet{Buchanan2006, Buchanan2009}.  In their sample of 60 objects, the oxygen-rich objects are dominated by RSGs, and they arrived at a total count of 22 RSGs. Four objects have the physical appearance of an RSG, but their apparent brightness is such that if they were placed at the LMC distance, they would have unrealistic luminosities.  Hence, the authors concluded that these objects (MSX LMC 412, 1150, 1677 and 1686) are foreground objects. Without distinguishing between AGB stars and RSGs, \citet{Sloan2008} presented 15 IRS spectra of oxygen-rich evolved stars in the LMC, and describe the mineralogical spectral features in their spectra. They specifically mention the presence of crystalline silicates, and the currently unidentified 13 and 14 $\mu$m features. \citet{Groenewegen2009} compiled a sample of several observing programs targeting RSGs and AGB stars in the LMC. They found that LMC RSGs have mass-loss rates very similar to Galactic RSGs \citep[as determined by ][]{Verhoelst2009}, a finding backed up recently by \citet{Goldman2017}. Note, though, that \citet{Mauron2011} found that LMC mass-loss rates do not show the same behaviour as Galactic rates for RSGs. \citet{Jones2012} compiled all 71 IRS spectra of LMC RSGs as part of their larger sample to study the emergence of crystalline silicates in oxygen-rich dust shells. They found that 24\% of their sample of RSGs in the LMC, SMC and Milky Way show crystalline silicates, versus 56\% for O-rich AGB stars. 

The more deeply-embedded RSGs have very red SEDs, which made them excellent targets for far-infrared follow-up. Indeed, the MIPS-SED sample of the \SSp program included  two of the most extreme RSGs: \objectname{WOH G64} and \objectname{IRAS 05280-6910} \citep{vanLoon2010}. \objectname{IRAS 05280-6910} was also detected as a point source in the first data release of the HERITAGE \emph{Herschel} survey of the LMC \citep{Boyer2010}, with a radiative transfer model reproducing the 10 $\mu$m silicate feature in absorption. A further two extreme RSGs were detected with \emph{Herschel}, the aforementioned \objectname{WOH G64} and \objectname{IRAS 05346-6949} \citep{Jones2015b}, with \objectname{WOH G64} considered to be possibly the largest star known \citep{Levesque2009}, having an extremely dense circumstellar dust shell \citep{Ohnaka2008}. For further background on these extremely embedded RSGs, \citet{Jones2015b} provide a detailed literature study.


Three {\tt LBV}s were observed with the IRS in the LMC: \objectname{S Dor}, \objectname{R 71}, and \objectname{R 110}. This category typifies how ancillary information from the literature was used to classify the sources. All were placed in the {\tt OTHER} category, and none of their IRS spectra look alike. \objectname{S Dor}, the prototype LBV,  shows a largely featureless, cool dust spectrum with a few weak forbidden lines. It also has (blue) continuum emission for $<$ 8\,$\mu$m  either from warm dust or perhaps the star itself. \objectname{R 71} also has a cool dust spectrum, but has prominent crystalline silicate features \citep{Voors1999}, as well as PAH emission \citep{GuhaNiyogi2014}. \objectname{R 110} has weak blue continuum emission with a rich set of PAH features and a sharply rising continuum beyond 20\,$\mu$m  indicative of cold dust. 


Several Wolf-Rayet ({\tt WR}) stars were observed with the IRS in the LMC, and they are grouped together here based on their known optical classifications. 
All show emission lines in their IRS spectra, either fine-structure lines or hydrogen recombination lines. At least two have emission from crystalline silicates at 23 and 33\,$\mu$m. All have rising spectra at longer wavelengths ($>$ 14 $\mu$m), indicative of cool dust. One, \objectname{Brey 13}, also has a falling continuum in the 5--14 $\mu$m regime from either warmer dust or a stellar continuum; this star also shows the strongest hydrogen recombination lines.


B[e] stars are young supergiants characterised by strong Balmer line emission, low excitation permitted lines (e.g., Fe II), forbidden [Fe II] and [O I] optical lines, and a strong infrared excess from dust emission. Supergiant B[e] stars have a luminosity $\geq$ 10$^4$\, L$_\odot$ \citep{Lamers1998}.  
The mid-infrared spectra of B[e] supergiant stars show emission from silicate dust and occasionally PAHs \citep[see][]{Kastner2010}. This circumstellar dust emission originates in flattened structures that were originally thought to be expanding along the equatorial plane \citep[e.g.,][]{Zickgraf1986} but are instead, at least in some cases, circumstellar disks orbiting the B[e] supergiants. Such a disk might be the result of mass that was ejected by the B[e] star during a prior red supergiant phase and subsequently trapped into circumbinary orbit by a lower-mass companion star \citep{Kraus2010,Kastner2010}.

Our sample includes nine {\tt B[e]} stars.  Optical spectroscopy indicates that eight are supergiant B[e] stars \citep{Lamers1998, Gummersbach1995, Kastner2010}, and the ninth, \objectname{IRAS 04530-6916} (OBJID 239), is a star with B[e] emission characteristics \citep{vanLoon2005}. However, as  \citet{vanLoon2010} note, this object also has the characteristics of an H\,{\sc ii} region (red continuum with strong PAH emission, amorphous silicate absorption, and low-excitation atomic emission lines), so its classification is less certain \citep[see also the discussion in][ on this source]{Sloan2008}.  

\cite{Kastner2006} find the infrared excess luminosities of RMC 126 and RMC 66 to be about 5\% and 18\% of their measured bolometric luminosities, such that the stellar flux contributes negligibly to the total emission at IRS wavelengths. Though many permitted and forbidden emission lines are seen in optical spectra of B[e] stars, almost no such lines are seen in the mid-infrared. However, \citet{Kastner2006} note the IRS spectrum of RMC 126 shows a 7.46 $\mu$m Pfund-alpha emission line, whilst \objectname{HD 38489} (OBJID 715) exhibits some low-excitation atomic emission lines (e.g., [Ne III] and [S III]).


Our sample also includes two B supergiants ({\tt BSGs}) and a yellow supergiant  (YSG).
The BSGs (OBJID 50 and 69) were discussed in Paper I; these stars have spectra which rise longward of 15 $\mu$m due to dust emission and their SEDs are double-peaked falling to a minimum at $\sim$8\,$\mu$m. Surprisingly, far-IR dust emission from \objectname{BSDL 923} (OBJID 69) was detected by \citet{Jones2015b}; this could again be the Pleiades effect \citep{Sheets2013,Adams2013} or cold dust emission due to a circumstellar disk or torus surrounding the central star. 

\objectname{HD 269953} (OBJID 713) is a well studied YSG. It is a confirmed post-RSG object \citep{Oksala2013} and is amongst the brightest ($M_{\rm bol}$ $\sim -$9.4) stars in the LMC. Its circumstellar dust emission, first characterized by \citet{Roche1993}, displays prominent emission bands at 10 and 18 $\mu$m due to amorphous silicate dust which may arise from a circumstellar disk. Also present in its spectrum are crystalline silicate features at 11.3 and 23 $\mu$m, although these are difficult to characterise at longer wavelengths due to fringing effects in the spectrum.

\subsection{The oxygen-rich AGB sample}

\begin{figure}  
  \includegraphics[trim=3.0cm 7cm 0cm 16cm, width=4.2in]{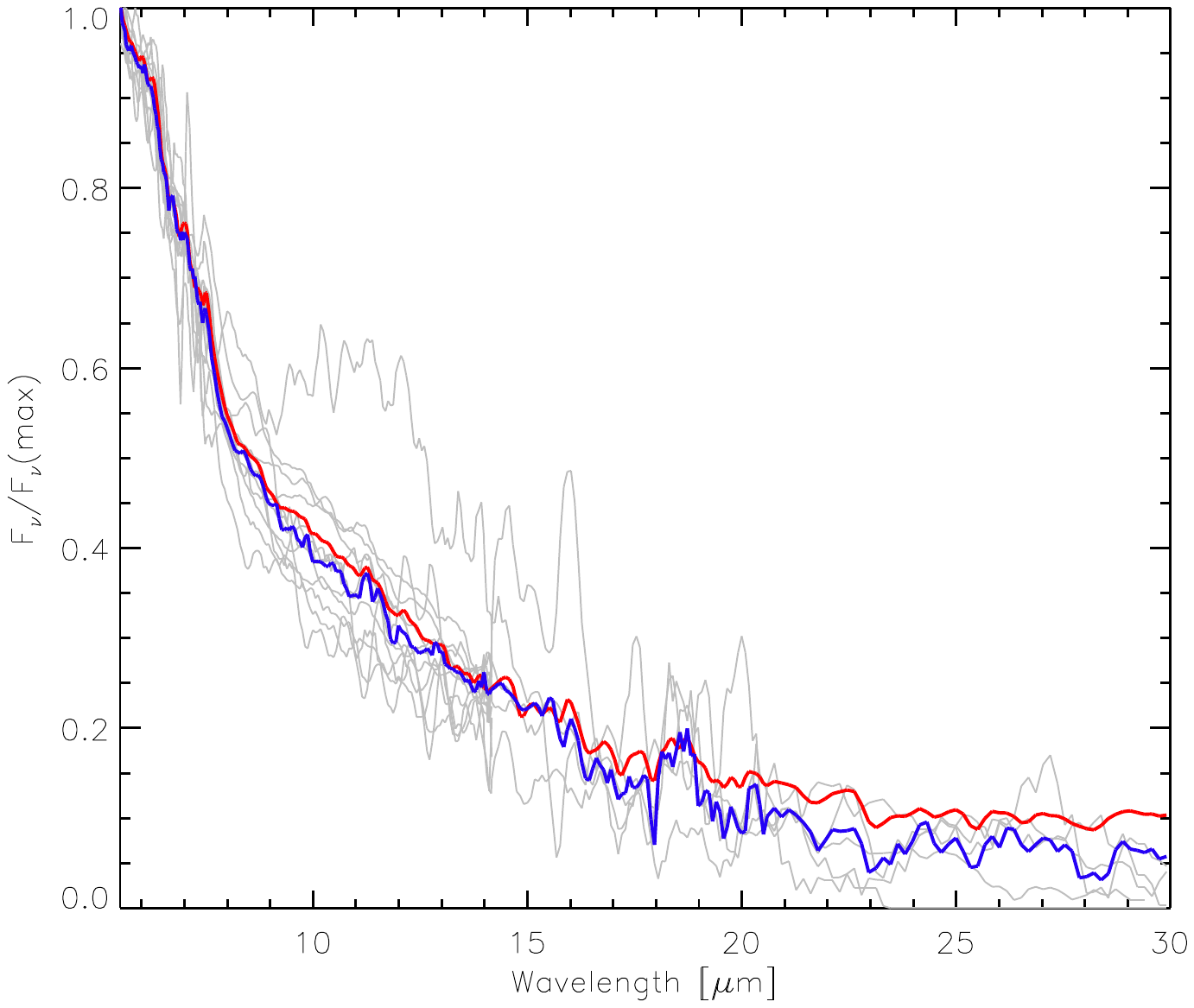}
  \includegraphics[trim=3.0cm 7cm 0cm 9cm, width=4.2in]{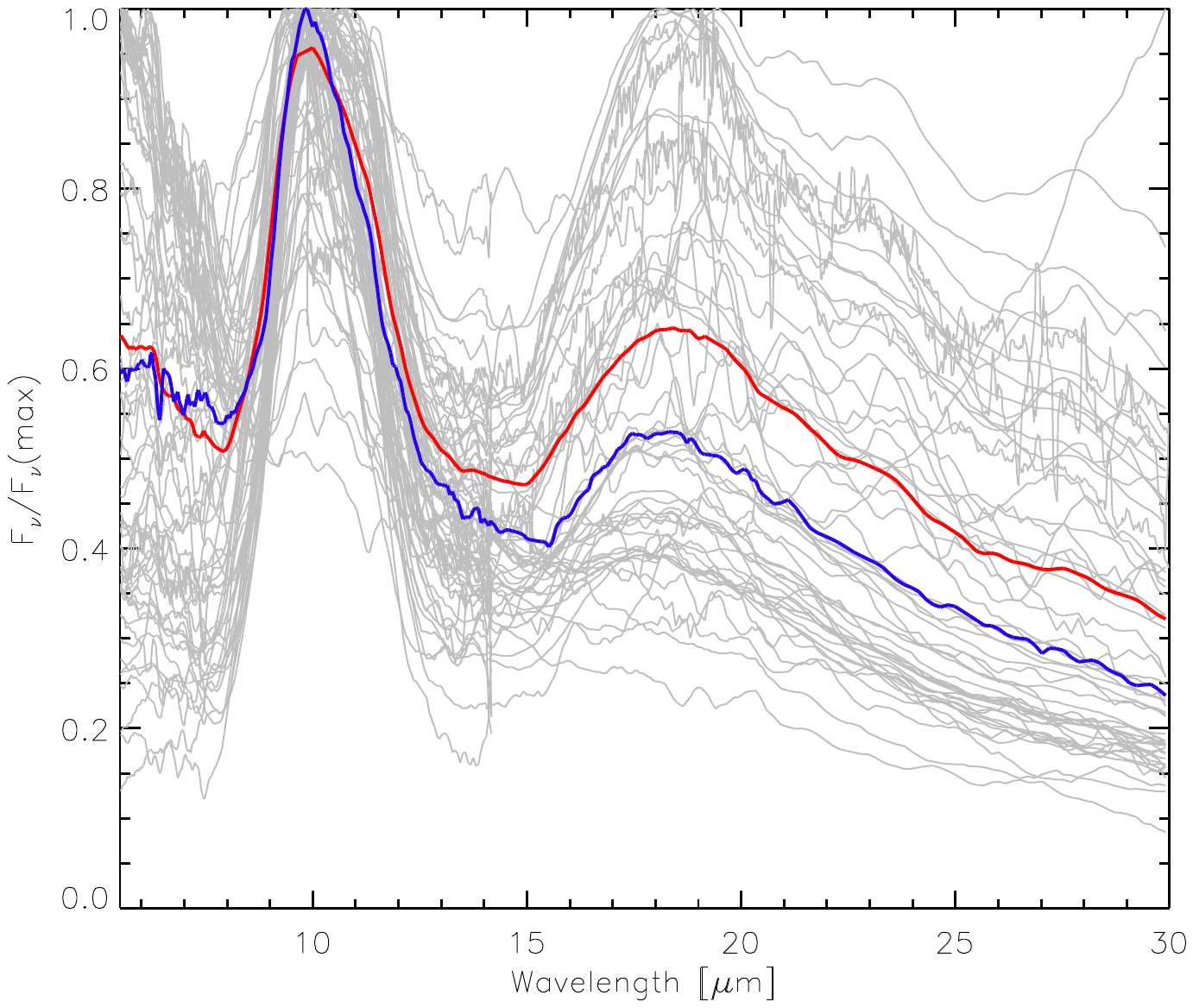}
  \includegraphics[trim=3.0cm 3cm 0cm 9cm, width=4.2in]{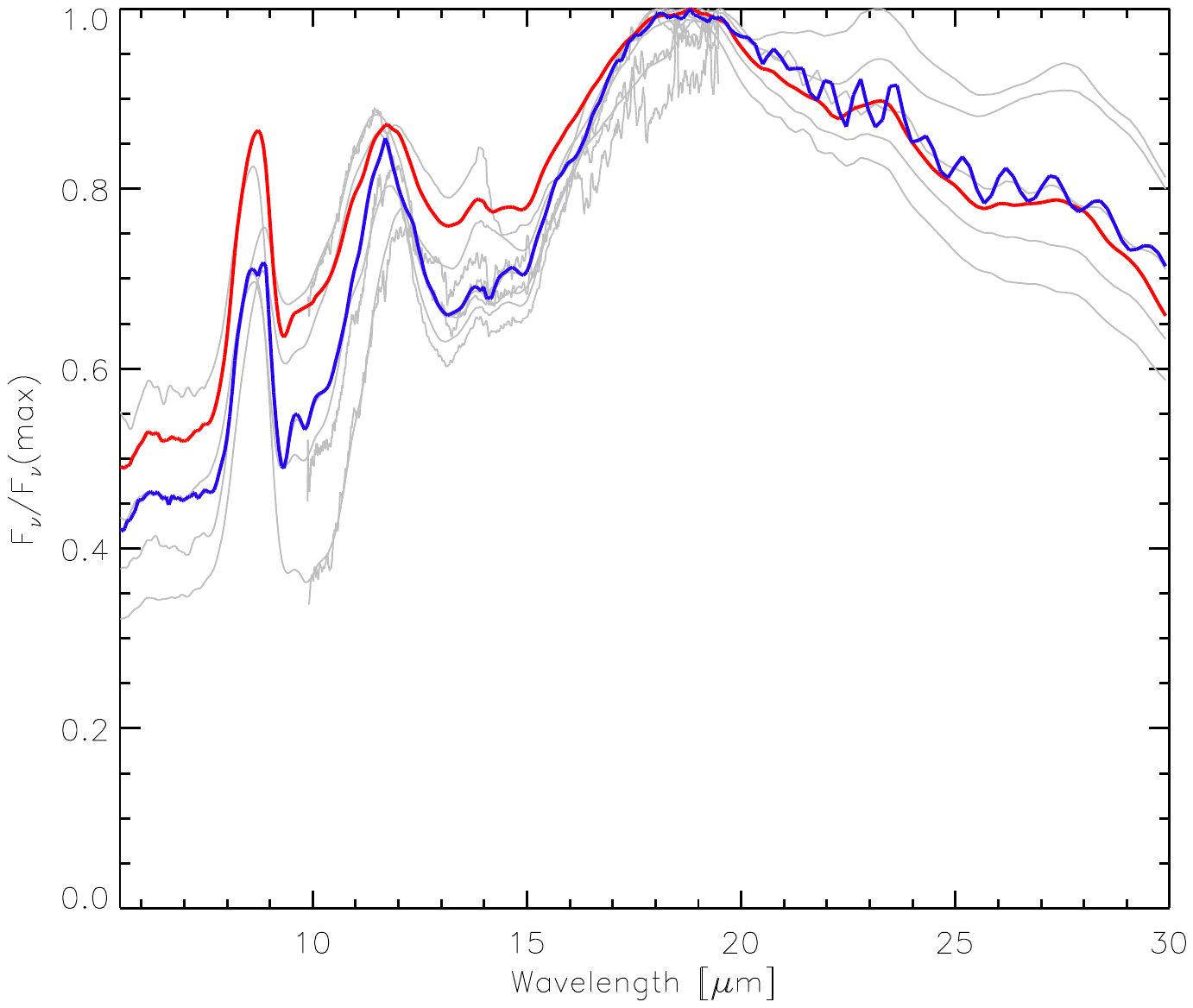}
\caption{Averaged spectra for the {\tt O-AGB} class, split into sub-classes showing SiO molecular features (top), amorphous silicate emission (middle) and self-absorption (bottom). Crystalline silicates are present in the second and third subclasses. \label{fig:O-AGBavg}}
\end{figure}

Seventy-eight unique {\tt O-AGB} stars have been observed with the IRS in the LMC field; we have separated these sources into three groups based on the opacity of their circumstellar envelope at 10 $\mu$m as shown in Figure~\ref{fig:O-AGBavg}.

There are sixteen early-type O-rich AGB stars (O-EAGB) in the LMC sample 
which do not show any evidence of dust features in their infrared spectra. O-EAGB stars have just recently initiated the thermally-pulsing-AGB phase of their evolution and exhibit long-period variability. The variable early-type O-rich AGB stars are presumably more evolved than genuine early-AGB (E-AGB) stars, which do not exhibit long-period variability and have not yet started helium shell burning. The onset of pulsations in an AGB star coincides with the beginning of significant dust formation and a slight infrared excess may be visible in the SED. O-EAGB stars can be identified from oxygen-rich molecular absorption features in their IR-spectra, notably the absorption feature at 8 $\mu$m due to the SiO fundamental vibrational mode. Five of these sources have been characterised by \citet{Sloan2008} who find that the strength of the 8 $\mu$m absorption band from SiO decreases with lower metallicity.


Fifty-five sources have amorphous silicate emission features at 10 and 18 $\mu$m, although there is a considerable spread in the peak strength, shape and width of these features. These AGB stars span a wide range in mass-loss rate \citep[$10^{-7}--10^{-5}~ \Msun\ {\rm yr}^{-1}$;][]{Groenewegen2009, Riebel2012, Jones2012} and the variations in the IRS spectra can be attributed to the amount of dust in the circumstellar envelope and the presence of other dust components, such as simple metal oxides. 
The dust composition of these sources is best fitted by amorphous silicates (with an olivine stoichiometric ratio), with appreciable amounts of amorphous alumina and additional, small contributions from metallic iron \citep{Jones2014}.  Alumina makes up less than 50\% of the dust composition and metallic iron is present at approximately the 4\% level with the exception of OBJID 182; it remains unclear why this source lacks metallic iron. 

As the mass-loss rate increases and the dust shell becomes optically thick, the 10 $\mu$m feature goes into absorption. There are no spectra of LMC sources where the 10 $\mu$m is completely in absorption, although six sources (OBJID 121, 204, 253, 599, 603, 694) have self-absorbed silicate features at 10 $\mu$m.  A number of other O-AGB spectra have 10 $\mu$m features which might be starting to demonstrate signs of self-absorption. In these sources, the 10 $\mu$m feature has a relatively flat profile at the peak of the 10 $\mu$m dust feature. All sources with a 10 $\mu$m feature in self-absorption have crystalline silicate features at 23, 28 and 33 $\mu$m \citep{Jones2012}.

In total, crystalline silicate features are apparent in 20 of the O-rich AGB star spectra. These sources have dust production rates $\gtrsim 10^{-8}$ M$_{\odot} {\rm yr}^{-1}$. Like Galactic O-AGB stars \citep[c.f.][]{Sylvester1999, Kemper2001}, crystalline silicates are more prevalent in O-AGB stars with higher opacity dust shells. However, the composition of the crystalline silicate dust grains in the LMC sources is different from Galactic O-AGB stars, with Mg-rich pyroxenes (e.g.~enstatite: MgSiO$_3$) more prevalent than olivines \citep[e.g.~forsterite: Mg$_2$SiO$_4$;][]{Jones2012}.

A narrow 13 $\mu$m feature is conclusively seen in OBJID 265, 326 and 407 and may also be present in a number of other sources. Definitively detecting this dust feature is challenging as such low-contrast emission features are difficult to separate from the noise. The 13 $\mu$m feature has also been reported in Galactic O-AGB stars \citep{Sloan1996}. This feature is caused by an Al–-O stretching mode from corundum (Al$_2$O$_3$) or spinel (MgAl$_2$O$_3$) grains \citep{Posch1999, Fabian2001, Sloan2003, Lebzelter2006}.

There are two particularly intriguing O-rich AGB stars in the sample, corresponding to OBJID 6 and 390. Both of these sources have a prominent 10 $\mu$m amorphous silicate feature which arises from the Si--O stretching mode. However, the 18 $\mu$m silicate feature due to the  Si--O--Si bending mode in the SiO$_4$ tetrahedral is absent from the spectra. If the silicate dust has recently formed then the apparent absence of an 18 $\mu$m amorphous silicate feature may be attributed to a lack of contrast with the local continuum \citep{Jones2014}. Alternatively, \cite{Gielen2011} propose that dust composed of only small 0.1 $\mu$m Mg-rich olivine and alumina grains can result in the suppression of the 18 $\mu$m feature. 

The unusual nature of \objectname{SSTISAGEMC J051333.74-663419.1} (OBJID 6) is also evident in its optical spectrum. Although indicative of an F8--G0 giant, the optical spectrum also shows very strong H\,{\sc i} (6563 \AA) and He\,{\sc i} (5876 \AA) emission, and broad Ca\,{\sc ii} absorption lines \citep{vanAarle2011}, suggesting that the star resides in a binary system with a hot companion.

\subsection{The carbon-rich AGB sample}


\begin{figure*}
\begin{multicols}{2}
\includegraphics[trim=3.0cm 7cm 0cm 8cm, width=4.2in]{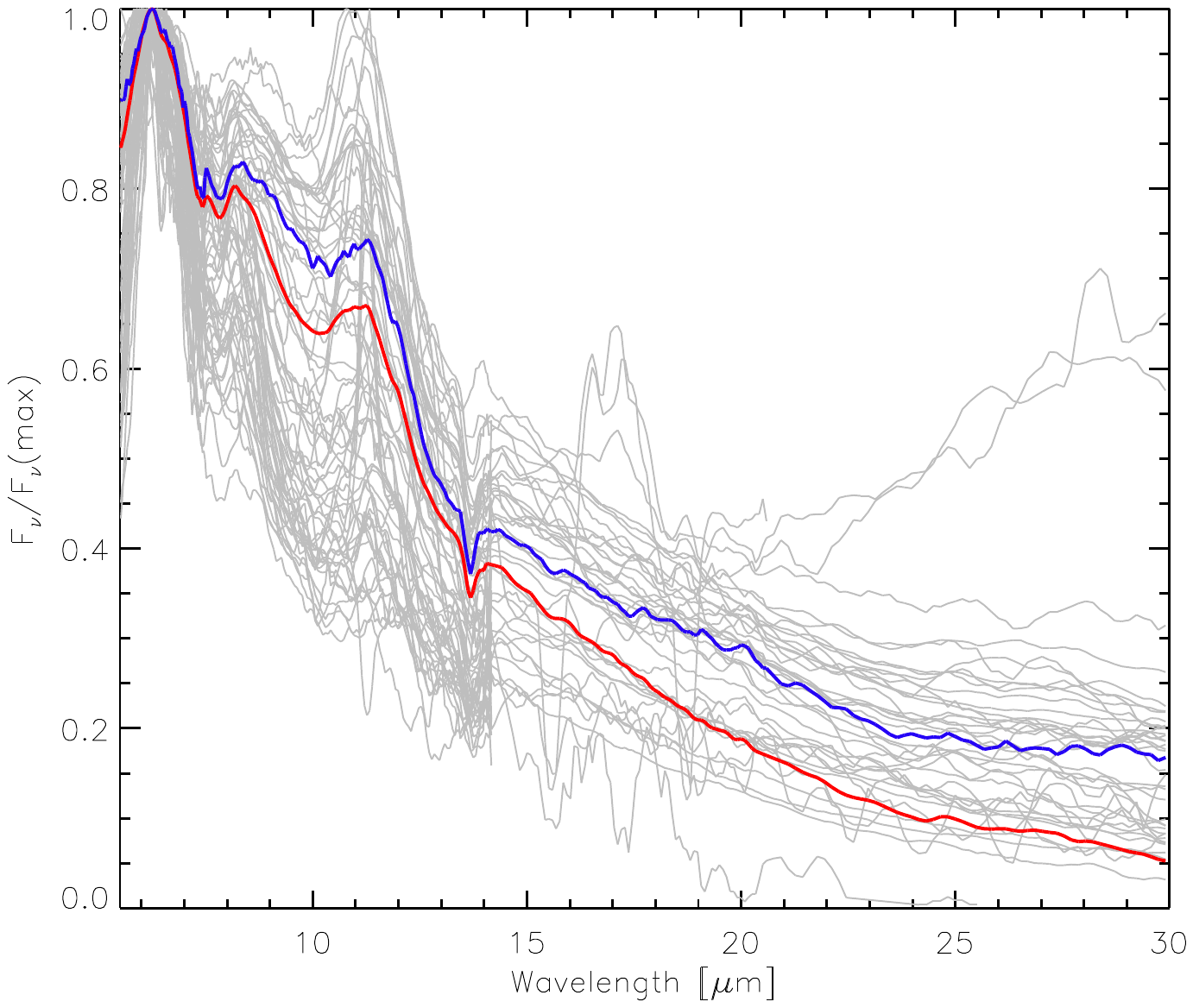}
\includegraphics[trim=3.0cm 7cm 0cm 8cm, width=4.2in]{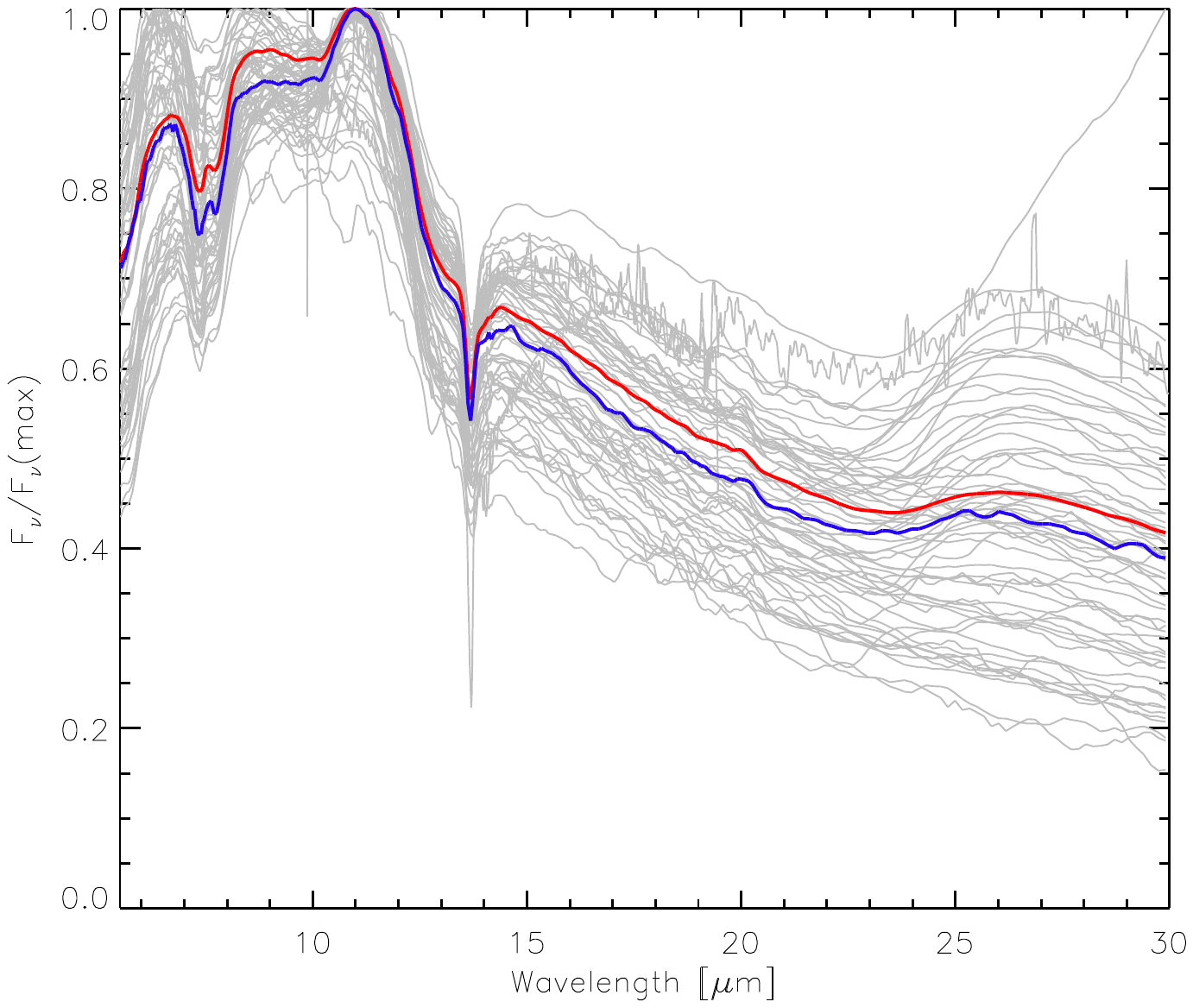}
    \end{multicols}
\begin{multicols}{2}
\includegraphics[trim=3.0cm 3cm 0cm 8cm, width=4.2in]{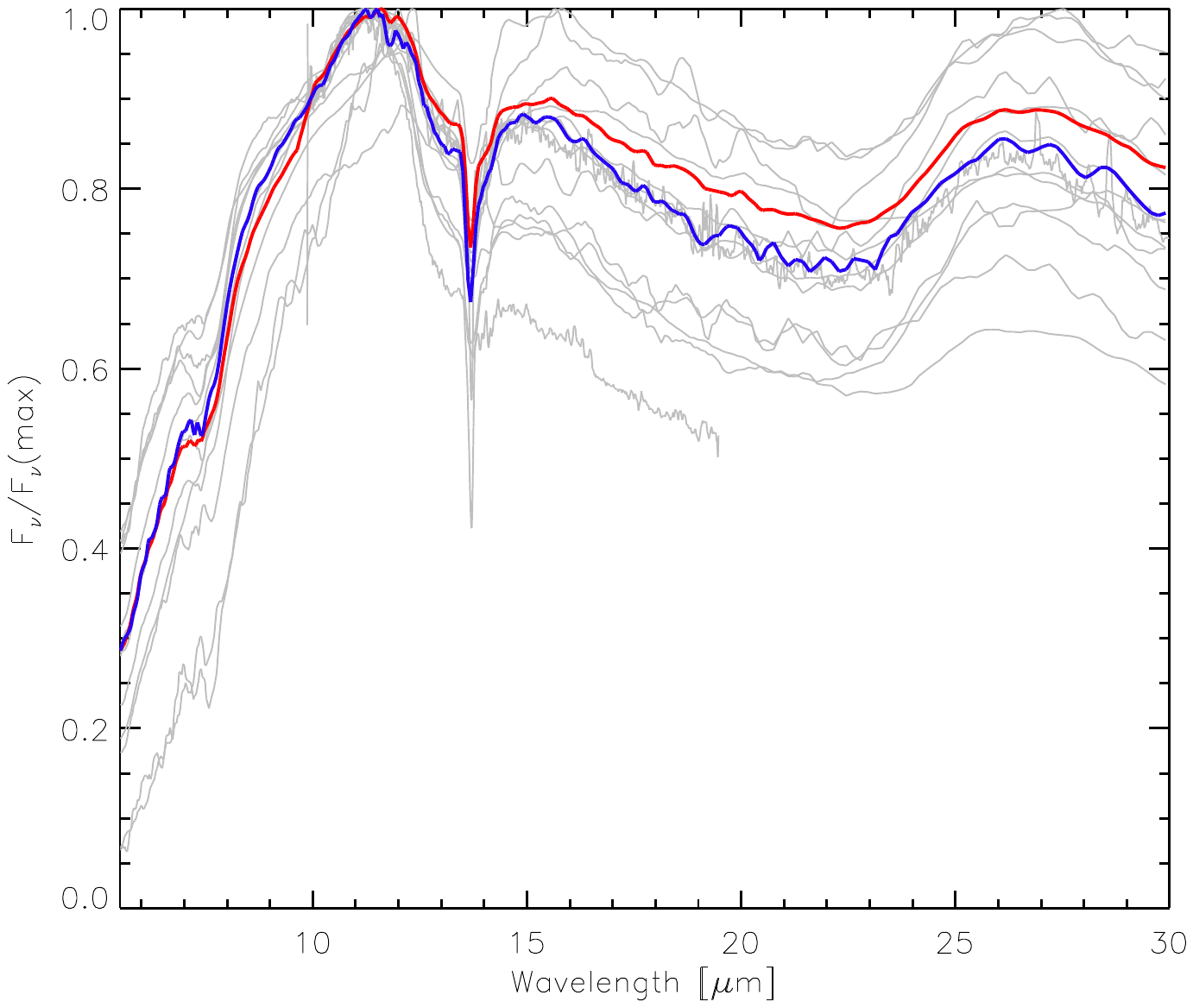}
\includegraphics[trim=3.0cm 3cm 0cm 8cm, width=4.2in]{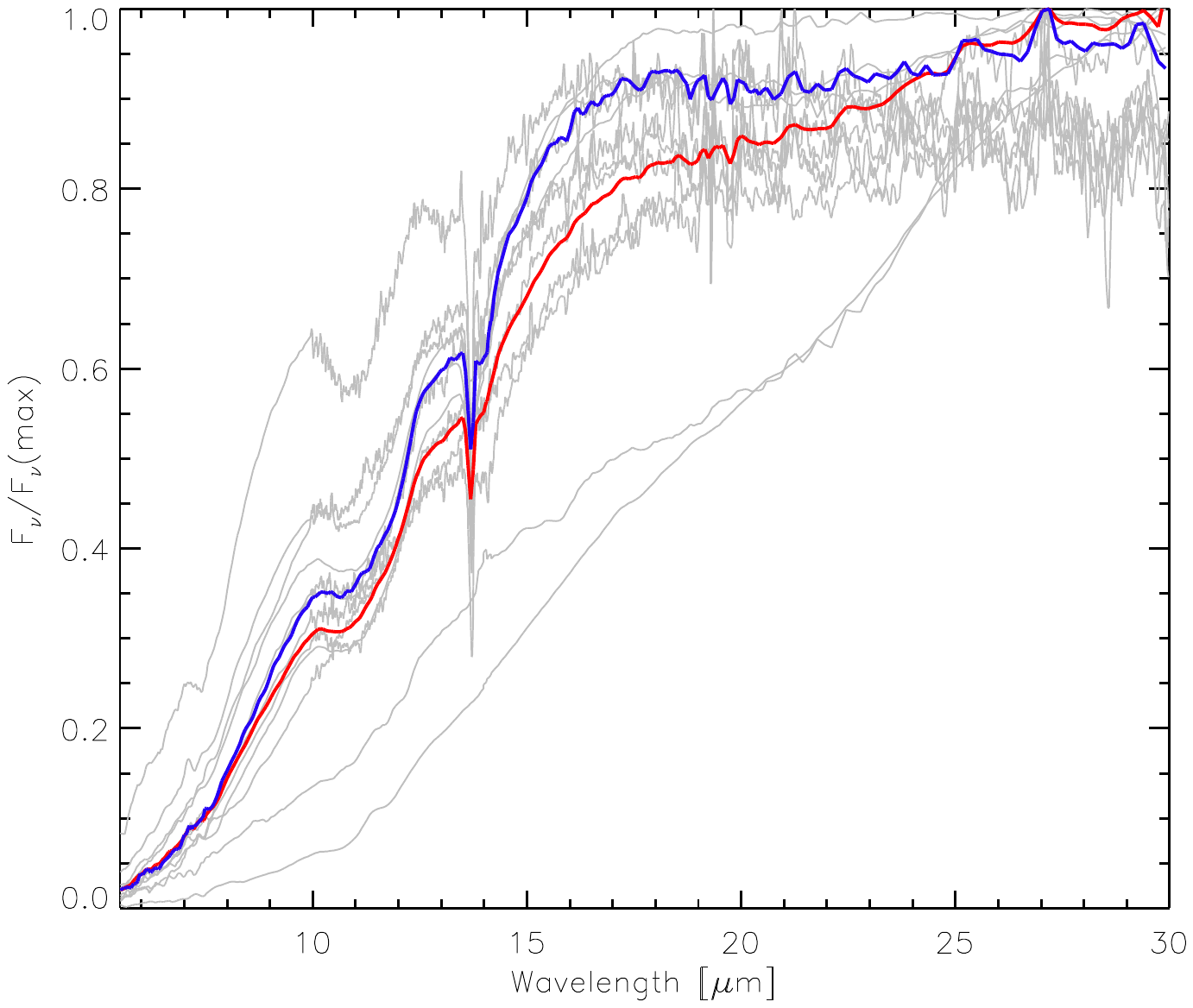}
\end{multicols}
\caption{Averaged spectra for the C-AGB subclasses. Top: the first, or bluest, (CE0--2; left) and second (CE3--4; right) {\tt C-AGB} spectral subclasses. The spectra are dominated by the emission in the 5--12\,$\mu$m region. Both the C$_2$H$_2$ bands at 7.5~\mum\ and 13.7~\mum\ are evident. 
Bottom row: Averaged spectra for the third (VROs; right) and fourth, or reddest, (EROs; left) {\tt C-AGB} spectral subclasses. These spectra can be quite red, with a SiC feature at 11.3~\mum\ in (self-)absorption and a 30~\mum\ emission feature.  \label{fig:CAGBavg_all}}
\end{figure*}


There are 148 carbon-rich AGB stars ({\tt C-AGB}) in our sample.  We divided the spectra into four smaller groups, based on their overall shape, from blue to red.  This approach is equivalent to classifying by most mid-infrared colours, because they all tend to increase smoothly as the mass-loss rate increases.  Our classifications correspond closely to the carbon-emission, or `CE' classes introduced by \citet{Sloan2016}, which should not be surprising because their CE classes are based on the [6.4]$-$[9.3] colours derived from the IRS spectra.  Similarly, our classifications align with other colour-based classifications for the same reason, as discussed in Section~\ref{sect:compAGB}.

Figure~\ref{fig:CAGBavg_all} (top) presents spectra from the first and second groups from the C-AGB sample.  These correspond to CE0--2 and CE3--4 in the scheme by \citet{Sloan2016}, respectively. The majority of the C-AGB sample fall into the first group, which is relatively blue and dominated by emission in the 5--12~\mum\ region.  In the first group, the C$_2$H$_2$ absorption bands at 7.5~\mum\ and 13.7~\mum\ are both strong, while the 30~\mum\ emission feature is weak or absent.  In the second, the 13.7~\mum\ band is even stronger, and the 30~\mum\ emission is growing stronger as well.

Figure~\ref{fig:CAGBavg_all} (bottom row) presents the third and fourth groups, which have been described, respectively, as `Very Red Objects' (VROs) in Paper I and `Extremely Red Objects' (EROs) by \citet{Gruendl2008}.  \citet{Sloan2016} classify both groups as `CE5', although it is evident here that the two groups can be distinguished meaningfully.  The 30~\mum\ emission feature is strong in the VROs and is weaker in the EROs.  The SiC feature at 11.3~\mum\ is weak in the VROs, probably because it is in self-absorption, while in the EROs, it usually appears as a clear absorption feature.  The apparent weakness of the 7.5~\mum\ C$_2$H$_2$ feature is deceptive; \citet{Sloan2016} show that its equivalent width does not change across the classes from CE0 to CE5.

The primary difference from the first to the fourth group (or from CE0 to CE5) is the amount of carbon-rich dust in the source.  However, the last two groups represent a major break in behaviour from the first two.  \citet{Sloan2016} showed that the pulsation amplitude of the central star increases from CE0 to CE4, but then falls in CE5 as the sources grow progressively redder.  In addition, the reddest sources often show surprisingly blue colours at shorter wavelengths (such as $J$$-$$K$ or [3.6]$-$[4.5]).  Combined, these characteristics suggest that the stars are evolving off the AGB and the central star may be becoming visible through an increasingly asymmetric dust shell \citep{Sloan2016}. In that light, the EROs may be transition objects toward the C-PAGB group, and eventually the C-PN group. 
If these sources have asymmetric envelopes seen `edge-on', it would cause exacerbated absorption without necessarily a larger amount of dust present.

\subsection{The post-AGB and planetary nebula sample}

We classify this group into post-AGB versus PN, which is an evolutionary sequence, and O-rich versus C-rich, which depends on the preceding evolution, giving four distinct groups. PNe differ from post-AGB in having strong ionic emission lines, but these may not always be obvious in the low-resolution IRS spectra, so that some known PNe are classified from the {\em Spitzer} spectra as post-AGB stars.  Dust features in general provide a strong indication for the C-rich versus O-rich separation, but both groups can show PAH emission. PNe without distinguishing dust features are classified as O-rich, and this group will include some objects where the gas is carbon-rich. 

\subsubsection{C-rich Sources}

\begin{figure}
  \includegraphics[trim=3.0cm 7cm 0cm 8cm, width=4.2in]{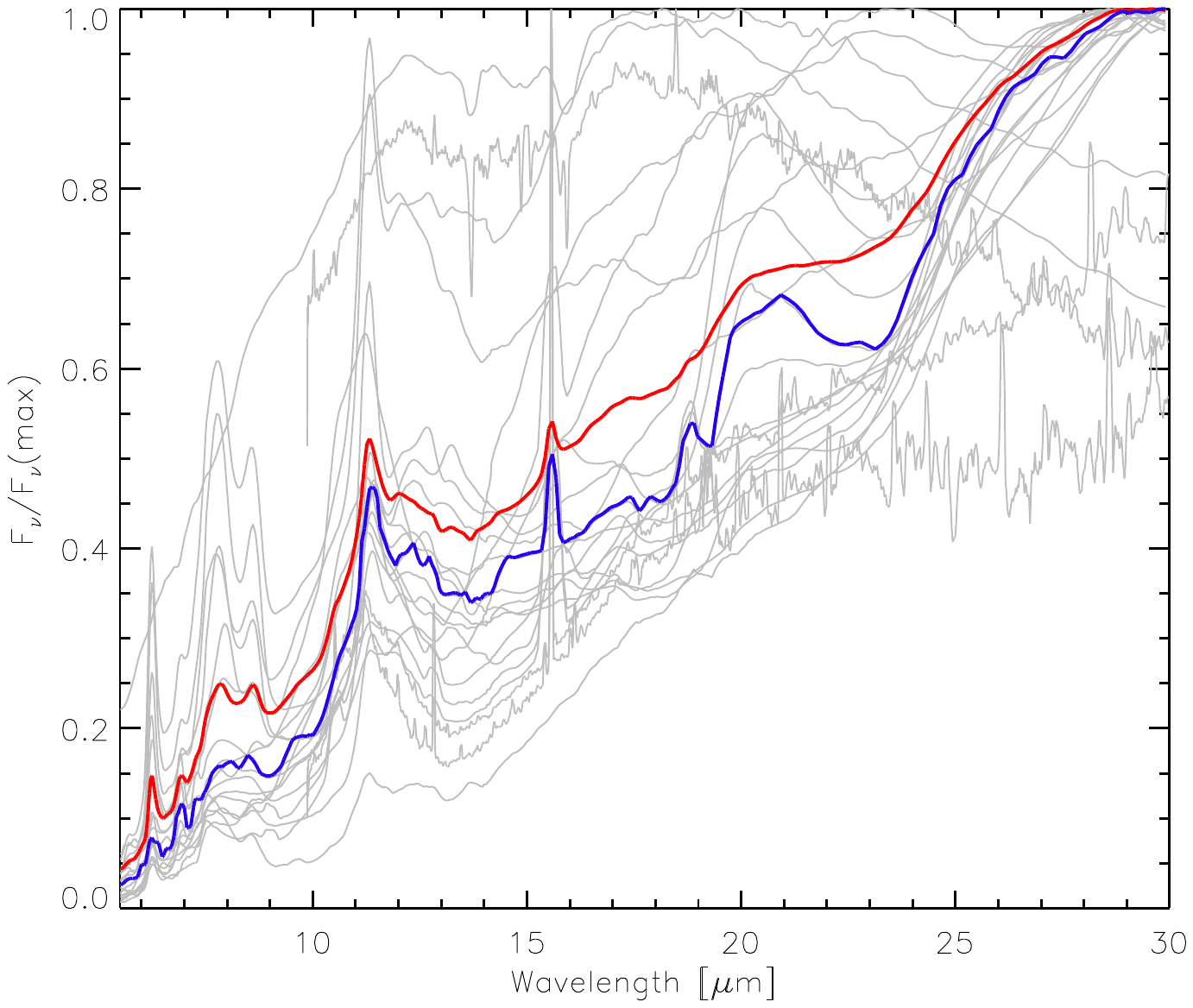}
  \includegraphics[trim=3.0cm 2cm 0cm 8cm, width=4.2in]{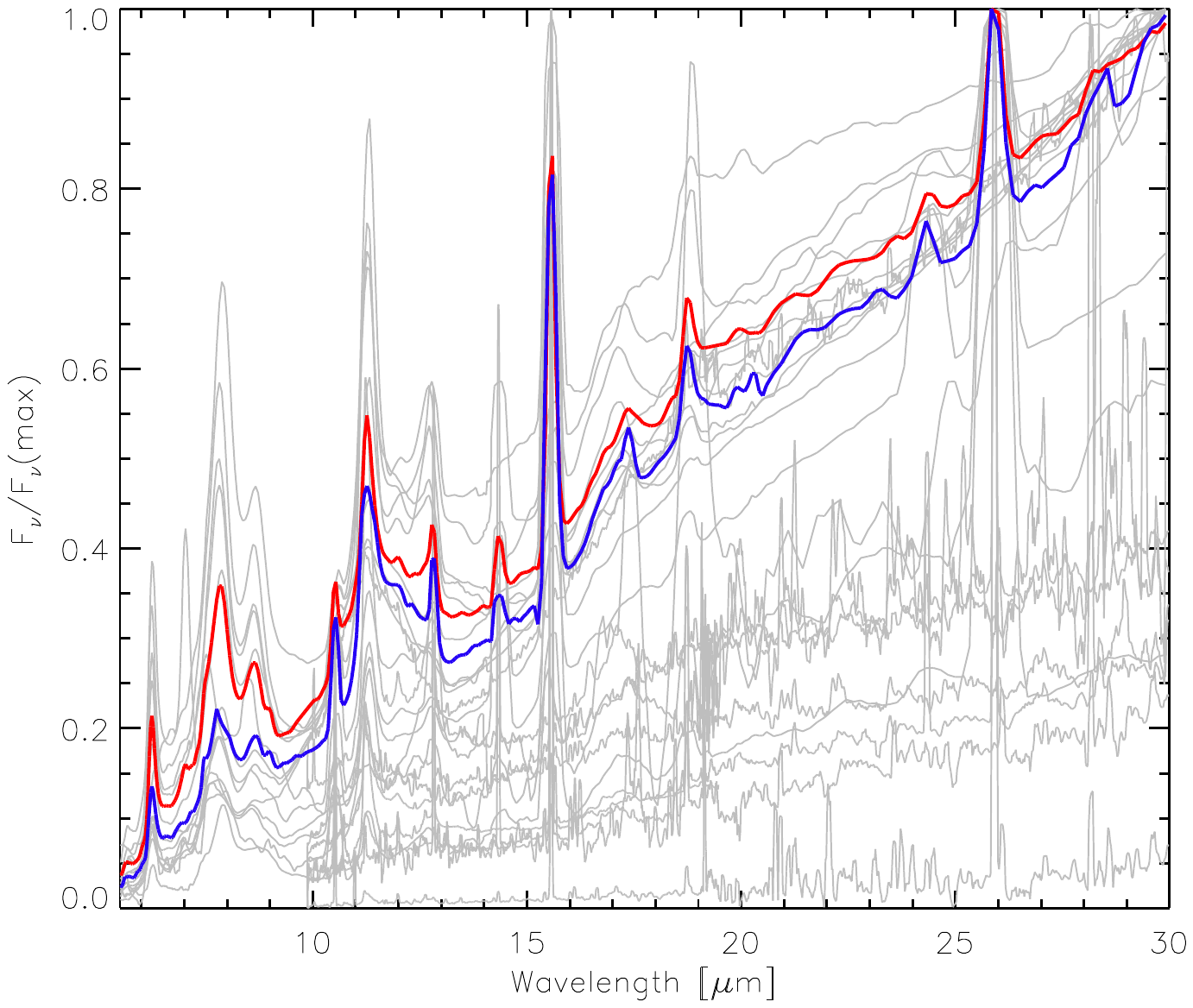}
\caption{Averaged spectra for the {\tt C-PAGB} (top) and {\tt C-PN} (bottom) classes.  In the top panel, the 21 $\mu$m feature only found in carbon-rich post-AGB stars can be clearly seen. \label{fig:cpagbavg}}
\end{figure}

We have identified 19 C-PAGB stars and 13 C-PNe in our sample. 
The sample of carbon-rich post-AGB objects includes three optically identified planetary nebulae (SMP LMC 8, 11 and 58) in addition to sixteen other post-AGB objects earlier in their evolution off the AGB. All of the SMP LMC object spectra are discussed by \citet{BernardSalas2009} and by \citet{Sloan2014}. The other post-AGB objects are discussed by \citet{Volk2011}, \citet{Sloan2014} and \citet{Matsuura2014}.  A few of these have optical spectroscopy and are A-type or F-type stars.

The C-PAGB stars include eight sources with a 21 $\mu$m feature \citep{Kwok1989}. This feature is only found in carbon-rich post-AGB stars \citep{Hrivnak2009}; the carrier remains unidentified.  Most sources also show a prominent, broad feature at 30 $\mu$m, seen exclusively in the spectra of AGB stars and PNe.  This is commonly attributed to MgS \citep{Goebel1985,Hony_02_carrier,Lombaert_12_Observational},  but this is disputed \citep{Zhang_09_Magnesium, Messenger2013} and other carriers have been suggested \citep[e.g.][]{Otsuka2014}.  The integrated intensity cannot be well measured from IRS spectra as the band extends beyond the available spectral range. If MgS is the correct identification, it must form a coating on the grains; there is some evidence for this from the fact that it only forms at low dust temperatures \citep{Zijlstra2006,Leisenring2008}. \citet{Matsuura2014} note that the composition of the PAHs in the spectra of carbon-rich evolved stars gradually changes as the central stars evolve from the post-AGB phase to a planetary nebula. This effect is more significant than any effects due to metallicity in determining the stars' spectral appearance.

The strengths of the features that are seen in the average spectrum vary widely from source to source (Fig.~\ref{fig:cpagbavg}).  Only features due to PAHs are seen in all the spectra, whereas the SiC, 21, and 30 $\mu$m  features are absent in some spectra.  As a result, while the average spectrum is somewhat representative it nonetheless shows some differences from any of the individual spectra.  In particular the 21 $\mu$m feature objects generally have higher temperature continuum shapes than the average spectrum, due to a number of steeper continuum sources (i.e., generally more evolved sources) in the sample -- including most of the optical planetary nebulae.


Three C-PAGB stars are classified in the optical as PNe. The most peculiar object, \objectname{SMP LMC 11} (OBJID 227), also known as \objectname{LHA 120-N 78}, shows an infrared spectrum that is very atypical of planetary nebulae or post-AGB objects in general \citep{BernardSalas2006}. It shows some features  normally associated with  C-rich AGB stars: strong molecular bands of C$_2$H$_2$ and HCN are visible, on a fairly low temperature continuum, although not as low in temperature as is typical of post-AGB objects and planetary nebulae. But it also shows hydrocarbons up to benzene in absorption \citep{BernardSalas2006}, with prominent bands 6.8 and 14.8 $\mu$m: these are never seen in AGB stars. SMP LMC 11 has similarity to the Galactic source CRL 618, and likely contains a dense carbon-rich disk, seen edge-on. A comprehensive discussion of its circumstellar chemistry is given by \citet{Malek2012}. 

The other two sources optically identified as planetary nebulae (SMP LMC 8 and 58) have similar spectra, showing the 11.3 $\mu$m SiC emission feature on a lower temperature continuum and the nebular forbidden lines ([Ne\,{\sc iii}] 15.55 $\mu$m, [Ne\,{\sc ii}] 12.81 $\mu$m, and others).  The spectral shapes diverge beyond 20 $\mu$m.  \objectname{SMP LMC 58} shows a rather weak 30 $\mu$m feature, while \objectname{SMP LMC 8} does not show this feature.  From the high resolution spectra \objectname{SMP LMC 8} and \objectname{SMP LMC 58} would correctly be classified as planetary nebula due to the relatively strong nebular emission lines that are visible.  However, the contrast between the emission lines and the continuum is much less in the low resolution {\em Spitzer} spectra, and they are instead classified as C-PAGB stars according to the decision-tree.


PNe in the LMC are expected to be predominantly carbon-rich, because the sub-Solar metallicity moves the stellar mass range over which AGB stars become carbon rich \citep[e.g.][]{Renzini1981b,Karakas2014} to lower masses and it increases the C/O ratio.  The main way to identify whether a PN is carbon-rich is from the presence of molecular and dust features. This works best for young, compact PNe since evolved PNe lose their dust features \citep[e.g.][]{Stanghellini2007}. It is more difficult to measure carbon abundances from optical emission lines only, and for PNe without dust, the C/O ratio is not always well known.

The distinguishing carbonaceous features are PAHs, fullerenes, SiC and the 30 $\mu$m feature. Molecular bands (apart from PAHs) are rare. PAH bands are common. The features are probably due to a combination of aromatics and aliphatics, rather than pure PAHs \citep{Kwok2011, Sloan2014}. The PNe show evidence for structural evolutionary changes in the PAHs, related to photo-processing \citep{Matsuura2014} with increasing aliphatic content \citep{Sloan2014}.

PAHs are not confined to carbon-rich objects. In oxygen-rich PN in the Galactic bulge, PAH bands have been seen to arise in dense tori \citep{Guzman2014}, and thus bipolar PNe have stronger PAH features. \citet{Stanghellini2007} find that the presence of a dense torus is less important for carbon-rich PNe than it is for oxygen-rich PNe.  PAH formation rates increase strongly for higher C/O \citep{Guzman2011}, leading to stronger PAH bands for LMC PNe.  One should not expect LMC PNe to be carbon-copies of their Galactic counterparts.

Fullerenes (C$_{60}$ and C$_{70}$) are relatively common in LMC PNe \citep{GarciaHernandez2011}. Their presence is indicated by four narrow emission bands but two of these can be merged with strong nebular lines at the lowest IRS resolution. A 5th band at 6.49 $\mu$m \citep{Sloan2014} may also be due to fullerenes \citep{Brieva2016}. PNe with fullerene emission tend to show small, photo-processed PAHs \citep{Otsuka2014}, although fullerenes can also be found in PNe without PAH emission \citep{Sloan2014}. Fullerenes are only seen in PNe with cool central stars (up to 40\,000 K). At higher stellar temperatures the  bands disappear, whilst PAH emission can remain, indicating that fullerenes are more easily destroyed \citep{Sloan2014}. 

The 11.3-$\mu$m SiC band in LMC PNe is fairly common, and can be extraordinarily strong (dubbed the `big-11' feature by \citealt{Sloan2014}), sufficient to affect the IRAS 12-$\mu$m flux \citep{BernardSalas2009}. In contrast, SiC is rarely seen in Galactic PNe, in spite of the higher Si abundance. \citet{Sloan2014} show that there is some PAH emission included in the SiC band, but SiC accounts for 88\%\ of the strength of the `big-11' feature. The extreme strength may be due to photo-excitation, or because at a high C/O ratio SiC forms on the surface of carbon grains, rather than as the first condensate \citep{Sloan2014}.  The SiC features disappear when lines with ionization potential above 55 eV appear \citep{BernardSalas2009}, and this seems to be the photo-dissociation energy for SiC.  The `big-11' sources can also show a broad band at 16-22 $\mu$m, which is unidentified. It is called the '18-$\mu$m shoulder' by \citet{Sloan2014}.


\subsubsection{O-rich Sources}

\begin{figure}
  \includegraphics[trim=3.0cm 7cm 0cm 16cm, width=4.2in]{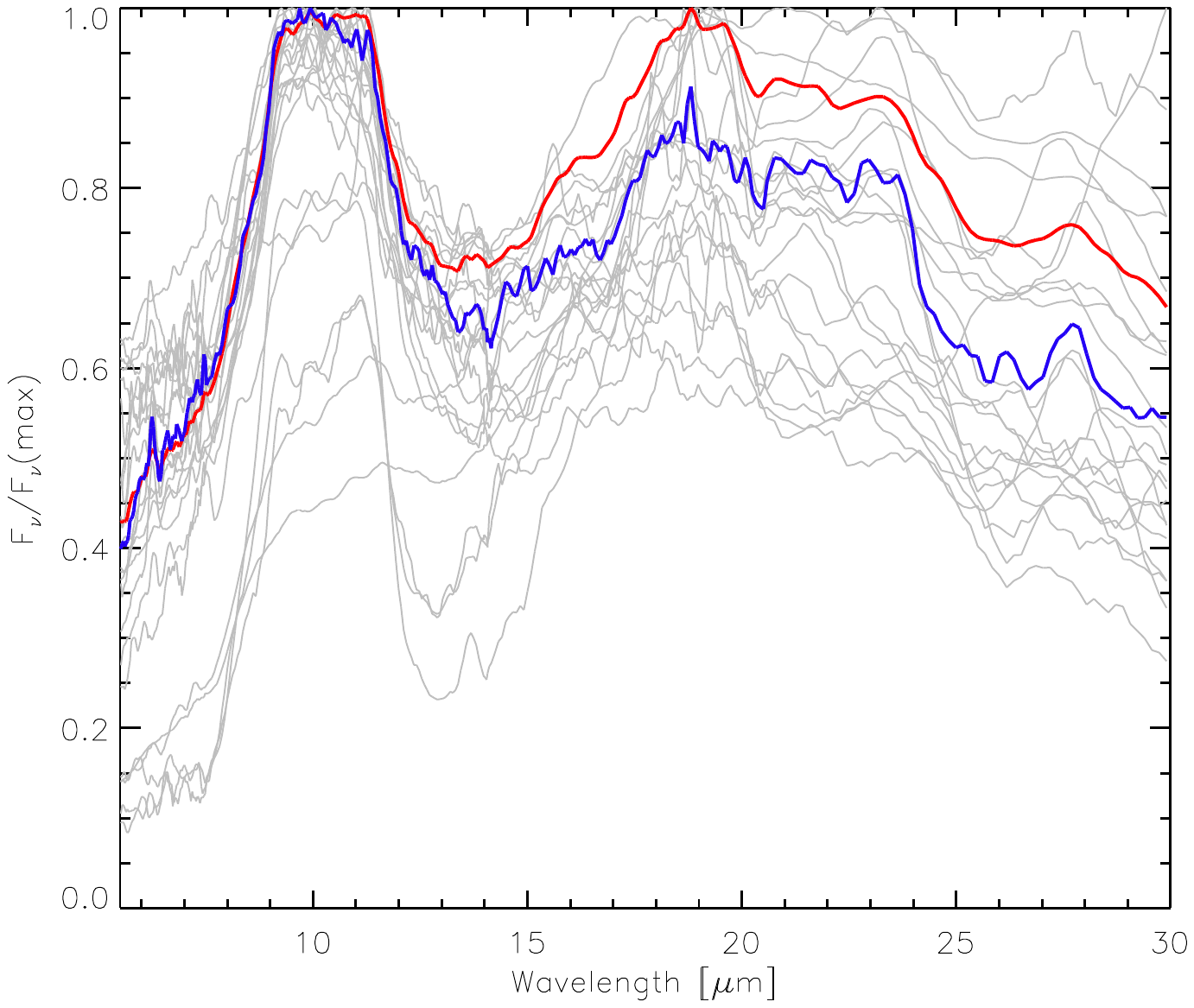}
  \includegraphics[trim=3.0cm 7cm 0cm 9cm, width=4.2in]{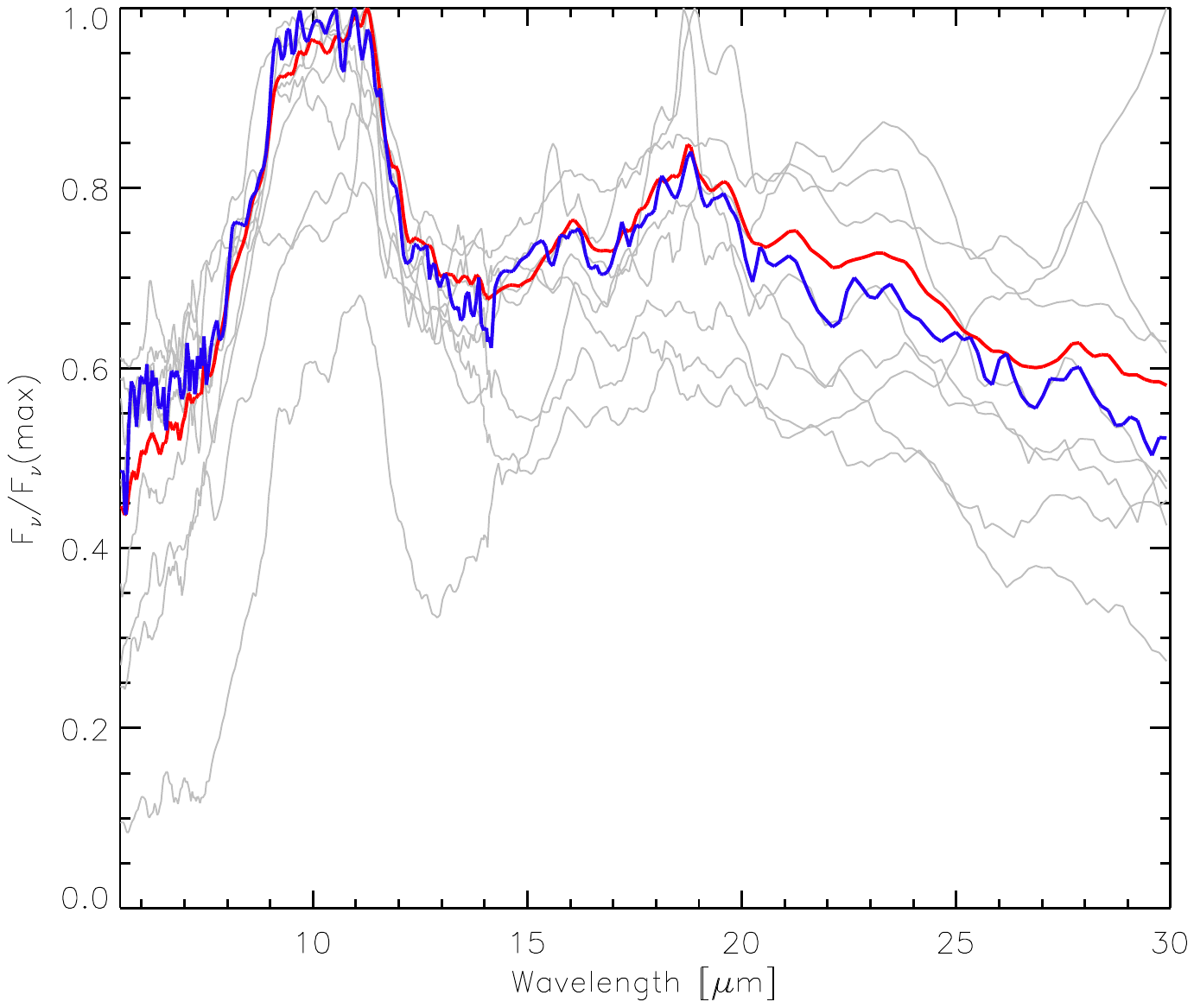}
  \includegraphics[trim=3.0cm 3cm 0cm 9cm, width=4.2in]{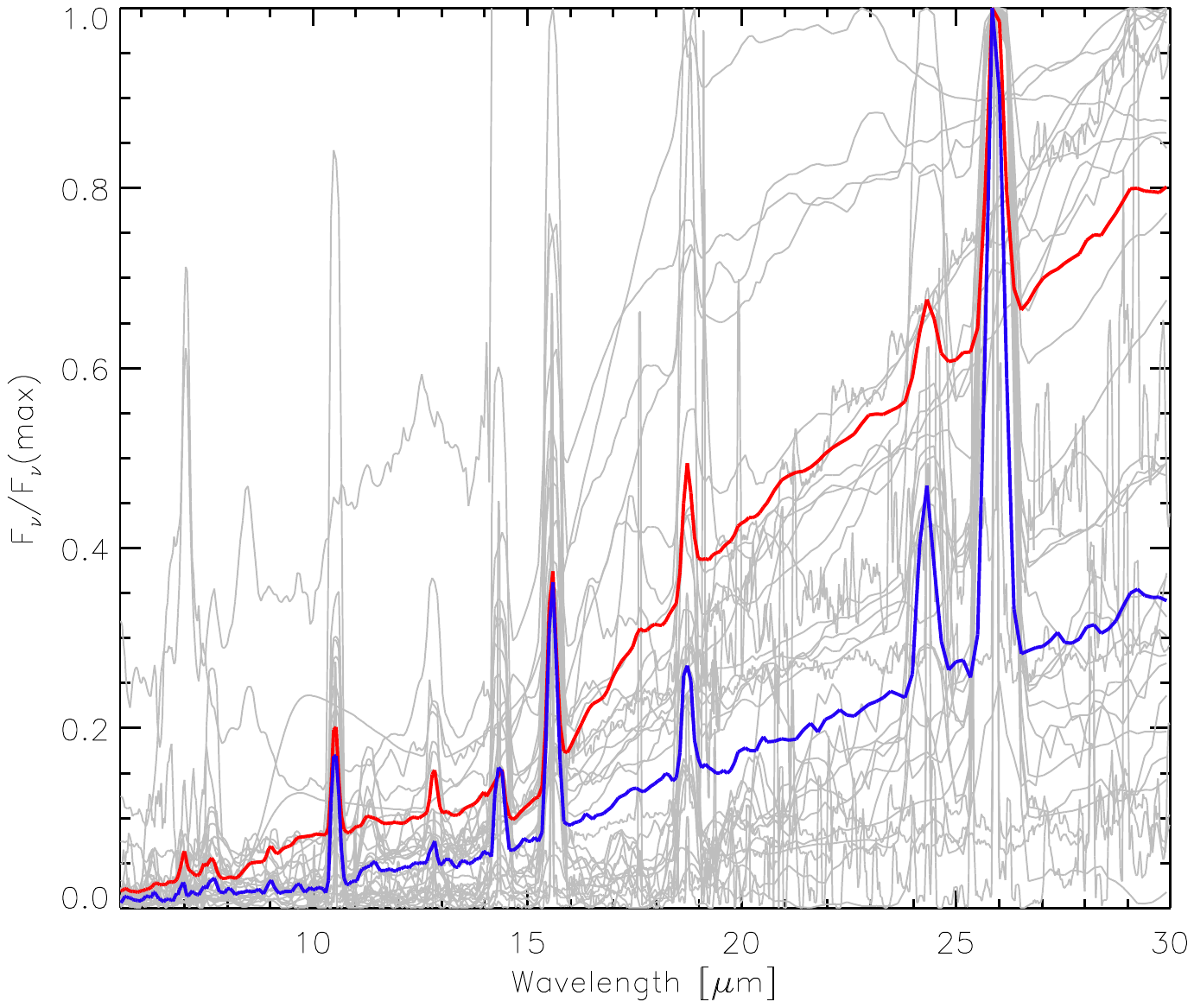}
\caption{Averaged spectra for the {\tt O-PAGB} (top, excluding RV Tau objects), the {\tt RV Tau}  stars (middle), and the {\tt O-PN} (bottom) classes.  \label{fig:opagbavg}}
\end{figure}

There are 23 O-PAGB and RV Tau-type stars in our sample and 28 O-PNe. The {\tt O-PAGB} group (Fig.~\ref{fig:opagbavg}) shows strong amorphous and crystalline silicate features superimposed on a low temperature continuum.
The 10 $\mu$m amorphous silicate feature is seen in all the spectra with some variation in the width; this feature has a relatively flat-topped peak compared to the O-AGB stars  (Fig.~\ref{fig:O-AGBavg}). In addition weak PAH features are seen, such as the 6.3 $\mu$m  emission feature. There is a wide variation in the relative feature strengths of both the 10 and 18 $\mu$m amorphous silicate features and the various crystalline silicate features.

The majority of the O-PAGB objects are discussed by \citet{Gielen2011}. That paper interprets the flat-topped silicate band as due to a combination of amorphous silicate, crystalline fosterite (mainly at 11.3 $\mu$m) and crystalline enstatite (9 $\mu$m). The 11.3 $\mu$m/9.8 $\mu$m flux ratio  indicates the degree of crystallinity, which is high in these sources. The peak-to-continuum ratio is on average around 1.5, which is low and indicates larger grains; a few of our sources show much higher ratios. The degree of crystallinity and grain size can result from  dust processing in  stable circumstellar disks \citep{Gielen2009}.

The O-PAGB sample includes nine {\tt RV Tau}-type stars, which were identified from Harvard variable lists, MACHO light curves or the OGLE survey \citep[i.e.,][]{Leavitt1908, Payne-Gaposchkin1971,Alcock1998,Soszynski2008}.  The mean spectrum from the RV Tauri group is rather similar to the mean spectrum from the other O-PAGB objects ({Fig.~\ref{fig:opagbavg}), showing strong amorphous and crystalline silicate dust features. The general continuum level in the mean spectrum appears to show a slightly higher temperature than is the case for the O-PAGB objects, hence the 10 $\mu$m feature appears stronger with respect to the 18 $\mu$m feature in this spectrum. The majority of the crystalline silicate features seen in the RV Tau stars are also seen in the O-PAGB mean spectrum, but the 16.1 $\mu$m feature does not appear to be present in the RV Tau stars. The 11.3/9.8 $\mu$m ratio is slightly higher in the average RV Tau spectrum, indicating a higher degree of crystallinity, and the peak-to-continuum is a bit lower, which may mean larger grains. In the model of \citet{Gielen2011}, this indicates a higher degree of dust processing in the disk, perhaps through a more evolved disk.


Two of the objects in the O-PAGB sample are known optical planetary nebulae. \objectname{SMP LMC 64} (OBJID 115) \citep[Paper I]{Henize1956, Sanduleak1978} is one of the youngest known planetary nebulae, and has unusually high densities and temperature \citep{Dopita1991}.  The spectrum of this object only covers the 5--14  $\mu$m  range. It  shows a fairly strong amorphous silicate feature and a weak 12.8 $\mu$m  [Ne\,{\sc ii}] nebular emission line.
The IRS spectrum of \objectname{SMP LMC 64} was not classified as a PN spectrum using the flow chart because the nebular lines are weak compared to the continuum, and only the one low excitation line is detected. 
The second optically-detected planetary nebula in the sample, \objectname{[L63] 31}
\citep{Lindsay1963, Egan2001} shows amorphous silicate features at 10 and 18 $\mu$m  plus weak PAH emission features (6.2, 7.6, 11.3 $\mu$m) that may be interstellar in origin.  There is no sign of nebular emission lines in the IRS spectrum, hence it is classified in the O-PAGB group.


The 28 objects classified as oxygen-rich planetary nebulae ({\tt O-PN} class) in this spectroscopic sample all have prominent emission lines and a rising continuum toward longer wavelengths.  The median of the spectra reveals a number of highly ionized emission lines indicative of the very hard radiation field expected from the central star of a PN, including:  [S\,{\sc iv}] 10.51 $\mu$m, [Ne\,{\sc v}]  14.32  and 24.28 $\mu$m, [Ne\,{\sc iii}] 15.55 $\mu$m, [S\,{\sc iii}] 18.71 $\mu$m,  [O\,{\sc iv}] 25.91 $\mu$m.  The continuum of the median spectrum shows a gentle rise from 6 to 30 $\mu$m and no broad band features are evident.  The continuum at these wavelengths is dominated by emission from warm dust; temperatures have been measured to be in the range 60 to 190 K for a subsample of these PNe \citep{Stanghellini2007}.

The median spectrum, however, over-simplifies the rich diversity of the O-PN spectral features.  The relative strengths of the above-listed prominent spectral lines varies significantly over the sample of 28 objects, related to differences in the central star temperatures. In addition, some low-ionization lines that appear in individual spectra are not apparent in the median spectra, including  [Ne\,{\sc ii}] 12.8 $\mu$m and [Ar\,{\sc ii}] 6.98 $\mu$m.  These two lines seem more prominent in those O-PNe, e.g. OBJID~305 and 732,  which have brighter continuum emission in the 6--30 $\mu$m wavelength range, consistent with the expectation that cooler central stars have less evolved nebulae.

The continuum level compared to the line strengths varies by over an order of magnitude. There are some O-PNe, e.g.~OBJID 363, that have essentially no continuum. In the sources with brighter continua,  the silicate emission features are evident at $\sim$10  and 18 $\mu$m, such as in the PNe observed in the spectra, SSID 4143 and 4735. The change in the continuum level is caused by the evolution of the PN.  As a PN evolves,  the circumstellar ejecta expand away from the central star, cooling the dust and causing the peak of the dust continuum to shift to longer wavelengths.  The IRS spectra are sampling the Wien edge of the modified blackbody of the dust continuum, so a small change in temperature creates a large change in the continuum level. 

We comment on three particular O-PNe.  \objectname{SMP LMC 21} (OBJID 331), which \citet{Stanghellini2007} notes as having O-rich dust, shows evidence for crystalline silicates  at 28--31 $\mu$m.   It has a quadrupolar morphology and evidence for very high excitation as seen in {\em Hubble (HST)} observations \citep{Shaw2001}. \objectname{SMP LMC 27} (OBJID 340), reveals an almost complete ring in the IRAC 4.5 $\mu$m image, and a complete ring in 8 $\mu$m surrounding the central compact PN \citep{Hora2008}.  The IRS spectrum of this central compact PN has a featureless continuum, as do most of our sample, with prominent emission lines.  {\em HST} observations of \objectname{SMP LMC 27} reveal an intermediate excitation nebula with a quadrupolar morphology for the inner compact structure \citep{Shaw2001}, which is present within the halo detected by IRAC.  
The IRS spectrum of \objectname{SMP LMC 9}  (OBJID 219) shows PAH features at 7.7, 11.3 $\mu$m and a plateau at 12--14 $\mu$m, while {\em HST} reveals a morphology with a bipolar core with barrel-shaped structure in very high excitation lines \citep{Shaw2001}. This source may have gas-phase abundances with C/O$>1$ \citep{Leisy2006}. 


\begin{figure}
 \includegraphics[trim=3.0cm 7cm 0cm 16cm, width=4.2in]{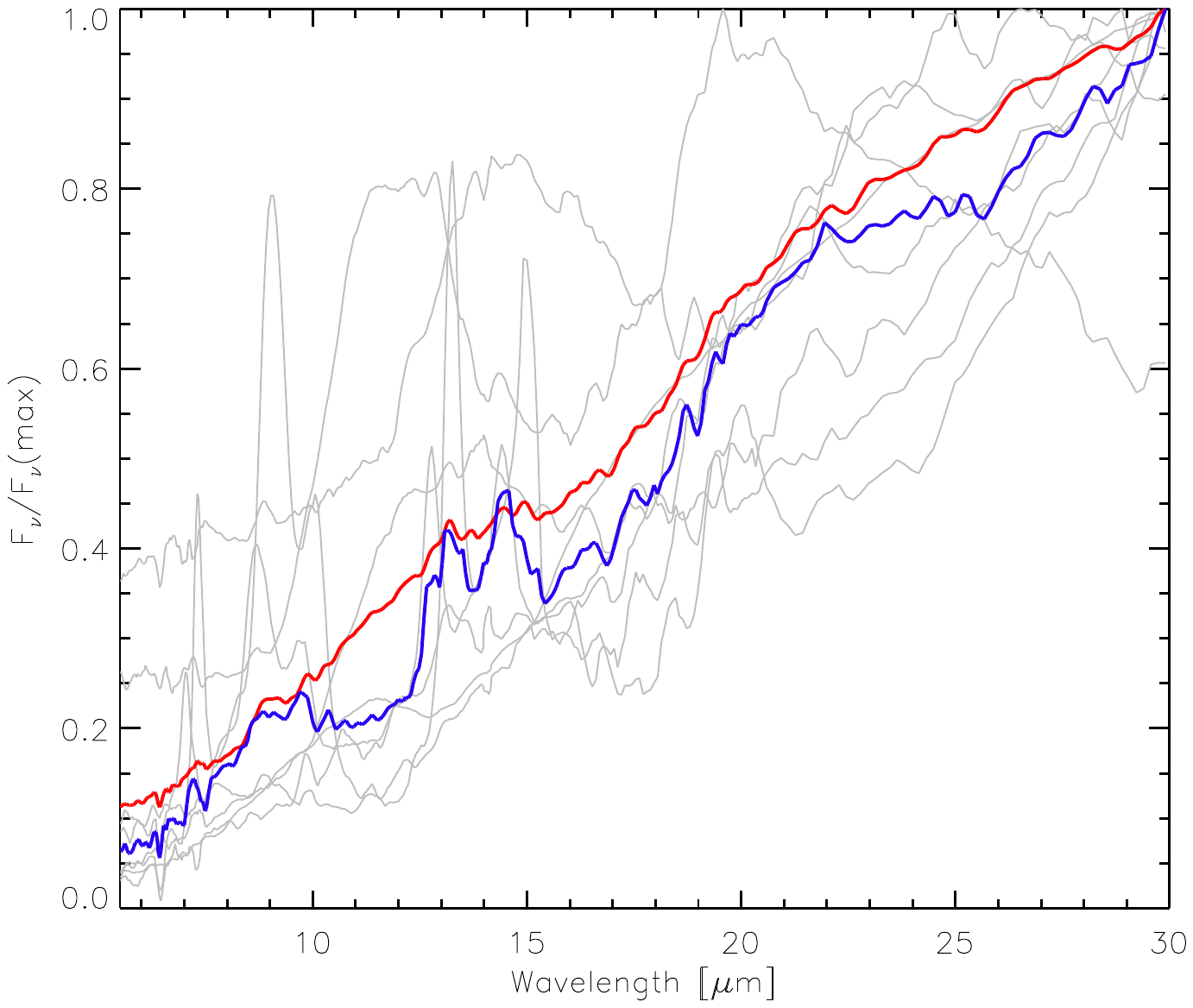}
 \includegraphics[trim=3.0cm 7cm 0cm 9cm, width=4.2in]{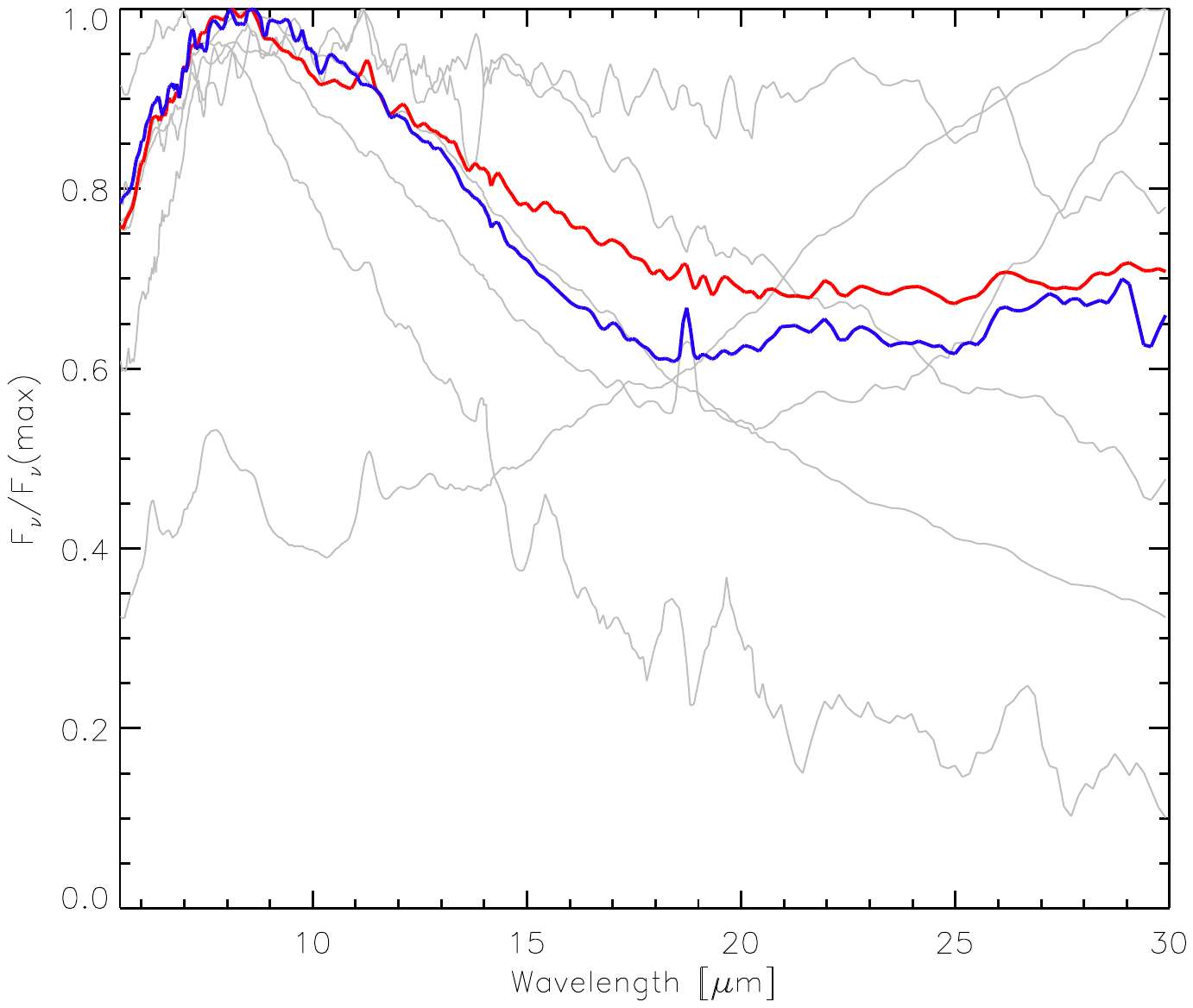}
 \includegraphics[trim=3.0cm 3cm 0cm 9cm, width=4.2in]{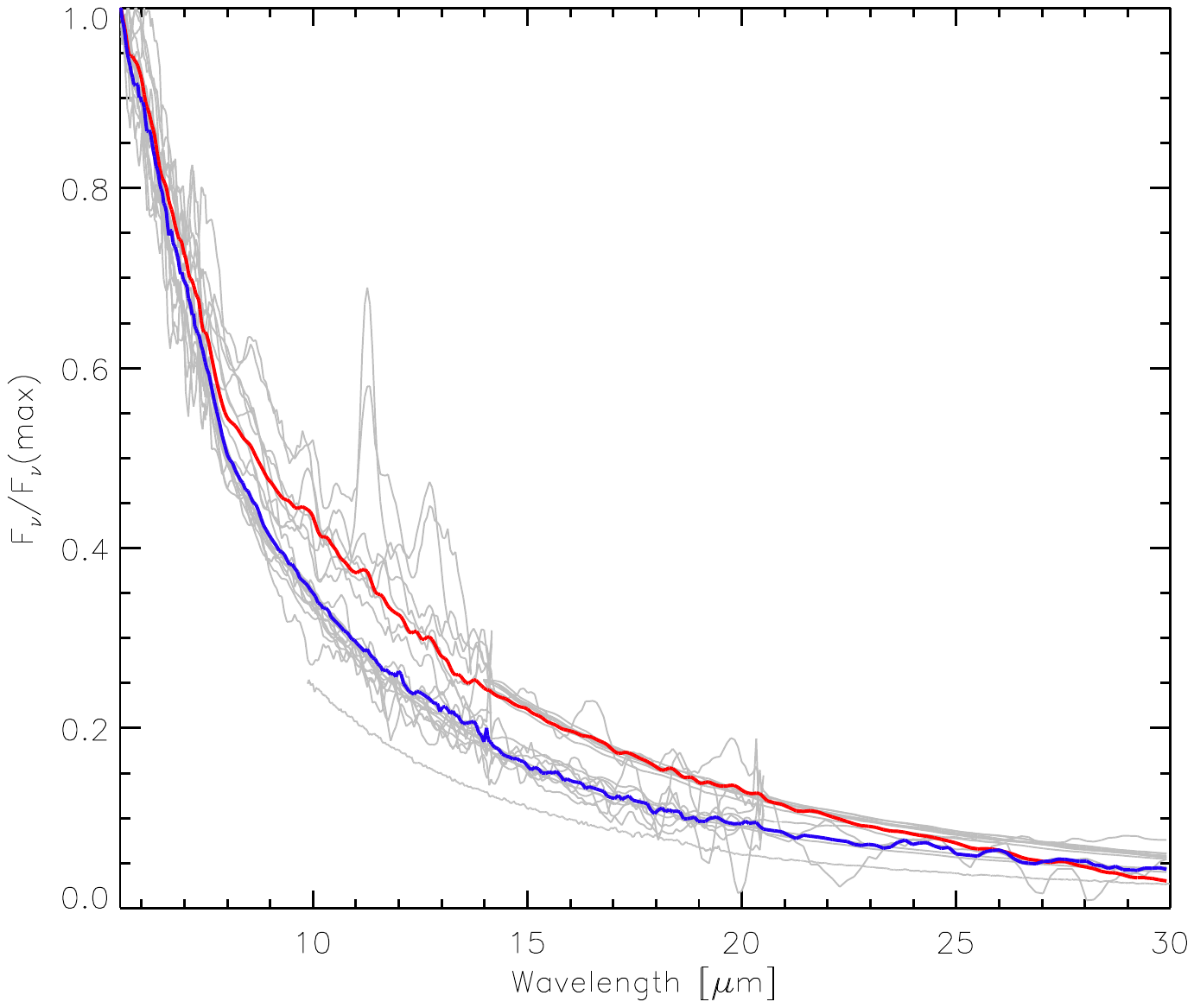}
\caption{Averaged spectra for the {\tt Gal}, {\tt RCrB} and {\tt Star} classes.  \label{fig:miscavg}}
\end{figure}

\subsection{R Coronae Borealis stars}

R Coronae Borealis ({\tt RCrB}) stars are a rare type of carbon-rich,  hydrogen-poor evolved star that irregularly undergoes extreme drops in  brightness in the visible \citep[e.g.,][]{Clayton2012}. Observations of RCrBs  with the infrared spectrometers on {\em ISO} and {\em Spitzer} have found that  their infrared SEDs are often dominated by graybody emission from (roughly)  single-temperature dust. This leads to a very distinct shape for their IRS  (and previously, SWS) spectra, peaking between about 5 and 10\,$\mu$m  \citep[e.g.,][]{Kraemer2005,GarciaHernandez2011}. Some RCrBs show emission  from cooler dust, instead, often along with strong PAH features  (Fig.~\ref{fig:miscavg}). Of the seven sources classified here as RCrBs or  having RCrB-like IRS spectra, six show  the roughly graybody shape, with emission peaks around 7--8\,$\mu$m.  The seventh, \objectname{HV 2671}, has a spectrum with strong PAHs and cool  dust \citep{Clayton2011}, and would nominally be a {\tt YSO-3} according to  the decision tree. However, it is also known to be an unusual type of RCrB  star with a high effective temperature \citep{Alcock1996, DeMarco2002}, which  may also influence its dust properties \citep{Clayton2011}. 

Four of the sources were known to be RCrB stars prior to the IRS  observations and one, \objectname{MSX LMC 1795}, has since been confirmed as a likely RCrB (G. Clayton et al. 2017, in prep.). The two RCrB candidates,  \objectname{RP 1631} and \objectname{MSX LMC 527}, have spectra that are  similar to the known RCrB stars but have some differences, and they both  lack supporting observations such as optical light curves or spectra.  Based on its infrared photometry, MSX LMC 527 has been classified as either  a YSO \citep[e.g.][]{Whitney2008}, and extreme (carbon-)AGB star  \citep[e.g.][]{Srinivasan2009, Boyer2011, Riebel2012}, or miscellaneous \citep{Seale2009}.

\subsection{Supernova remnants and novae}

There are nearly 80 known supernova remnants (SNRs) in the LMC \citep{Badenes2010}, and the IRS observed two of them, including several epochs for SN 1987A. The two SNRs do not seem to have any common spectral features, apart from the presence of strong atomic emission lines, including [Ne {\sc ii]}, [Ne {\sc iii}] and [Fe {\sc ii}]. Silicate dust features are seen in the spectra of the remnant of SN 1987A (see Fig.~\ref{fig:Novaspec}), which was observed over multiple epochs (OBJID 631; SSID 4577 -- 4580) to study the long-term IR  evolution of SN 1987A \citep{Dwek2010, Arendt2016}.  An amorphous carbon or metallic iron component may also be present in the spectra from 5--8 $\mu$m \citep[Fig.~5 of ][]{Dwek2010}.  The dust in the equatorial ring was formed by stellar winds during mass loss episodes from the progenitor's star, and is now undergoing dust processing due to shocks produced by the SN blast wave \citep{Bouchet2006}.
The newly-formed SN dust in the ejecta \citep[see e.g.,][]{Matsuura2011} is too cold and therefore too faint to be detected with the {\em Spitzer} IRS.

The other SNR in our sample is \objectname{SNR B0540-69.3} (OBJID 711).  \cite{Williams2008} published the IRS spectrum of the remnant and they found that the IR excess due to dust rises longward of 20 $\mu$m. A similar excess has also been observed in the Crab Nebula \citep{Temim2006}. The spectrum contains emission lines which may be slightly broadened, suggesting that they may be produced in the expanding SN ejecta.  

The spectrum of \objectname{Nova LMC 2005-11a} (OBJID 361) is unusual. It has H recombination line emission at 7.5, 12.4, 16.2, 16.9 and 27.8 $\mu$m, and structure at 16.5 $\mu$m which may possibly be due to  an artifact in the data reduction. The underlying dust continuum is relatively featureless and likely to be carbonaceous (and warm), similar to what is seen in Galactic dust-producing novae. This is the only nova observed with the IRS in the Magellanic Clouds. The spectrum was taken 16 days after the eruption.

The two Supernovae remnants and \objectname{Nova LMC 2005-11a} were identified based on literature information. Following the decision tree, these spectra were  classed as {\tt Other}.

\begin{figure}
  \includegraphics[trim=2.0cm 2cm 0cm 4cm, width=6.5in]{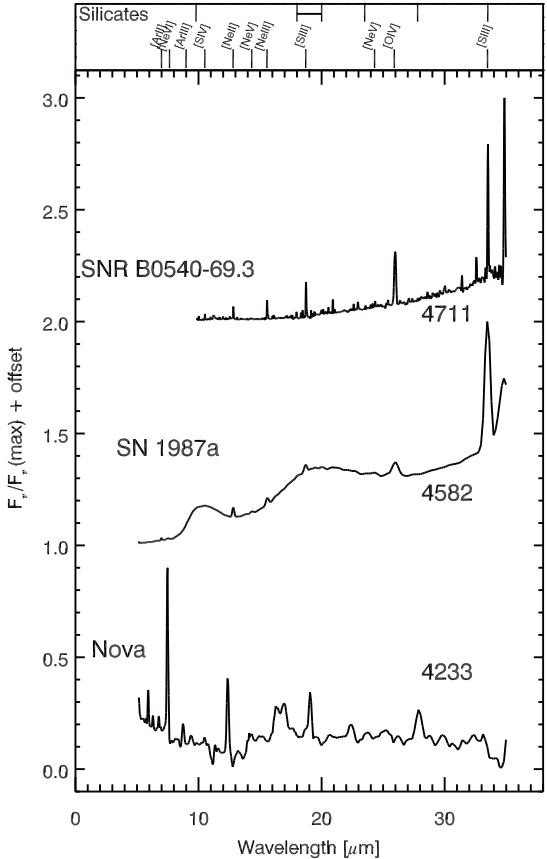}
\caption{Spectra of the two SNRs and the Nova. The spectra are labelled with their SSID number. At the top, the wavelengths of the silicate features and spectral lines are labelled.}  
\label{fig:Novaspec}
\end{figure}

\subsection{Dustless stars} 

Our spectral catalogue contains 30 sources which resemble a stellar photosphere (class {\tt STAR}), with no significant excess emission due to dust or gas. These sources have effective stellar temperatures ($T_{\rm eff}$) ranging from 2\,673 to 8\,000 K and have no long-period variability reported. The spectra of the stellar photospheres are shown in Figure~\ref{fig:miscavg}. 

\subsection{The sample of galaxies}

The LMC IRS sample includes eight galaxies and one candidate galaxy. Seven of these have positively-identified emission features redshifted by $z=0.14$--0.43 $\pm$ 0.1; the rest-frame spectra of these galaxies are shown in Fig.\ 11 of Paper I.

For two of these, \citet{Kim2012} estimated photometric redshifts, both of which are much smaller than our spectroscopic estimates from the IRS spectra: \objectname{SSTISAGEMC J055143.27$-$\allowbreak684543.0} (OBJID 193) has $z_{\rm IRS}=0.43$ compared to $z_{\rm phot}=0.29$, and \objectname{SSSTISAGEMC J053253.36$-$\allowbreak660727.8} (OBJID 138) has $z_{\rm IRS}=0.27$ compared to $z_{\rm phot}=0.10$. Two other galaxies, \objectname{SSTISAGEMC J054524.23$-$\allowbreak683041.4}  (OBJID 184) and \objectname{SSTISAGEMC J043727.61$-$\allowbreak675435.1} (OBJID 2) had both already been identified as galaxies, by \citet{Gruendl2009} and \citet{vanLoon2010}, respectively. Both spectra are dominated by PAH emission and continuum emission steeply rising towards longer wavelengths.

Three galaxies have silicate emission: \objectname{SSTISAGEMC J053730.59$-$674041.6} (OBJID 154; $z=0.28$) resembles a star-forming galaxy that also contains colder dust, while \objectname{SSTISAGEMC J055143.27$-$\allowbreak 684543.0}  (OBJID 193; $z=0.43$) has a flatter mid-IR spectrum and a featureless {\em K}-band  spectrum \citep[2.09--2.36\,$\mu$m;][]{Blum2014}. It is \objectname{SSTISAGEMC J053634.77$-$\allowbreak 722658.6} (OBJID 151), however, which turned out to be the most intriguing. \citet{Hony2011} argued that the lack of far-IR emission and the optical and near-IR faintness indicate that the mid-IR emission is dominated by the AGN torus; this galaxy has the largest silicate-to-continuum ratio known. Subsequently, \citet{vanLoon2015} obtained optical spectra with the Southern African Large Telescope (SALT), refining the redshift to $z=0.1428$ and discovering strong, broad H$\alpha$ emission from near the AGN and relatively narrow stellar absorption in the spectrum of an early-type galaxy. The $M_{\mathrm{BH}}/M_\star$ mass ratio they derived is extreme, and makes it a modest-mass, old galaxy with an over-massive central black hole.

One additional galaxy, \objectname{SSTISAGE1C J051618.70$-$\allowbreak715358.8} (OBJID 78) does not
exhibit any emission features, which is one of the reasons for it to have been classed as a high probability YSO by \citet{Whitney2008} and as an AGB star by \citet{Gruendl2009}. However, the IRS spectrum unmistakably displays a broad silicate absorption feature centred at 14 $\mu$m. This yields a
redshift $z=0.4$; due to the flat bottom of the feature, from 13--15 $\mu$m and the well-documented variation in the peak wavelength of the silicate feature this redshift is not more accurate than $\sim\pm0.1$ but the galaxy nature is firmly established by this measurement. In fact, this object is \objectname{IRAS 05170$-$7156}, which \citet{deGrijp1987} had already identified as a candidate
AGN on the basis of the blue IRAS [25]--[60] colour indicating warmer dust than is typically seen in galaxies. It was subsequently listed in the  PSCz catalogue of quasars compiled by \citet{Saunders2000}, but no optical confirmation of the nature of the object was made, and no redshift had been determined until now. We propose that we are seeing a dusty AGN torus edge-on, filled in by free--free continuum emission, and we predict that an optical spectrum would resemble a Seyfert 2 (narrow-line) as we expect the broad-line region to be obscured from optical view.

The candidate galaxy, \objectname{OBJID 411} (\objectname{[WSI2006] 456)} is in fact a YSO, of a type we classify as {\tt YSO-1}, i.e.\ an early, embedded stage. This source was categorized as a YSO candidate by \citet{Whitney2008} but became the single target of IRS program 30180 ``An Enigmatic Source Towards The LMC'' (PI: G. Fazio) when it was noted to be very bright at 24\,$\mu$m while faint in the IRAC bands. In their IRS proposal, they proposed that it might be a redshift $z\sim2$ starburst galaxy or a dust-enshrouded LBV in a proto-planetary nebula phase. The IRS spectrum, however, shows clear emission at 7.7\,$\mu$m which may be attributed to PAHs, and possibly silicate dust absorption just shortward of 10\,$\mu$m, CO$_2$ ice absorption at 15\,$\mu$m and a dust continuum bulging around 20--30\,$\mu$m. Hence we classify this as a definitive YSO within the LMC.

\subsection{Unknown objects}
\label{sect:UNK}

There are five objects of unknown spectral class in our catalogue, sources we are unable to classify due to low S/N ratio data, or unidentifiable spectral features. There is also little to no information on the nature of the corresponding point sources in the literature. The spectra of the objects of unknown type are shown in Fig.~\ref{fig:unknspec}. Three of the unknown objects are listed and examined in Paper I. 

\objectname{SSTISAGEMC J054546.32−673239.4} (OBJID 185) was classified as an O-AGB star in Paper I. It is a dual chemistry source with both silicate dust emission and weak C$_2$H$_2$ absorption at 13.7 $\mu$m. The 10--12 $\mu$m emission is rather unusual and it may have a weak contribution from both SiC and silicate dust at this wavelength. Furthermore, the 18 $\mu$m feature might not be due to silicates, instead the structure may be due to absorption from carbon-rich dust similar to \objectname{SMP LMC 11} or \objectname{MSX SMC 049}. Due to the unusual dust chemistry this source exhibits we re-classify OBJID 185 to {\tt UNK}.

The spectrum of \objectname{IRAS 05413-6934} (OBJID 729) exhibits features at 21 and 30 $\mu$m. While it may also contain PAH features at 6--9 $\mu$m, its spectrum is too noisy to be certain. If these features are verified then \objectname{IRAS 05413-6934} is likely to be a carbon-rich post-AGB star.

\begin{figure}
  \includegraphics[trim=2.0cm 2cm 0cm 4cm, width=6.5in]{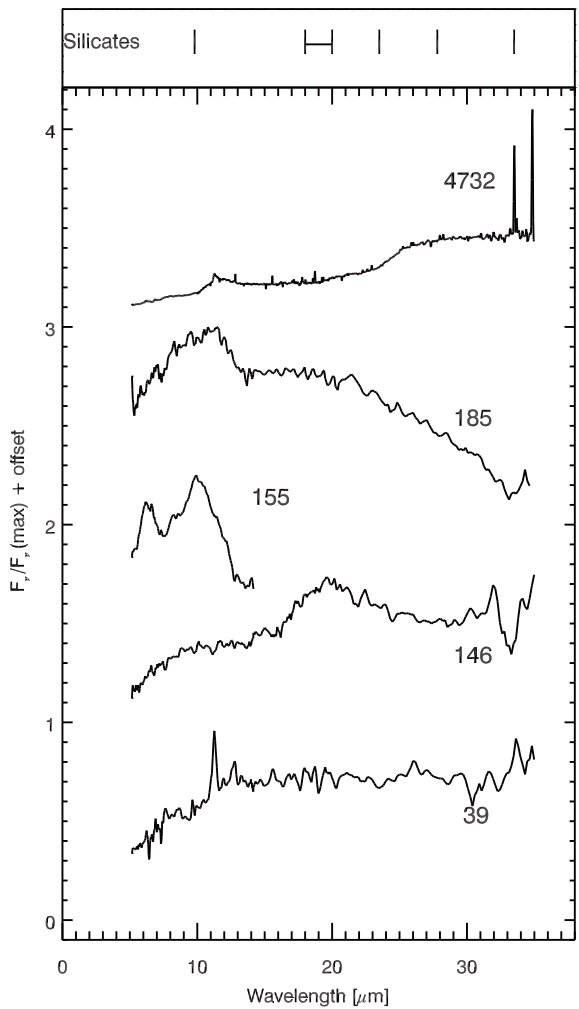}
\caption{Spectra of the unidentified objects. The spectra are labelled with their SSID number. At the top, the wavelengths of the silicate features are given. }  
\label{fig:unknspec}
\end{figure}


\section{Comparison with photometric classifications}
\label{sec:photCompare}

Photometric and spectroscopic classifications each have their own advantages and disadvantages: photometric methods are significantly less time-consuming to apply to large samples, whereas spectroscopic methods are more reliable, especially when combined with a photometric SED over a large wavelength range, as we have in this case. This greater reliability means that we can test photometric classifications with spectroscopic classifications in order to gauge their accuracy. There have been a number of photometric surveys of the LMC, or surveys which have included the LMC, and while we do not try to test the results of all these surveys, there have been several in recent years which include a classification of the point sources detected.\footnote{A note about catalogue matching: Source matching is not a trivial matter, especially in crowded regions of the LMC. Where possible we have used the SAGE and 2MASS associations to match sources in the papers discussed below. In some cases this has not been possible (e.g., when the authors have decided to use their own identifiers). In these cases, we have matched according to position, using a 3\arcsec\ radius, and where possible, matching IRAC photometry as an added check. }
In the following subsections we look at stages of evolution and classes of object that we encounter, and the seminal photometric classifications for each group.

\subsection{Young Stellar Objects}
\label{sect:compyso}

There have been three LMC-wide photometric classification efforts of YSOs in the past decade: \citet{Whitney2008}, \citet{Gruendl2009}, and \citet{Seale2014}, while four detailed studies focused on individual star formation regions: \citet{Chen2009, Chen2010}, \citet{Romita2010}, and \citet{Carlson2012}. Other studies producing lists of candidate YSOs as a by-product were performed by \citet{Kastner2008} and \citet{Buchanan2009}. We discuss these publications below.

\subsubsection{\citet{Whitney2008}} 

The work of \citet{Whitney2008} significantly expanded on the number of known YSOs in the LMC. Using {\em Spitzer} photometry, they identified over 1\,000 YSO candidates, with a bias towards young high- or intermediate-mass objects. They were less sensitive to the most massive YSOs, though, due to the limitations of the SAGE Point Source Catalogue (PSC) upon which their source list was based. An extensive set of colour--magnitude selection rules (based on theoretical modelling) was applied to select candidate YSOs from the SAGE PSC, resulting in a source list of 1197 objects. Further screening with more stringent colour--magnitude criteria and removal of known contaminants led to a list of  458  ``highly probable'' YSOs (YSO\_hp). Other classes of object were also identified in the process of separating contaminants: evolved stars and PN, and they will be discussed separately below.

The two panels in Figure~\ref{fig:YSO_col_comp} show the [8.0] versus [4.5]--[8.0] and [8.0]--[24] CMD, used in the \citet{Whitney2008} classification scheme. The objects from the IRS sample are over-plotted on a Hess diagram of the entire SAGE PSC. In this figure the `YSO\_hp' sources from \citet{Whitney2008} are shown in light blue. 
Of the 458 YSO\_hp sources, we have matched 92 to IRS objects. Of these 92, 79\% have been identified as YSOs or H\,{\sc ii} regions. The majority of the pollutants are evolved stars (14\%) and galaxies (4\%). The purpose of the cull that brought the original list of YSO candidates down to the `YSO\_hp' list was to exclude evolved stars; \citet{Whitney2008} reported moderate success with this, but it is clear from Figure~\ref{fig:YSO_col_comp} that contaminants still remain. 
The list of 739 YSO candidates with the more relaxed criteria contains 54 {\em Spitzer}-IRS matches, of which  63\% match our classification of YSO or H\,{\sc ii} region. Again, the majority of contaminants are AGB and post-AGB stars (35\%).

\begin{figure*}
\centering
\includegraphics[trim=0.0cm 0cm 0cm 0cm, width=6.5in]{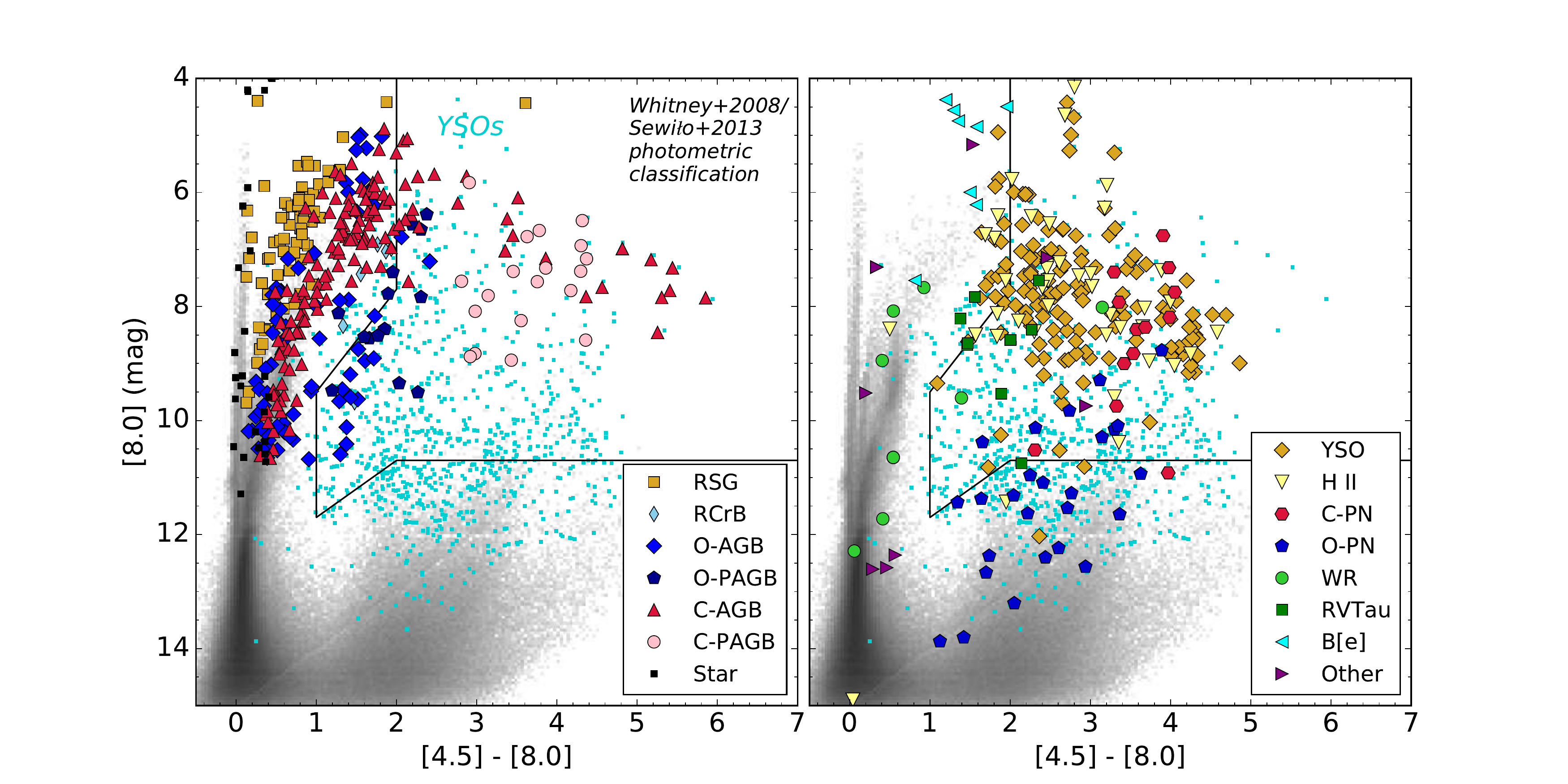}
\includegraphics[trim=0.0cm 0cm 0cm 0cm, width=6.5in]{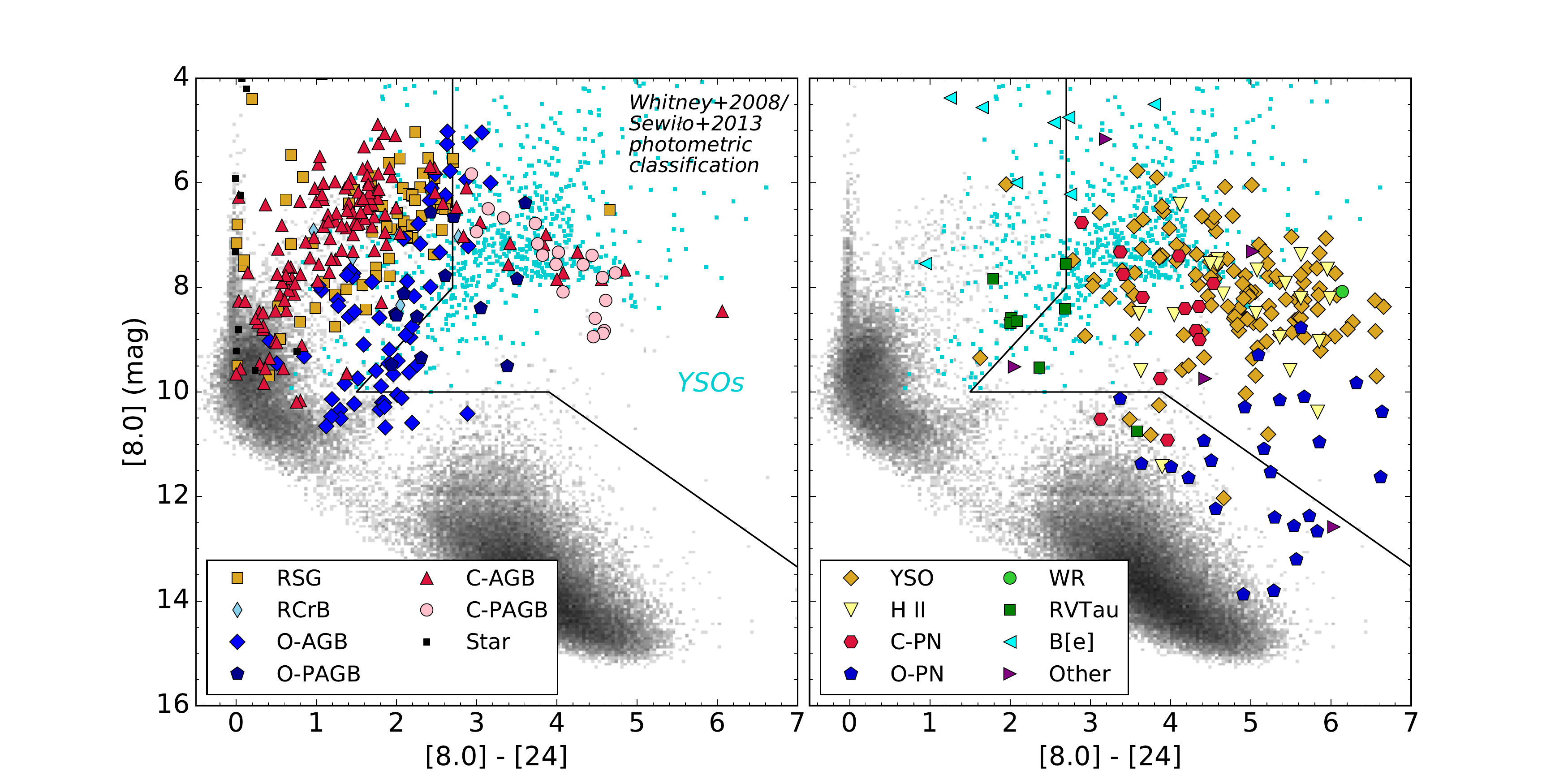}
\caption{The [8.0] versus [4.5]--[8.0] CMD (top) and the [8.0] versus [8.0]--[24] CMD (bottom) used in the \citet{Whitney2008} classification of YSO candidates. The cuts used by \citet{Carlson2012} and \citet{Sewilo2013} to select YSOs are shown by solid black lines. The grey-scale in the background shows a Hess diagram of the SAGE LMC catalogue, with light blue dots representing the photometrically selected YSOs from \citet{Whitney2008}. The spectral sample is plotted as coloured points and separated by spectral class according to the legend.}  
\label{fig:YSO_col_comp}
\end{figure*}

\subsubsection{\citet{Gruendl2009}}

\citet{Gruendl2009} identified approximately 2\,900 point sources in the LMC using {\em Spitzer} IRAC and MIPS data. Almost 1200 of these sources are proposed as likely YSOs, whilst $\sim$1100 of these are thought to be background galaxies. They adopt relaxed selection criteria for different classes: stars and evolved stars are separated with a colour selection of $[4.5]-[8.0]>2.0$, whilst galaxies are isolated with $[8.0]>14-([4.5]-[8.0])$. This source selection procedure is less strictly defined than that of \citet{Whitney2008}, which would have eliminated 850 of the $\sim$1200 YSO candidates. Contrarily, the \citet{Gruendl2009} selection misses fainter and more evolved YSOs that are included in the \citet{Whitney2008} sample. Approximately 20--30\% of the \citet{Whitney2008} sources would have been classified as background galaxies by the colour cut of \citet{Gruendl2009}.

Based on an analysis of the morphology, SED (optical to mid-IR), and immediate environment of each source, \citet{Gruendl2009} identified three different qualities of YSO candidates:  ``Definite YSOs'', ``Probable YSOs'' and ``Possible YSOs''. We have found 282, 17, and 4 IRS sources in these groups, respectively. An extremely high (99\%) percentage of the objects in the ``Definite YSOs'' group are spectroscopically identified as YSOs. The only two contaminants in the selection are an oxygen-rich AGB star, \objectname{MSX LMC 610} (OBJID 569) and the source corresponding to spectrum SSID 4732 which has an {\tt UNK} spectral type. Thus we can say that the method of ``Definite YSOs'' selection employed by \citet{Gruendl2009} is extremely reliable. In the two other groupings with lower confidence, we find 72\% of the ``Probable YSOs'' are also YSOs in our evaluation. The contaminants in this grouping are mostly evolved stars and PNe.
None of the four ``Possible YSOs" are YSOs in our classification.

\subsubsection{\citet{Carlson2012}}

\citet{Carlson2012} presented alternative colour selection criteria for the identification of YSOs in star-forming regions in the LMC. This selection focused on lower mass ($\lesssim 4\ M_{\rm \odot}$) YSO candidates that may have been missed by earlier surveys. 1\,045  low- to high-mass YSO candidates, of which 918 are entirely new, were identified via colour-selections and SED fitting in nine active star-forming regions in the LMC; we find 106 matches to our spectral catalogue. The \citet{Carlson2012} catalog of YSO candidates contains a complete list of YSO candidates for each H\,{\sc ii} region, including those identified as YSO candidates in other studies \citep{Chen2009,Chen2010,Romita2010} that do not fulfill all their criteria or are missing from the SAGE PSC. 

The colour-selection described by \citeauthor{Carlson2012} was very successful at identifying YSOs and compact H\,{\sc ii} regions in all nine star-forming regions with a  99\% accuracy. However, Figure~\ref{fig:YSO_col_comp} suggests that this accuracy would be reduced if these colours cuts were applied on a global scale rather than in verified star-forming regions. 
Twenty-five of the 105 sources spectroscopically identified as YSOs and compact H\,{\sc ii} regions were included in the list of those `well-fit' by YSO models3. Nine of these we class as {\tt H\,{\sc ii}} regions; we class four as {\tt YSO-1}, two as {\tt YSO-2}, seven as {\tt YSO-3} and three as {\tt H\,{\sc ii}}/{\tt YSO-3}. All the YSO classes were determined to be Stage I YSOs by \citet{Carlson2012}. 
The final list of `well-fit' YSO candidates has no contamination from other evolutionary classes, while the YSO candidates identified via colour selection include one evolved star contaminant.

\subsubsection{\citet{Seale2014}}

Using {\em Herschel} far-IR photometry from the HERITAGE survey, \citet{Seale2014} identified $\sim$35\,000 far-IR point sources in the SAGE field. These represent the dustiest populations of sources in the LMC. Using colour selections, 3\,518 sources were thought to be very young, embedded YSOs or dusty ISM clumps, while 9\,745 are background galaxy candidates. 
\citet{Seale2014} identified {\em Herschel} YSO candidates by requiring detections in at least three {\em Herschel} and SAGE MIPS 24 $\mu$m bands, and applying the criteria that would ensure that the measured photometry is not contaminated by the emission from neighboring sources.  All sources identified in literature as non-YSOs or as {\em Herschel} background galaxy candidates were removed.  \citet{Seale2014} classified sources brighter than 100 mJy in the 250 $\mu$m band as ``probable YSO candidates'' and those fainter as ``possible YSO candidates''. 

Of  the $\sim$35\,000 far-IR sources, we have matched 246 objects to {\em Spitzer} spectra.  The 246 sources in common contain 228 sources spectroscopically identified as YSOs or H\,{\sc ii} regions. Of these 228 spectroscopically confirmed YSOs or H\,{\sc ii} regions, \citet{Seale2014} classified 204 (83\%) as either ``probable'' (191) or ``possible'' (13) YSOs. The remaining 24 spectroscopically confirmed YSOs were not assigned a class by \citet{Seale2014} as they were too faint.

\subsubsection{\citet{Kastner2008} and \citet{Buchanan2009}}

There are 46 matches between the YSOs and compact H\,{\sc ii} regions in our sample and the sample of \citet{Kastner2008} and \citet{Buchanan2009} (see Section.~\ref{sec:2MASS-MSX_Class}). Of these, 42 sources (91\%) were thought to be H\,{\sc ii} regions, and we also find that the vast majority  (31/42) are indeed compact H\,{\sc ii} regions or class {\tt YSO-3/HII} intermediate objects. The other 14 sources in common between the two catalogues were classified by us as {\tt YSOs 1-4}; according to \citet{Kastner2008} they were thought to be H\,{\sc ii} regions (11 objects), are not assigned a class  (2 sources) and one source was classified as C-AGB.

\subsection{Red supergiants and Asymptotic Giant Branch stars}
\label{sect:compAGB}

Due to their importance in late stages of stellar evolution and their feedback of material into the dust budget of the LMC, several authors have used mid-IR colours to identify RSG and AGB stars in the Magellanic Clouds. 
Seminal work using {\em Spitzer} photometry includes the studies by \citet{Blum2006}, \citet{Kastner2008}, \citet{Matsuura2009}, \citet{Boyer2011} and \citet{Riebel2012}.
We discuss each of these publications below. For completeness, we also discuss works which do not focus on identifying AGB stars, but that nevertheless contain a number of evolved star candidates in their catalogs. 

\subsubsection{\citet{Blum2006} and \citet{Boyer2011}}

As a first step toward the identification of AGB stars in the LMC, \citet{Blum2006} identified about 32\,000 colour-selected evolved stars brighter than the tip of the RGB, which includes 17\,500 oxygen-rich, 7\,000 carbon-rich, and 1\,200 `extreme' AGB (x-AGB) stars. In Paper I we compared the initial \SSp sample of 197 sources to the classifications of \citet{Blum2006} and confirmed that the `extreme' AGB stars are predominantly carbon-rich. 

In this paper we compare our classifications to those of \citet{Boyer2011}. This  classification scheme supersedes that of \citet{Blum2006} for mass-losing evolved stars in the Magellanic Clouds as it incorporates (and improves upon) both the \citet{Blum2006} and \citet{Cioni2006} photometric classifications. This refined classification scheme uses the $K_{s}$ vs.~$J-K_{s}$ colours (shown in Figure~\ref{fig:Boyer_col_comp}) to identify evolved stars which are brighter than the the tip of the red giant branch, and to separate carbon- and oxygen-rich AGB stars, and RSGs. For sources with a dust excess the $J-[8.0]$ and $[3.6]-[8.0]$ CMDs were used to identify evolved stars which are experiencing a superwind. 
Finally objects which have a rising continuum from 8 to 24 \mum\  ($[8]-[24] > 2.39$ mag) were classified as far-IR (FIR) sources. This class was added as a mechanism to remove potential contamination from YSOs, compact H\,{\sc ii} regions, PNe and background galaxies from the evolved star population, though some evolved stars remain in the FIR category.


\begin{figure*}
\centering
\includegraphics[trim=0.0cm 0cm 0cm 0cm, width=6.5in]{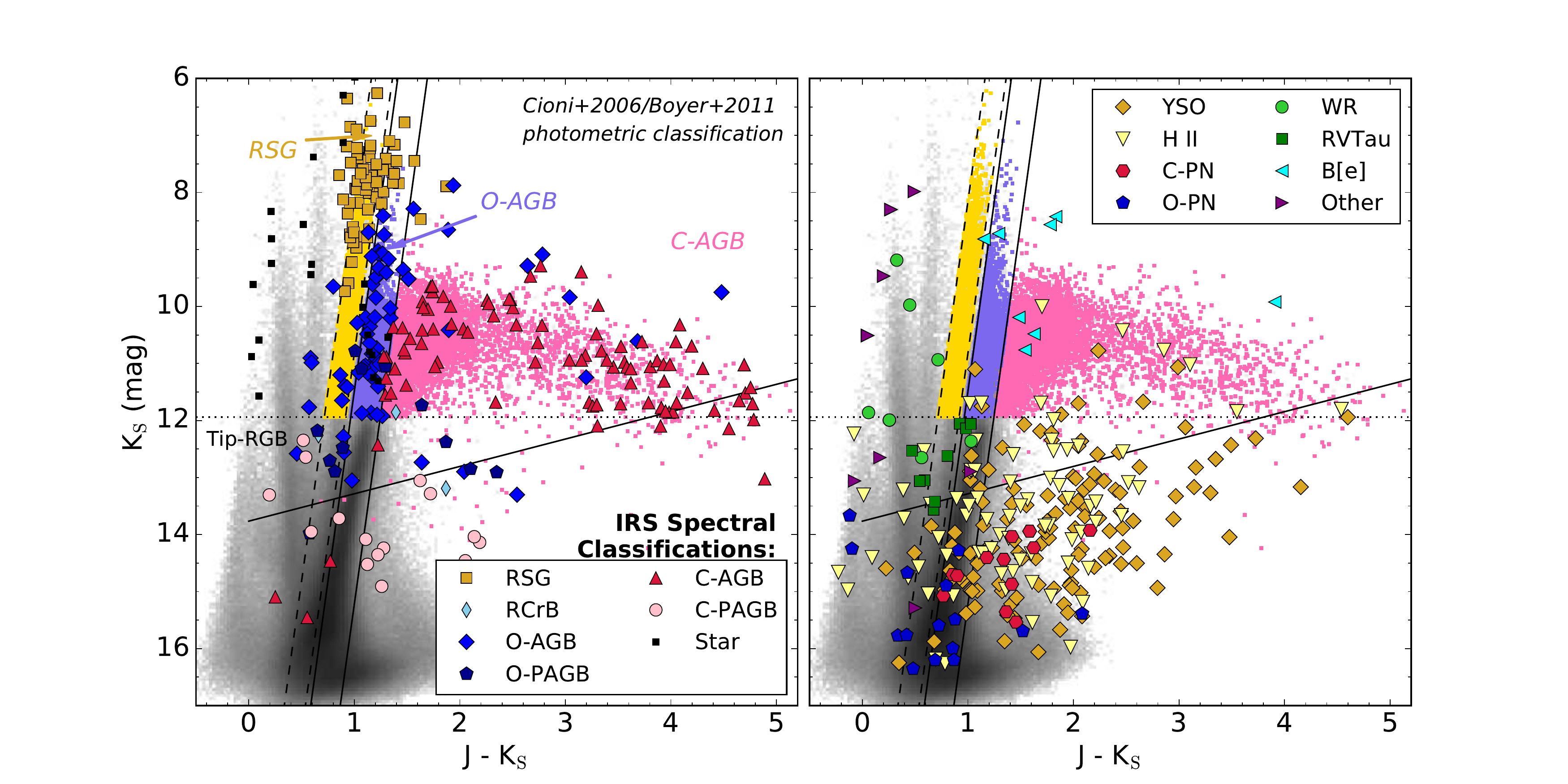}
\includegraphics[trim=0.0cm 0cm 0cm 0cm, width=6.5in]{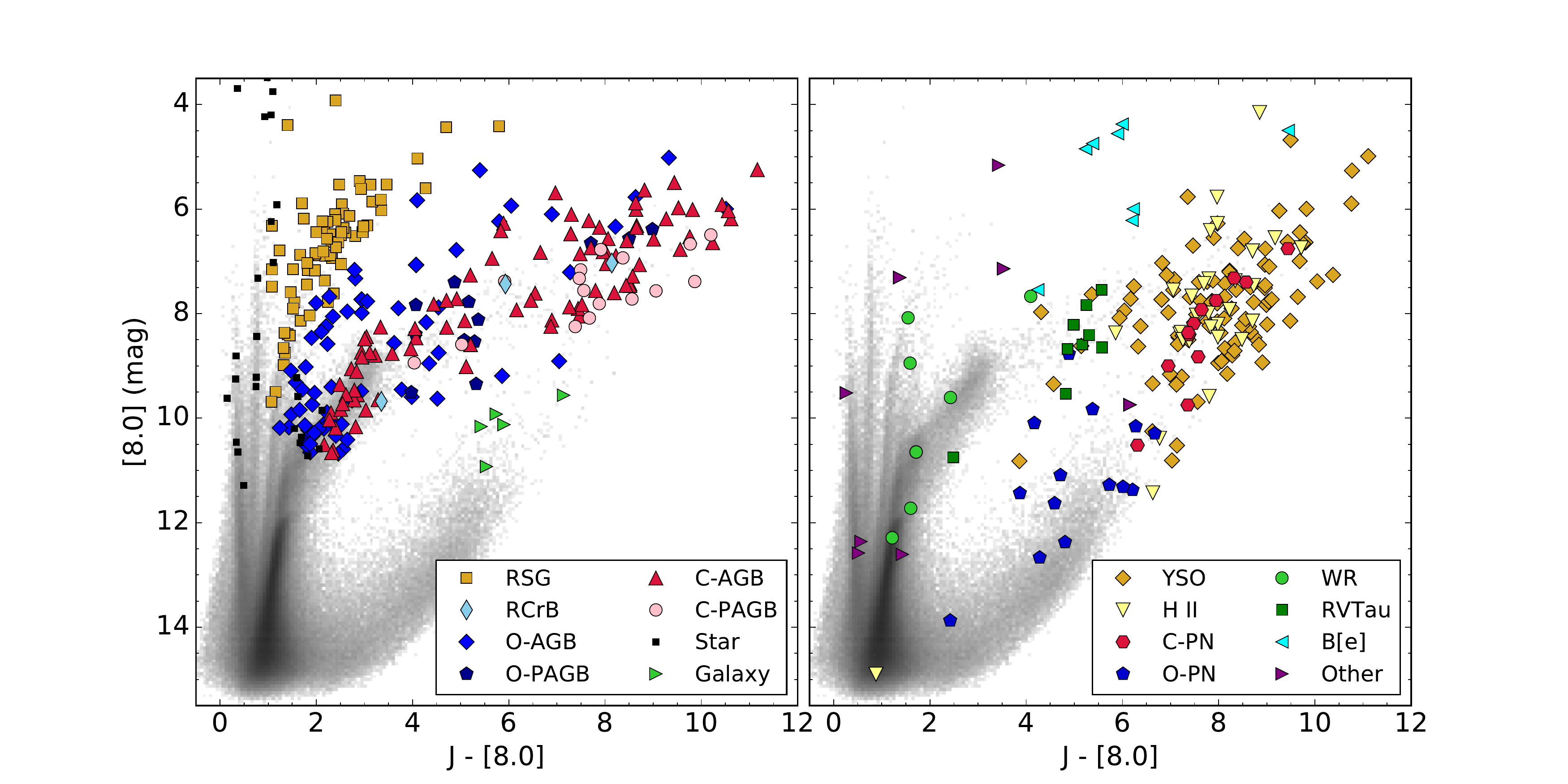}
\caption{The $K_{s}$ vs.~$J-K_{s}$ (top) and $[8.0]$ vs.~$J-[8.0]$ (bottom) CMDs showing the distribution of the {\tt O-AGB}, {\tt RSG} and {\tt C-AGB} classes plotted over the evolved star sample of \citet{Boyer2011}. The lilac and yellow dots represent the RSG and O-AGB object as they are classified via the $K_{s}$ vs.~$J-K_{s}$ CMD. For simplicity, the pink points include both C-AGB and x-AGB sources.
For clarity the CMD has been duplicated to show the locations of the various object classes. The IRS spectrum classification symbols are as for Fig.~\ref{fig:YSO_col_comp}.}  
\label{fig:Boyer_col_comp}
\end{figure*}


Most (134/148) of the spectroscopically identified carbon stars were classified as either a C-AGB star (23 objects), an x-AGB star (100 objects), or an FIR source (14 objects) by \citet{Boyer2011}. Two IRS sources were misidentified as O-AGB stars by \citet{Boyer2011}. Ten carbon stars were not included in the \citet{Boyer2011} catalog because they lacked either 2MASS magnitudes or 3.6~\micron\ magnitudes or because they fell outside the employed colour boundaries.


Objects that we classify as {\tt O-AGB} stars fall across the range of evolved star photometric classes for the LMC. Twenty-five IRS sources were correctly identified as O-AGB stars by \citet{Boyer2011}; all the early-type O-AGB stars (with molecular absorption features in their spectra) lie within this class. Twelve of the spectroscopically-confirmed {\tt O-AGB} stars are matched to their aO-AGB photometric class; 
confirming that this photometric class represents a subset of oxygen-rich AGB stars \citep{Boyer2015c} that are just beginning to form a significant amount of dust, rather than an indication of S-type AGB stars (in which the ratio of carbon to oxygen is close to 1). Four sources were placed in the x-AGB category and an additional 13 were classified as FIR sources by \citet{Boyer2011}. Sources in these latter categories are highly enshrouded by dust and have strong 10 and 18 \mum\ features due to amorphous silicates.  Four IRS {\tt O-AGB} sources fall in the C-AGB region of colour space and five fall in the RSG region. Finally, twelve {\tt O-AGB} stars (24\% of the IRS sample)} are not included in the \citet{Boyer2011} sample either because they lack 2MASS magnitudes or because they fall outside the 2MASS colour-selection.

Near-IR colours are most successful for identifying O-AGB stars which appear as stellar photospheres with an SiO inflection at 8 \mum\ or with mild dust emission. Dustier {\tt O-AGB} sources show a wide colour spread that was also seen in the SMC (Paper II), and indicates that once thermally-pulsing O-rich sources start to form significant amount of dust they cannot be easily isolated in colour space. Dusty O-rich AGB stars are rare at low metallicity compared to carbon stars, but their numbers are expected to increase in more metal-rich systems. Depending on {\it Spitzer} colours from studies of the Magellanic Clouds to classify dusty AGB stars would in that case lead to large uncertainties in estimates of the total amount of carbon-rich and oxygen-rich dust production. 

One region of colour-space where one might expect significant overlap is the O-AGB/RSG boundary. 
This disagreement between the O-AGB/RSG classification can be in part explained by the overlap in bolometric magnitude between massive O-AGB stars undergoing hot bottom burning and the RSGs. Furthermore, the lowest mass RSGs become RSGs at a bolometric luminosity below the classical AGB limit. They will be a little warmer than O-AGB stars at the same bolometric luminosity, but that may be hard to tell from infrared photometry alone. Variability may also play a role in the cross-contamination between the two samples.


The 71 {\tt RSGs} in our spectroscopic sample fall mostly within the RSG (33), O-AGB  (7), x-AGB (1) and FIR (12) regions of colour-space, giving a $\sim$75\% agreement. Sources in these regions are thought to have an oxygen-rich chemistry, thus the chemical type is more reliably identified than the stellar types.  One {\tt RSG} (\objectname{IRAS 05389$-$6922}; OBJID 660) was classified as a C-AGB based on its photometric colours. Like the O-AGB stars, RSGs with high optical depths have red $[8.0]-[24]$ colours. This is due to the strong amorphous silicate dust feature at 18 \mum\ and (in some instances) silicate self-absorption suppressing the flux at shorter wavelengths. Thus O-AGB stars and RSGs undergoing significant mass-loss will reside in the FIR photometric class rather than the O-AGB region of colour space. In total, 39 spectroscopically confirmed {\tt AGB} stars and {\tt RSGs} fall within the FIR category of \citeauthor{Boyer2011}. The other 104 IRS sources in this region are spectroscopically classified as {\tt YSOs} and compact H\,{\sc ii} regions (74); post-AGB stars (17); PNe (6); or are rare objects in the {\tt OTHER} category.

Similar to the O-AGB stars, $\sim$77\% (54/71) of the {\tt RSGs} in our IRS sample have a match in the \citep{Boyer2011} evolved star photometric catalogue. The missing {\tt RSGs} either have no 2MASS magnitudes or fall outside the \citep{Boyer2011} colour-cuts, with most (12) falling bluewards of the RSG cut shown in Figure~\ref{fig:Boyer_col_comp}.


Six stellar photospheres ({\tt STAR}s) were matched to the \cite{Boyer2011} catalogue. These are all thought to be O-AGB or aO-AGB stars according to their $J-K_{s}$ colours, raising the possibility that these sources are early-AGB stars which have not yet started to thermally pulsate or are OB stars with an IR excess, which is not thought to be related to the star itself \citep[see][]{Adams2013, Sheets2013}. 
None of these objects are found in the \citet{Vijh2009} catalogue, signifying that they are not mid-IR variables, which would be consistent with a very early-AGB stage.


Pollution of the \citeauthor{Boyer2011} x-AGB sample (145 objects in the IRS sample) is predominantly by YSOs and compact H\,{\sc ii} regions (31) along with 3 {\tt O-PAGB} stars. There is little overlap with the population of dusty O-AGB sources ($\lesssim$3\%). Five massive evolved stars also reside in the x-AGB region of colour--magnitude space, however 4 of these are {\tt RCrB} stars which have a carbonaceous chemistry. These rare objects are not included in colour selections.

\subsubsection{\citet{Buchanan2006, Buchanan2009} and \citet{Kastner2008}}
\label{sec:2MASS-MSX_Class}

\citet{Buchanan2006, Buchanan2009} and \citet{Kastner2008} used a 2MASS-{\em MSX} colour classification system to characterise the most luminous mid-IR point sources in the LMC. This classification scheme has been augmented by IRS spectra of 60 bright 8\,$\mu$m sources selected from the {\em MSX} catalogue \citep{Egan2001}. 

132 of the 250 most luminous point sources in the LMC at 8\,$\mu$m \citep{Kastner2008} have been observed with the IRS.  Unsurprisingly the brighter \citet{Kastner2008} sample (incorporating the updates of \citealt{Buchanan2009}) contains a large number of matches  (33) to the RSGs in our sample. A further eight massive stars (mostly comprised of B[e] supergiants) were also identified. The classifications of RSGs and massive stars have an  85\% agreement with our catalogues. 
The \citet{Kastner2008} sample was also very successful at identifying C-AGB stars. Only one of the 37 matches was incorrectly identified as an H\,{\sc ii} region, while nine of the twelve O-AGB stars were classified correctly, the other three sources were classified as carbon-rich AGB stars.

\subsubsection{\citet{Matsuura2009, Matsuura2013}}

\begin{figure*}
\centering
\includegraphics[trim=0.0cm 0cm 0cm 0cm, width=6.5in]{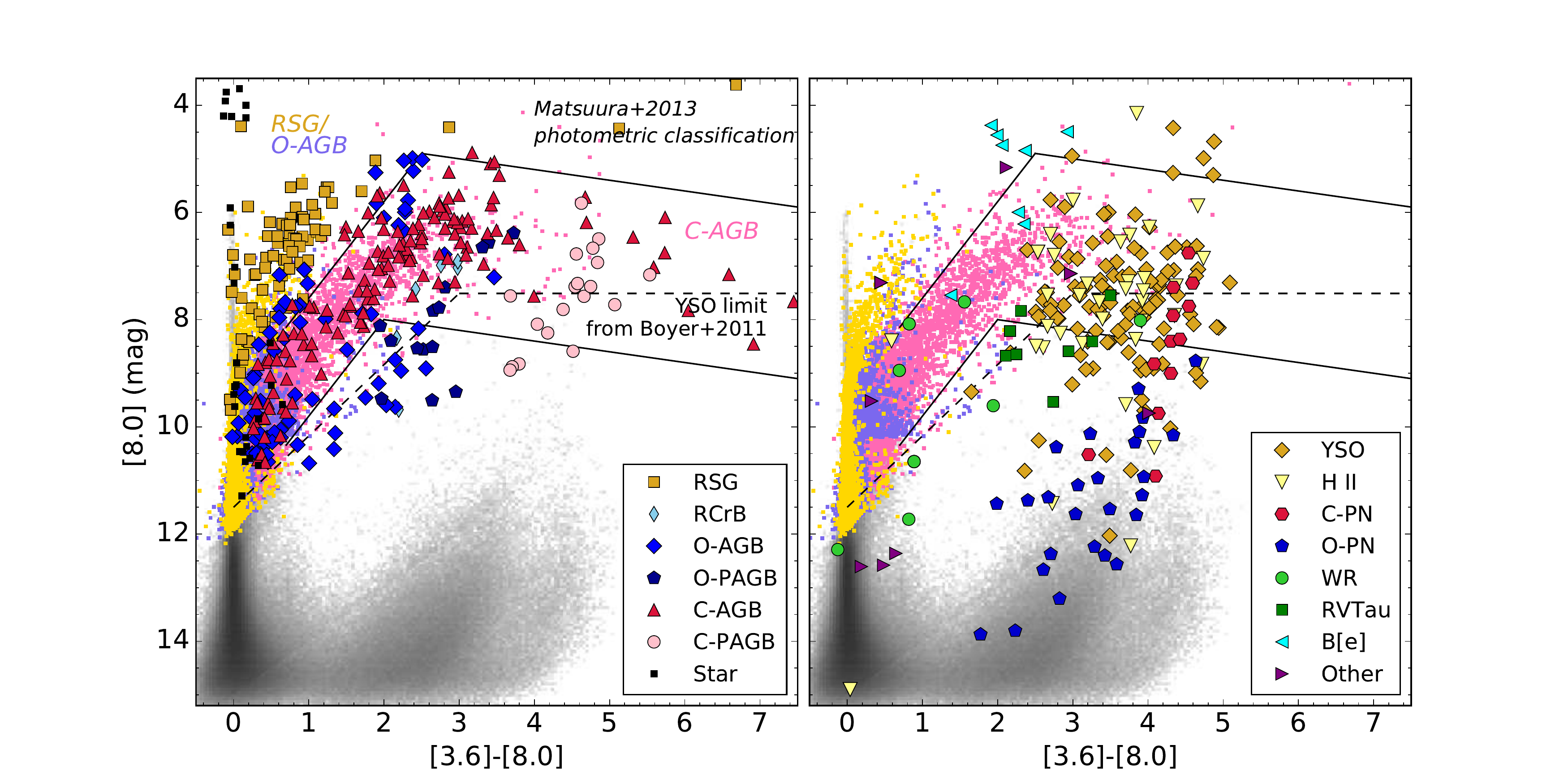}
\includegraphics[trim=0.0cm 0cm 0cm 0cm, width=6.5in]{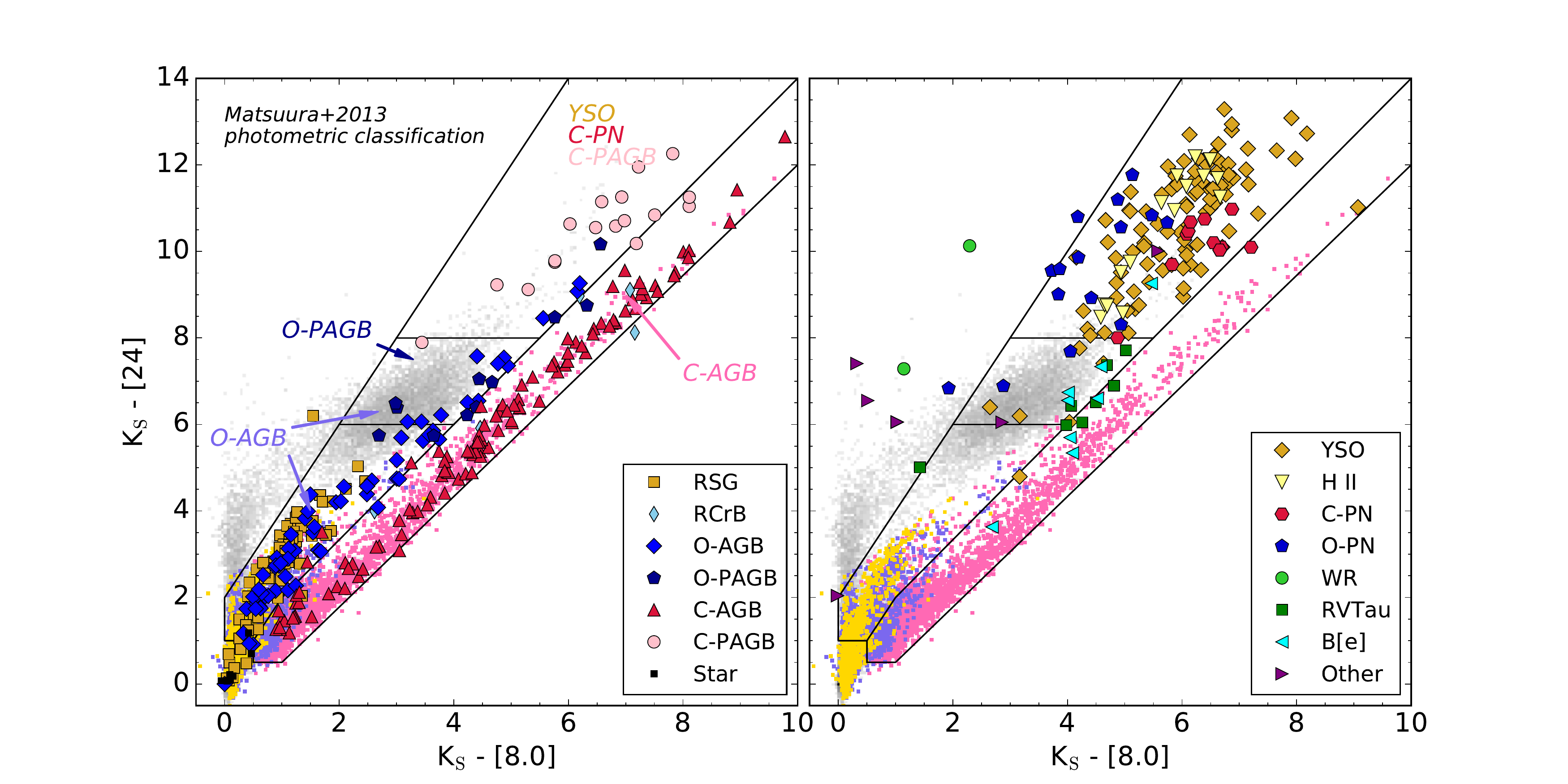}
\caption{The infrared colour--magnitude diagrams (CMDs) used in the \citet{Matsuura2009, Matsuura2013} photometric classification method. The solid lines in the [8.0] versus [3.6]--[8.0] CMD (top) are used to separate carbon and oxygen-rich evolved stars in the LMC. The dashed line shows the position of the YSO limit used by \citet{Boyer2011}. The $K_{s}-[24]$ vs.~$K_{s}-[8.0]$ colour--colour diagram (CCD; bottom) was an additional step introduced by \citet{Matsuura2013} to provide a cleaner separation between oxygen-rich stars and carbon-rich stars. The symbols and colours have the same meaning as in Fig.~\ref{fig:Boyer_col_comp}.}  
\label{fig:Matsuura_col_comp}
\end{figure*}

The colour classification scheme proposed by \citet{Matsuura2009, Matsuura2013} was used to identify mass-losing AGB stars and RSGs and determine their mass-loss rates.
We applied the [8.0] versus [3.6]--[8.0] CMD cuts described by \citet{Matsuura2009, Matsuura2013} to our sample in Figure~\ref{fig:Matsuura_col_comp} and compare their classifications to our own. 

This photometric classification was designed with redder sources in mind and is not intended to recover early-AGB stars and objects with low mass-loss rates (i.e.~sources with $[3.6]-[8.0] < 0.7$ mag). Furthermore, \citet{Matsuura2009, Matsuura2013} only make subdivisions according to the chemistry rather than the stellar type and thus do not distinguish between O-AGB stars and RSGs.  

Applying the $[3.6]-[8.0] > 0.7$ mag cut eliminates 358 sources, which are mostly comprised of stars, and oxygen-rich post-main sequence stars with mild to moderate dust emission.  
The C-AGB region defined by \citet{Matsuura2009} contains 283 IRS sources, of which 109 are C-AGB stars (38.5\%). Nineteen embedded O-AGB stars also occupy this section of colour space and one must be wary of contamination from these oxygen-rich evolved sources when applying mass-loss relations. This region also has significant pollution from YSOs and compact H\,{\sc ii} regions (110 sources), and, at redder colours, post-AGB stars and PNe (31 objects). Eleven massive stars are also included in this selection.  

The cuts used  by \citet{Matsuura2009} to isolate RSGs and O-AGB stars are more successful; sixty nine sources occupy the RSG/O-AGB region, of which 37 are RSGs and eight are massive O-AGB stars (65\% accuracy). Here the contaminants are ten C-AGB stars, seven massive stars (five B[e] stars, one WR and one YSG) and seven  YSOs/compact H\,{\sc ii} regions. 

A second cut was applied to the SMC sources in \citet{Matsuura2013} using the near-IR $K_{s}$  band and the mid-IR [24] magnitudes (Figure~\ref{fig:Matsuura_col_comp}, lower panel). This second classification was introduced to remove contamination from YSOs and reduce overlap between carbon- and oxygen-rich AGB stars. The second step takes precedence if the two colour classifications disagree. From Figure~\ref{fig:Matsuura_col_comp} it is clear that this second step provided a vast improvement to the reliability of the colour-classifications, although dusty O-AGB stars, and O-PNe are still overlooked in the selection.

\subsubsection{\citet{Riebel2012}}

Unlike the other works discussed thus far, \citet{Riebel2012} used the `Grid of Red supergiant and Asymptotic giant branch ModelS' ({\sc grams}; \citealt{Sargent2011, Srinivasan2011}) to identify evolved stars in the LMC and constrain their chemistry.
The initial sample is based on a subset of the full SAGE LMC catalogue, and was selected using the evolved star photometric colour-cuts of \citet{Blum2006} and \citet{Boyer2011}.

There are 263 sources in common with our spectral sample; the preliminary colour selection and fits with the {\sc grams} models successfully identified 82\% (243/295) of the spectroscopically confirmed evolved stars. 
Of these the selection and identification of {\tt C-AGB} stars was the most successful, excluding only fifteen sources and misidentifying another four as oxygen-rich.
The chemical type of all the {\tt RSGs} was correctly determined to be oxygen-rich by the {\sc grams} models, although 20 of the 71 {\tt RSGs} are not included in the \citet{Riebel2012} catalogue. Finally, 52 of the 73 (71\%) spectroscopically confirmed {\tt O-AGB} stars were included in the catalogue, with four incorrectly determined to be carbon-rich.
Consequently 52 evolved stars that are undergoing mass-loss were not included in the final dust budget estimate of the LMC by \citet{Riebel2012}; the excluded sources are predominantly oxygen-rich. Thus we would expect both the global and oxygen-rich dust-production rates for the LMC need to be revised upwards. 

Eight stellar photospheres ({\tt stars}) also fall in the sample, of which all but one were identified as oxygen-rich.
In total twelve sources not on the giant branches are included in the \citet{Riebel2012} directory. Five YSOs were all classified by \citet{Riebel2012} as extreme carbon stars. The other seven sources were determined spectroscopically to be {\tt RCrB stars}  (3), {\tt O-PAGB} stars (2) and a {\tt B[e]} star. The final source was classified as {\tt UNK}.

\subsubsection{\citet{Seale2014, Jones2015b}}

\citet{Seale2014} proposed that a small population of dusty objects in the late stages of stellar evolution may be present in the {\em Herschel} point-source catalogue for the LMC. \citet{Jones2015b} identified all the post-main sequence stars in this catalogue, including fifteen sources with IRS spectra. In the LMC, two are AGB stars, three are RSGs, and five are masssive stars (three B[e] stars; a BSG and a WR star).  The remainder are post-AGB stars and PNe.

\subsubsection{\citet{Whitney2008}}

\citet{Whitney2008} included a list of evolved stars which were contaminants in their YSO selection process. Many of these had been previously identified as carbon stars \citep[e.g., using the catalogues of ][]{vanLoon1999,vanLoon2006b}, and others were separated from the YSO populations using colour--magnitude cuts. \citet{Whitney2008} distinguished several different classes within their `evolved' group: AGB and post-AGB stars, Cepheids, Wolf--Rayet stars and emission line objects. 
Thirty-six of these `evolved' sources were matched to an IRS object, and 83\% of these are classified as evolved stars by us. The remaining 17\% are composed mainly of YSOs (11\%) and objects of unknown spectral type (6\%).

\subsubsection{\citet{Gruendl2009}}

As a by-product of their YSO classification, \citet{Gruendl2009} identified 117 evolved stars, which they classify into two groups: AGB and ERO. We find 53 of their evolved-star candidates in our spectroscopic sample. We have classified 48 (87\%) of these as RSGs, AGB stars or post-AGB stars, which substantiates \citet{Gruendl2009}'s photometric identification. The remaining five sources are evolved YSOs, or, in one case, a star.

The 14 sources designated as EROs by \citet{Gruendl2009} are thought to be extreme carbon stars. 
Eleven matches to the EROs are found in the IRS sample, ten of which are classified as {\sc C-AGB}. The remaining object (\objectname{IRAS 05346$-$6949}; OBJID 615) is classified as a {\sc RSG} due to a pronounced absorption feature at 10\,$\mu$m. 
Furthermore, the IRS spectrum of \objectname{IRAS 05346$-$6949} shows no trace of carbon-rich content, and thus we suggest that this object is misclassified as an extreme carbon star.

\subsubsection{\citet{Kamath2015}}

\citet{Kamath2015} identified 382 optically bright M stars in the LMC based on the presence of strong TiO and VO lines in their optical spectra.  Nineteen of these objects have IRS spectra, and all 19 are classified by us as {\tt O-AGB} stars (13 objects) or  {\tt RSGs} (6 objects).

\citet{Kamath2015} also compiled a list of carbon stars based on C$_2$, CN, and CH absorption lines in the optical spectra. Of the 55 carbon stars identified, four have IRS spectra, and two of these are classified as carbon stars by us. The remaining two objects are a YSO and an oxygen-rich AGB star. Interestingly, the latter source \objectname{SSTISAGEMC J053128.44$-$701027.1} (OBJID 130) has two clear features at 10 and 20\,$\mu$m in the mid-IR spectrum, giving strong evidence for a silicate-rich circumstellar dust. This object may be undergoing the transition from oxygen- to carbon-rich, i.e., it is a mixed-chemistry object, such as those surveyed by \citet{Smolders2012}.

\subsubsection{OGLE classifications}

125 of the 142 sources with matching OGLE information are classified a long-period variables (LPVs) by \citet{Soszynski2009}. The LPVs consist of 67 Mira variables, 34 semi-regular variables (SRVs), and 24 OGLE small-amplitude red giants (OSARGs). There is good agreement between the spectrosopic and OGLE chemical classifications: these classifications are based on $V-I$ vs.~$J-K$ colours and the extinction-free Wesenheit indices ($W_{I}-W_{JK}$) derived from the near-IR and visual magnitudes. This index is able to correctly identify 30 of 42  O-AGBs and 66 of 67 C-AGBs. Most of the remaining 16 LPVs have OGLE class AGB (the other two are classified as RGB stars); these include 6 {\tt STAR}s, 2 {\tt YSO-4}s, 2 {\tt O-PAGB}s, 3 {\tt C-PAGB}s, and three massive stars (1 {\tt RCrB} and 2 {\tt RSG}s). Of the 17 sources without LPV classifications, seven {\tt RV Tauri} stars and four {\tt RCrB} stars have matching OGLE classes.

\begin{figure*}
\centering
\includegraphics[trim=0.0cm 0cm 0cm 0cm, width=6.5in]{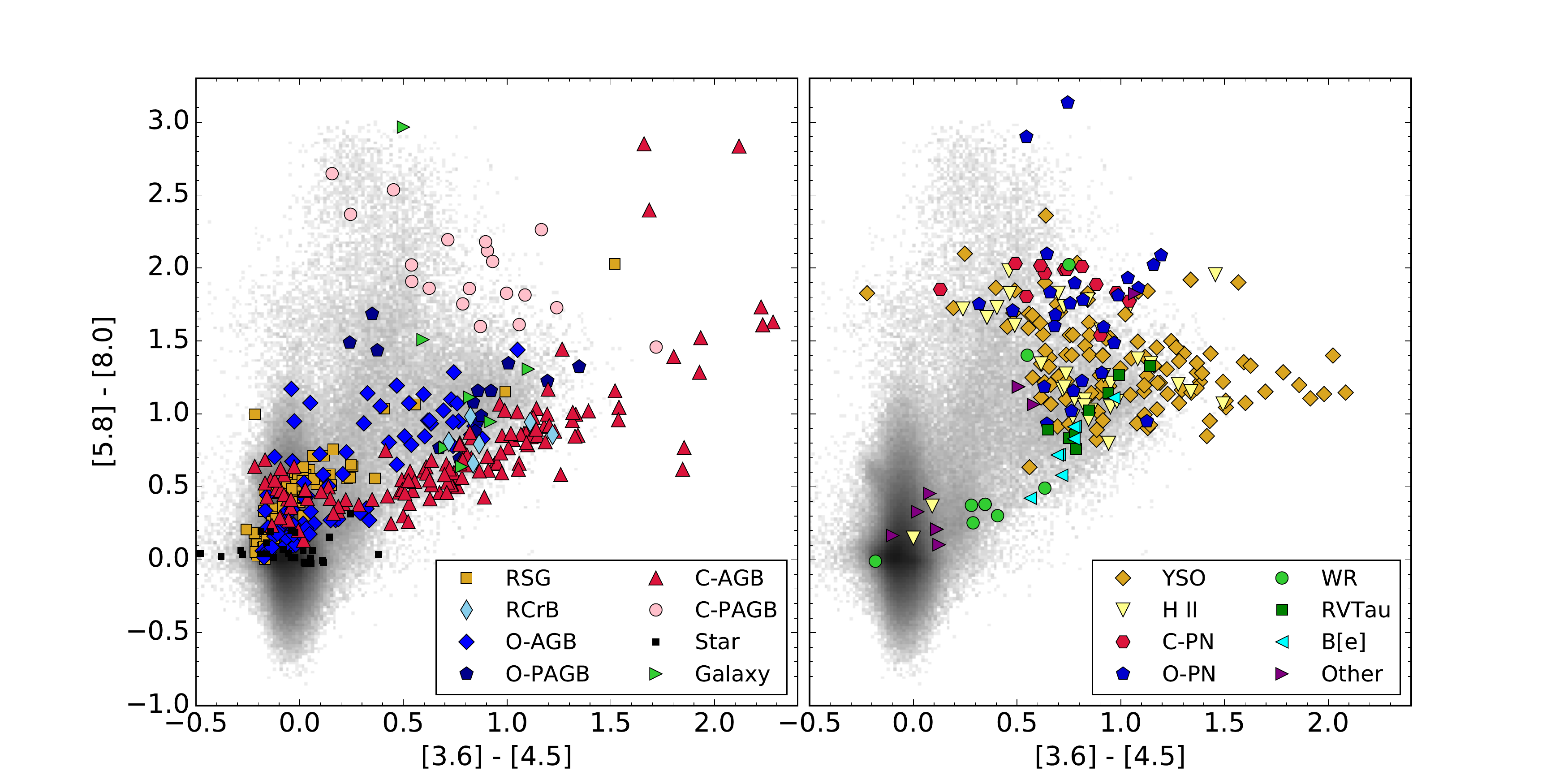}
\includegraphics[trim=0.0cm 0cm 0cm 0cm, width=6.5in]{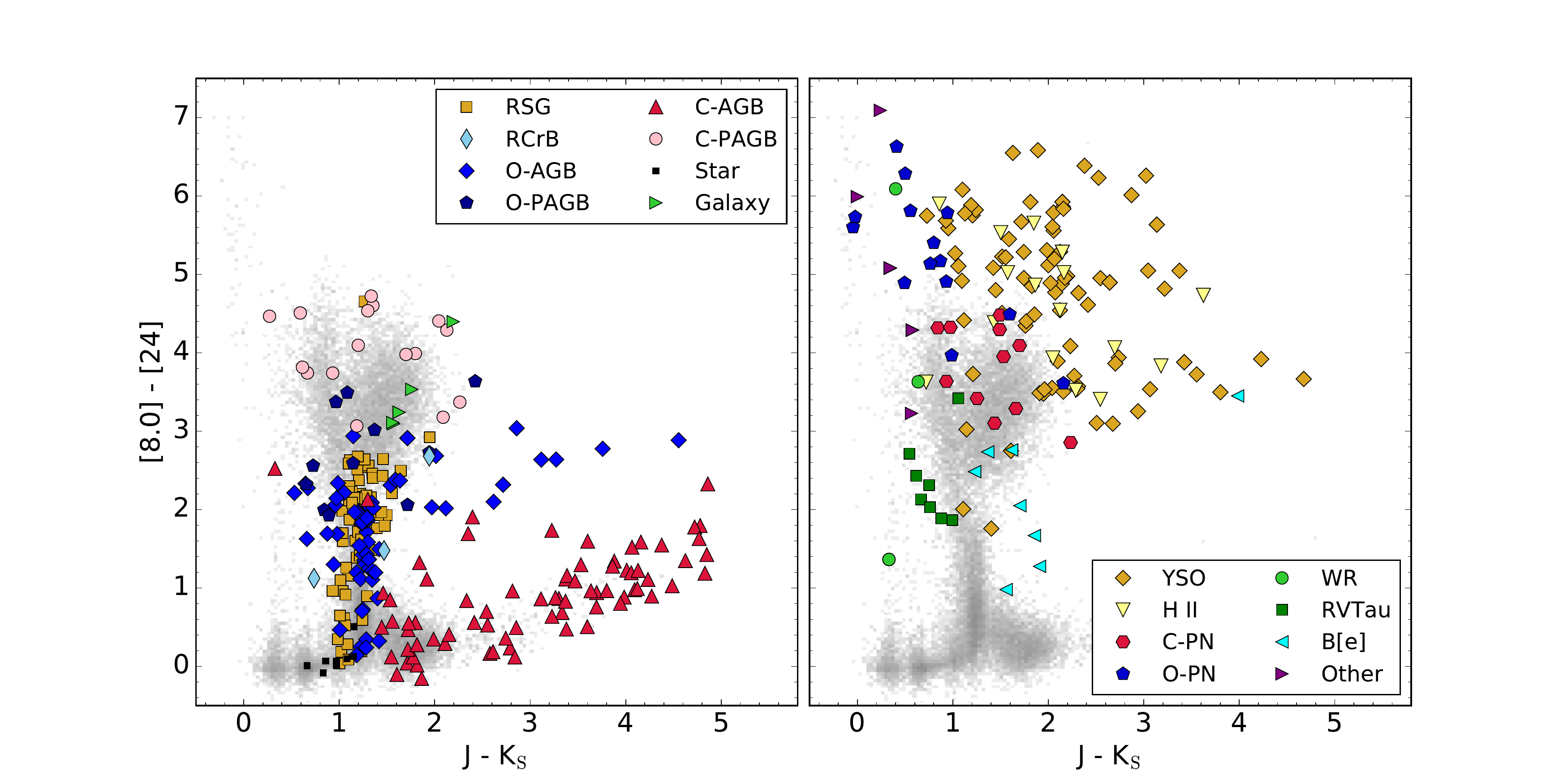}
\caption{Colour--colour diagrams (CCDs) showing the distribution of all the IRS sources in colour-space. The [5.6]$-$[8.0] vs.~[3.6]$-$[4.5] CCD is composed of the four IRAC bands.  This CCD is commonly used in stellar population colour classification schemes  (see e.g.~\citet{Whitney2008, Gruendl2009, Kastner2008, Buchanan2009} etc.) as it is distance independent. It can be directly compared to the SMC via Fig.~11 of Paper II. In this diagram, there is significant overlap at blue colours between sources that have a modest IR excess, however it is useful for separating populations enshrouded by dust.
The $J-K_{s}$ vs [8.0]$-$[24] CCD is noteworthy as it provided a clean separation between C-AGB stars and the other stellar populations. A similar distribution of sources was seen for the SMC (Fig.~12 in Paper II.)
The objects with spectroscopic classifications are marked with symbols as in Fig.~\ref{fig:YSO_col_comp}.}
\label{fig:CCDs_misc}
\end{figure*}

\subsection{Red supergiants and massive stars}
\label{sect:comprsg}

In this section we focus on the massive stellar population of the LMC. Nominally we include RSGs in this section, however due to the difficulty involved in separating RSGs from O-AGB stars using only photometry much of the discussion on RSG is included in Section~\ref{sect:compAGB}.

\begin{figure*}
\centering
\includegraphics[trim=0.0cm 0cm 0cm 0cm, width=6.5in]{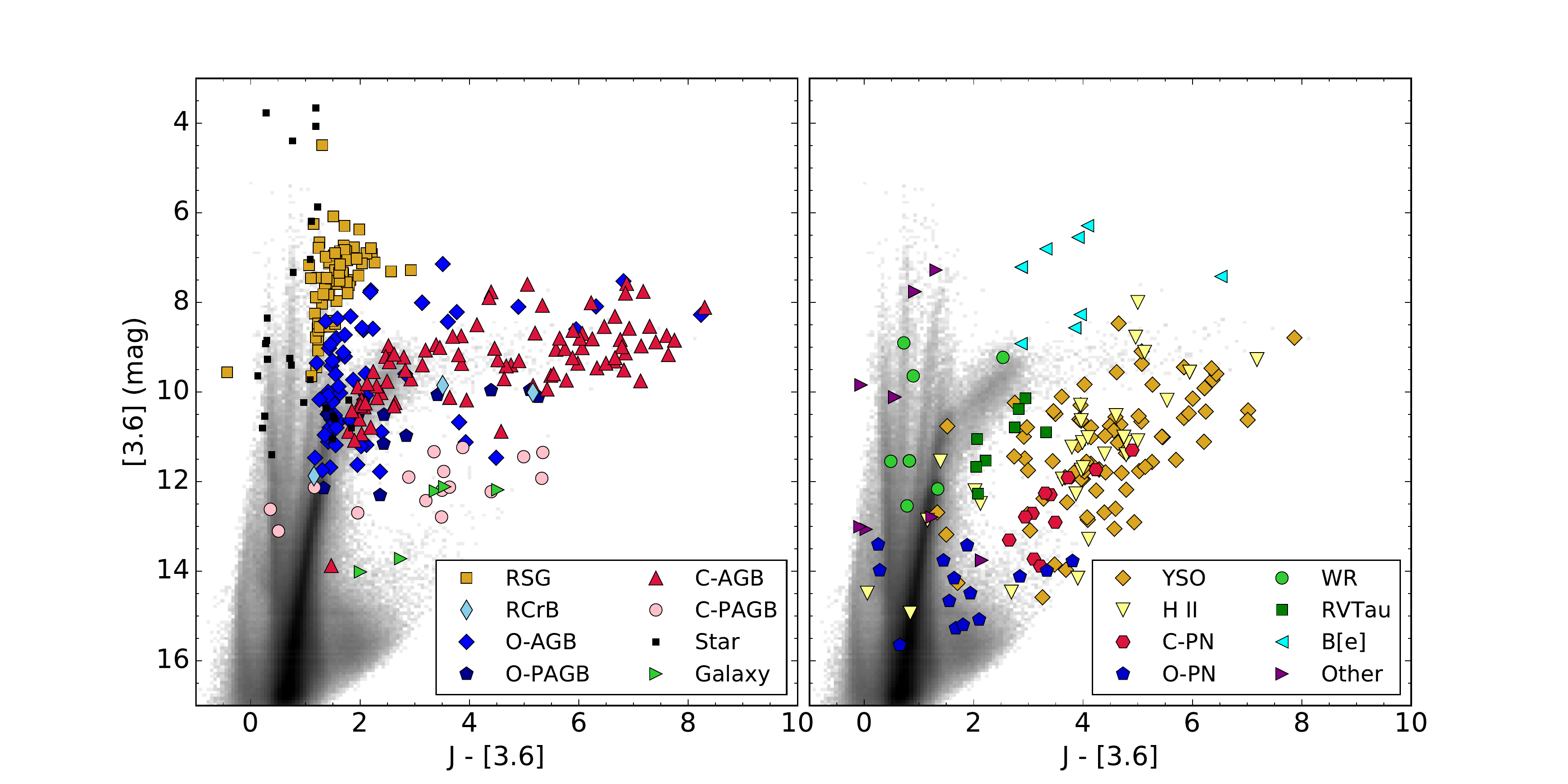}
\caption{The [3.6] versus $J-[3.6]$ CMD is very successful at separating RSGs from the other evolved classes. This selection is extremely clean and corroborates the optically selected massive star sample of \citet{Bonanos2009}.  Symbols are as in Fig.~\ref{fig:YSO_col_comp}.  
}  
\label{fig:CMD_misc}
\end{figure*}

\subsubsection{\citet{Bonanos2009}}

\citet{Bonanos2009} compiled the first major catalogue of spectral types and multi-wavelength photometry of massive stars in the LMC. Of the 1\,270 massive stars in this catalogue, we have matched 52 to IRS spectra. 
Twenty-six RSG were recovered, all with spectral type K or M. The one exception was \objectname{HD 269227} (OBJID 399) which is identified as a Wolf--Rayet star or an LBV candidate. This star has a nearby M2 supergiant which is not physically bound to \objectname{HD 269227} \citep{Schmutz1991} and likely contributes to (or accounts for all of) the IR flux. The presence of this nearby ($<$0.22'') RSG may also result in an incorrect match between our catalogue and that of \citet{Bonanos2009} for this source.

Eighteen massive post-main-sequence stars were matched to this catalogue; these are comprised of  seven WR stars, seven  B[e] stars, three LBVs and a BSG.  For the 73\% of the sources in common, the agreement is 100\%. The 18 stars include two in OB associations and eight in H\,{\sc ii} regions. Each of these ten was assigned an O or early-B spectral type with emission from an optically thin H\,{\sc ii} region. 

As noted by \citet{Bonanos2009}, massive stars are among the brightest sources at 3.6 \mum\  and occupy distinct regions in the [3.6] versus J$-$[3.6] CMD. Figure~\ref{fig:CMD_misc} shows that the AGB stars, RSGs, B[e] stars and LBVs can be cleanly separated using the J$-$[3.6] colour, which corroborates this finding. Furthermore, in Paper I a cut along $ J = 1$ and $[3.6] = 12-2(J-[3.6])$ was proposed to separate RSGs from the other stellar populations. This selection is extremely clean (as shown by the full spectral sample), and has only a few massive star contaminants.

\subsection{Post-AGB stars}
\label{sect:comppagb}

In this section we consider all galaxy-wide surveys of the LMC which focus on identifying post-AGB stars. Significant works in this field include \citet{vanAarle2011} and the followup paper by \citet{Kamath2015}.

\subsubsection{\citet{vanAarle2011}}

\citet{vanAarle2011} set out to produce a catalogue of post-AGB stars in the LMC, using the SAGE catalogue as a base, enhanced with optical spectra and photometry. Using the appearance of the SED, and in particular, the near-IR data, \citet{vanAarle2011} identified `shell' and `disk' sources. `Shell' sources have a distinctly double-peaked SED, with the longer wavelength peak corresponding to an expanding, detached shell of cooling dust, whereas the shorter-wavelength peak is stellar in origin. `Disk' sources have a broad excess in the near-IR, appearing as a shoulder on top of the stellar emission in the SED rather than a separate peak. This excess is indicative of a hot disk of dust, close to the star \citep{deRuyter2006}.
We would identify \citeauthor{vanAarle2011} `shell' sources as post-AGB stars, while their `disk' objects would mostly be classed as AGB stars or RSGs.

By enforcing a colour cut of [8]$-$[24]$>$1.384 mag, \citeauthor{vanAarle2011}  selected post-AGB candidates from the SAGE catalogue, and refined this list by requiring candidates to be optically bright (in the $U$, $B$, $V$, $I$, $R$ bands). It was assumed that supergiant and YSO contaminants would be removed by enforcing luminosity restrictions: YSOs have $L<$ 1\,000 L$_\odot$\ and supergiants $L>$ 35\,000 L$_\odot$, with post-AGB objects falling in between. Confirmation of some of the remaining post-AGB candidates was sought through low-resolution optical spectra. 

Agreement between the post-AGB catalogue of \citet{vanAarle2011} and our spectrum-based classification is poor. Of their $\sim$1\,400 post-AGB candidates,  179 have IRS spectra. Fifty-four of these are `disk' sources and 101 are `shell' sources;
the 24 remaining sources are not categorised as either `shell' or `disk' sources.
Of the `disk' sources, 38 are AGB stars and 11 are oxygen-rich post-AGB stars. Of the `shell' sources, the majority (68) are YSOs. Seventeen are planetary nebulae, and eleven are post-AGB stars. 
At best there is only a 17\% agreement between our classifications; from the 179 matches only 31 are post-AGB stars.

\begin{figure}
\includegraphics[trim=0cm 0cm 0cm 0cm, width=3.2in]{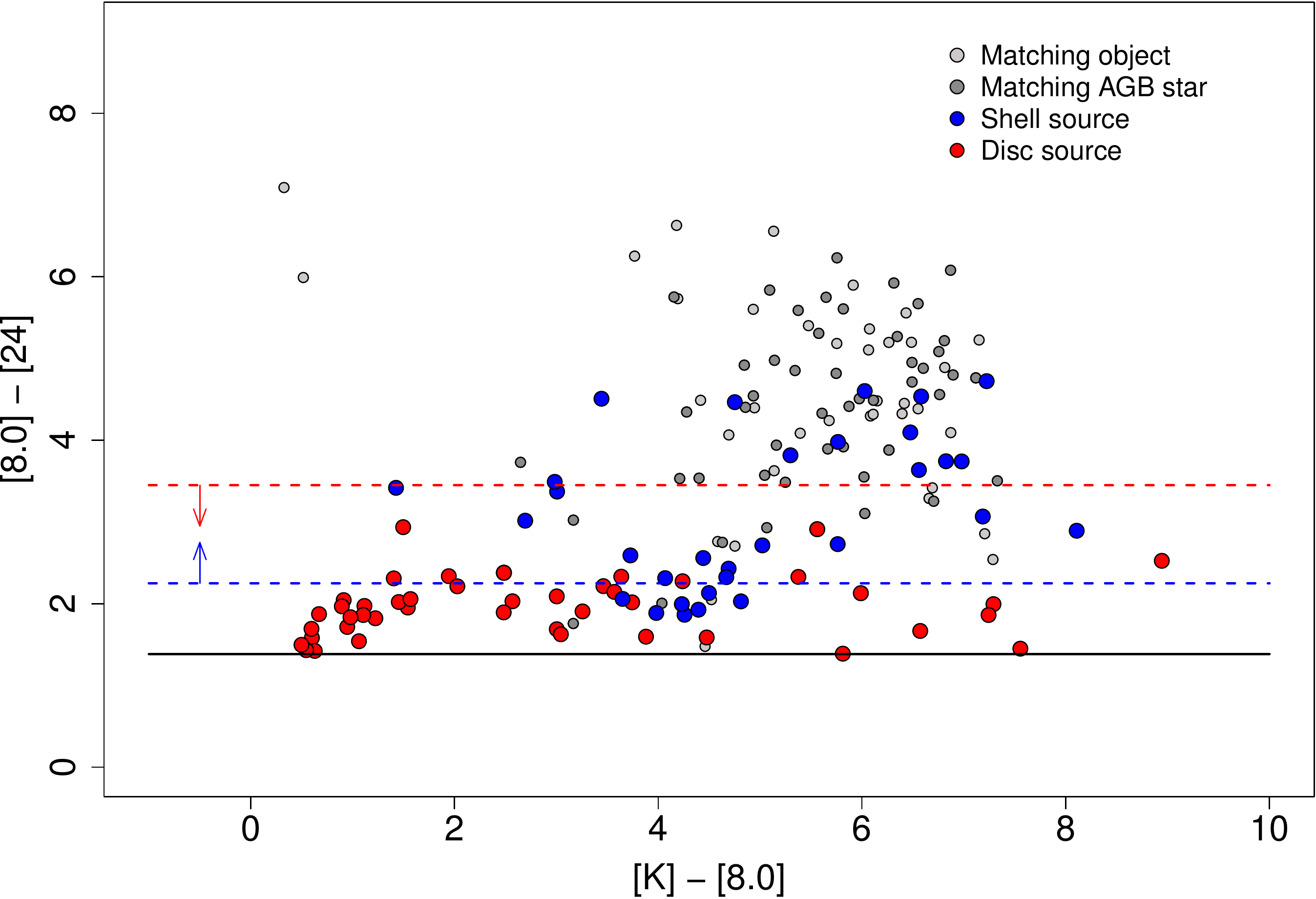}
\caption{The post-AGB star selection process of \citet{vanAarle2011} is  applied to the IRS sample (grey dots), disk sources (red dots) and shell sources (blue dots) are highlighted. Darker grey dots indicate AGB stars which are not deemed to be `disk' sources. The solid horizontal line marks the [8.0] $-$ [24] = 1.384 colour cut made as part of the selection process. The blue dashed line is the lower   limit for `shell' sources (post-AGB stars, in our classification),   according to \citet{vanAarle2011}. The red dashed line is the upper limit for `disk' sources (these are AGB stars with a large dust excess in our classification). The region between the two dashed lines may contain both `shell' and `disk' sources. \label{vAarle}}
\end{figure}

To understand the reason for this disparity, we compare the photometry of our LMC objects with the colour selection criteria employed by \citet{vanAarle2011} to identify `shell' and `disk' sources. Furthermore, high and low luminosity objects are discarded in an effort to minimise contamination by RSGs and YSOs.  Spectroscopically-confirmed dusty AGB stars are well constrained by the van Aarle colour limit for `disk' sources (dashed red line in Fig.~\ref{vAarle}). The limit for `shell' sources (dashed blue line), though, excludes a significant number of confirmed post-AGB stars with blue [8.0]$-$[24] colour. In addition, the cut at [8.0]$-$[24] = 1.384 mag excludes a few confirmed post-AGB stars (not shown in the Figure).
These selection criteria encompass a large number of sources which are not post-AGB stars (or indeed AGB stars), thus the level of contamination is high. The luminosity cuts made as part of the selection process to remove YSOs and supergiants exclude only a few YSOs at the low luminosity end (instead, it is mostly planetary nebulae which are excluded with L$_\star<$1\,000 L$_\odot$). At high luminosities, three `disk' sources are excluded (extreme O-rich AGB stars), as well as a large number of red supergiants and evolved YSOs.

\citet{vanAarle2011} also included a list of the $\sim$90 sources with optical spectra that they discarded from their catalogue. There are 20 matches to IRS spectra, two of which are classified as post-AGB by us (OBJID 256 and 776). This list is heavily dominated by RSGs and YSOs, while none of the matched sources are AGB stars.  Seven of these matches are `shell' sources and 13 are undetermined. There are no clear `disk' sources. 

\subsubsection{\citet{Kamath2015}}

The paper by \citet{Kamath2015} extended the analysis of \citet{vanAarle2011}.
\citet{Kamath2015} performed an optical follow-up, obtaining spectra for a large number of \citet{vanAarle2011} sources, but relaxing the requirement for objects to have L $>$ 1\,000\,L$_\odot$. They obtained 1,546 spectra of suitable quality, and identified 581 post-AGB, post-RGB and YSO candidates in the LMC. By modelling the spectra to obtain a log $g$ value and performing a further luminosity cut, the resultant list of post-AGB candidates numbered 35. 

Of these 35, eight objects have IRS spectra, and only three of these would we consider post-AGB stars. All three of the post-AGB star matches are oxygen-rich. The remaining five objects are comprised of two AGB stars (one carbon-rich,  one oxygen-rich), two YSOs (YSO-1, YSO-3) and an RCrB star.

\subsubsection{\citet{Matsuura2014}}

\citet{Matsuura2014} primarily selected post-AGB stars using a series of colour--colour and colour--magnitude diagrams (particularly the $K-$[8.0] and $K-$[24] colours; Fig.~\ref{fig:Matsuura_col_comp}). They also incorporated all spectroscopically-known post-AGB stars from \citet{BernardSalas2006,Gielen2009} and Paper I (see their Section 4), as well as known RV Tau stars from the light curves of \citet{WoodCohen2001}.  This spectral confirmation was used to revise the colour selections of \citet{Matsuura2009, Matsuura2013}, consequently there is a high degree of agreement between our samples.

\subsection{Planetary nebulae}
\label{sect:comppn}

There are over 300 confirmed planetary nebulae in the SAGE field of view. In the following section we focus on the papers by \citet{Reid2006} and \citet{Leisy1997} who have either compiled significant databases of known PNe or used IR colours to isolate these sources from other stellar populations in the LMC.

\subsubsection{\citet{Reid2006}}

The `Complete PN catalogue' of \citet{Reid2006} has long been a resource for planetary nebulae in the LMC. Potential PNe came from previous PN catalogues (169 objects) or a deep, high-resolution H$\alpha$\ survey of the LMC with the UK Schmidt Telescope, coupled with a targeted spectroscopic follow-up on the Anglo-Australian Telescope (460 objects). \citet{Reid2006} selected planetary nebulae based on the presence and strength of high-excitation lines (He\,{\sc ii}, [O\,{\sc iii}], [Ar\,{\sc iii}], [Ne\,{\sc iii}], often with strong [Ne\,{\sc v}] and [Ar\,{\sc v}]) and the shape, size and distinct boundary of the H$\alpha$\ emission (see their Section~4.1). Newly-identified PNe were rated in terms of certainty from `True" to ` Likely" to `Possible", largely dependent on the presence of obscuring stars.

We have found 54 IRS objects among the \citet{Reid2006} catalogue, which covers the central 25$\degr^2$  region of the LMC\footnote{There is no overlap between the more expansive 64$\degr^2$ survey of \citet{Reid2013} and our source list.}.  Thirty-three of these objects are from the `Previously known" part of the \citeauthor{Reid2006} catalogue, and all 33 are considered by us to be post-AGB stars or planetary nebulae: 18 oxygen-rich,  10 carbon-rich, and  5 post-AGB stars. 
The other part of the catalogue is for newly-discovered PN candidates. Of the  nine `True' PNe matched, only one source (\objectname{OBJID 153}) is classified as a PN by our methods. Of the remaining seven sources, four are YSOs or H\,{\sc ii} regions (two H\,{\sc ii}, a {\tt YSO-3}, and a {\tt H\,{\sc ii}/YSO-3}),  two carbon-rich AGB stars, one RCrB star, and one object with an IRS spectrum too noisy to classify. 
The H\,{\sc ii} region (\objectname{IRAS 05216$-$6753}, OBJID 453) shows no indication of high-excitation lines (only [S\,{\sc Iv}] and [Ne\,{\sc ii}] lines are evident), but has an unusually strong 10 $\mu$m silicate feature. Previous studies \citep[see][for a summary]{vanLoon2010} suggest a hot massive star is present. 
Of the three `Likely' and  nine `Possible' PNe,  none are considered PNe by us, although one object is a post-AGB star (OBJID 453, 256). The other eleven objects are evolved YSOs or H\,{\sc ii} regions.

A handful of revisions were made to their PN classifications in a follow-up publication, \citet{Reid2010}, and seven erroneous classifications were removed (two `True' PNe, one `Likely' PN and four `Possible' PNe). Most of these were reclassified as H\,{\sc ii} regions, in line with our classifications of those objects.

There are a number of missing PNe from the \citet{Reid2006, Reid2010} catalogues: of the 42 PN identified in our work, twelve do not appear in either of their catalogues. These well-known PNe fall outside of the initial survey area, and were not included in latter catalogues \citep[e.g.][]{Reid2013} as these focus on newly discovered PN candidates.

\subsubsection{\citet{Leisy1997}}

A detailed catalogue of 277 planetary nebulae in the LMC was complied by \citet{Leisy1997}. 
The \citet{Whitney2008} PNe sample was selected from this catalogue of LMC PNe candidates, as were the 56 PNe listed by \citet{Gruendl2009}. From the SAGE survey, \citet{Hora2008} detected 185 of these known LMC PNe in at least two IRAC bands. 
There are 44 PN candidates in common with our catalogue. Thirty-eight are matched to PNe in our sample, and a further six matches correspond to post-AGB stars. Of these, four are optically identified PNe (see Section~\ref{sec:classifications}), but are classified by the decision tree as post-AGB stars, possibly due to the low spectral resolution of the IRS spectra. Three of our IRS PNe are missing from the \citet{Leisy1997} sample. In total, the \citet{Leisy1997} agreement with the IRS sample is over 86\%.

\subsubsection{\citet{Kamath2015}}

\citet{Kamath2015}, in their search for optically-visible post-AGB objects, also compiled a table of 123 planetary nebulae. These objects were selected based on optical emission lines of Ne, Ar, C, O, and N. Of the 123 objects, 37 have IRS spectra. We find that 23 are classified as PNe (14 O-rich, 9 C-rich). The remaining objects are classified as carbon-rich post-AGB objects (four) and YSOs or H\,{\sc ii} regions (nine). Finally, one match (OBJID 39) has an {\tt UNK} classification due to the low quality of its IRS spectrum with only H\,{\sc i} lines clearly visible. 
All but one of the seven newly-discovered PNe in the list of \citet{Kamath2015} are classified as YSOs by us.

\subsection{Galaxies}

Background galaxies in the same field as the Magellanic Clouds have been studied by \citet{Kozlowski2013}, \citet{VeronCetty2010} and \citet{Cioni2013}. Follow-up works \citep[e.g.][]{Ivanov2016} have sought spectroscopic confirmation of the near-IR selected quasar candidates.   These distant galaxies are point-like at the resolution of {\em Spitzer} and present a significant problem when interpreting faint sources in CMDs. Our IRS sample has no targets in common with either the \citet{Cioni2013} or \citet{VeronCetty2010} catalogues.

\subsubsection{\citet{Kozlowski2013}} 

\citet{Kozlowski2013} confirmed 565 quasars in the region behind the LMC from the OGLE survey through studying their optical variability and mid-IR and/or X-ray properties. Two of these candidate quasars have IRS spectra, and both were examined in Paper I. We verify that \objectname{OBJID 42} is a galaxy, furthermore \citet{Kozlowski2013} gave a spectroscopic redshift for this galaxy of $z$=0.160, which agrees with our estimate. They also indicated that \objectname{OBJID 71} is a galaxy with a redshift of $z$=0.324. However, we have classified it as an H\,{\sc ii} region. Neither of these objects have optical variability nor associated X-ray sources (two of the three methods for discerning quasar status), and thus it is likely that both were classified as quasars based only on a colour--magnitude diagram (see their Fig.~1).

\subsubsection{\citet{Seale2014}}

Eight of our sources were identified by \citet{Seale2014} to be background galaxies based on {\em Herschel} data, but only one source (\objectname{OBJID 2}) shows redshifted features in its IRS spectrum. The other seven sources are post-main sequence stars, which includes three (post-)AGB and three massive stars \citep[see][for more detail on these objects]{Jones2015b}. The only other spectroscopically-confirmed galaxy included in the \citeauthor{Seale2014} sample is \objectname{OBJID 185}; we assigned this source an {\tt UNK} class.

\subsubsection{\citet{Gruendl2009}}

\citet{Gruendl2009} identified a significant number ($>$1\,000) of galaxies located behind the LMC, in the same region of the sky. The vast majority of these are not found amongst the IRS targets; however, five galaxies are matched, all from the targeted part of the \SSp survey (Paper I). The classifications concur, except for OBJID 78.

\subsection{Stars and foreground objects}

The {\tt STAR} class contains objects with a stellar photosphere but no variability, whilst the foreground subcategory denotes objects which are known to be Galactic stars.
It is difficult to identify and eliminate foreground sources from other stellar populations using only photometry. Stellar photospheres are not typically included in IR selection criteria as they are both bluer and fainter than typical IR stellar populations. Thus few catalogues exist with which to evaluate our classifications.

\subsubsection{\citet{Gruendl2009}}

 \citet{Gruendl2009} tabulated 291 stellar sources; although they do not explicitly state their selection criteria, 16 matches are found with the IRS catalogue. Of these 16, nine are high-certainty stars according to those authors, but none of these are stars in our consideration. We identify these sources as AGB or post-AGB stars. Seven of the 16 \citeauthor{Gruendl2009} stars are classified with less certainty; of these seven, we find that four are post-AGB objects and three are YSOs. Again, none are classified as stars by us. In fact the only stellar photosphere matched to our spectroscopic sample is classified by \citeauthor{Gruendl2009} as an AGB star. The lack of agreement with our IRS classifications may be due to the reliance by \citeauthor{Gruendl2009} on the IRAC [4.5]$-$[8.0] colour which is sensitive only to dusty sources, and hence will overlook normal stellar photospheric emission.

\subsubsection{\citet{Kastner2008}}

As foreground AGB stars and RSGs in the LMC are indistinguishable in 2MASS--{\em MSX} CCDs, \citet{Kastner2008} used measurements of the stars K-band amplitude to determine a classification. Two stars in our sample are identified by \citet{Kastner2008} to be Galactic Mira variables, \objectname{RS Men} (OBJID 407) and \objectname{MSX LMC 1686} (OBJID 787). \objectname{RS Men} saturates the SAGE photometry but appears point-like in the {\em Spitzer} IRAC image; it has a radial velocity of 140 km s$^{-1}$ and was assigned a distance of 4.75 kpc by \citet{Whitelock1994}. 
Similarly, \objectname{MSX LMC 1686} has a luminosity that is inconsistent with membership of the LMC \citep{Buchanan2006, Groenewegen2009}. Both these sources were flagged as foreground in Section~\ref{sec:foreground}.

\subsubsection{\citet{Groenewegen2009}}
\label{sec:groenewegen2009}
In their Section 5.1, \citet{Groenewegen2009} identified seven likely foreground objects based on their derived luminosities and mass-loss rates. Of these, we identify three (\objectname{RS Men}, \objectname{MSX LMC 1686}, and \objectname{MSX LMC 946}) as foreground objects using their radial velocities in Section \ref{sec:foreground}. The remaining four (\objectname{HD 269788}, \objectname{MSX LMC 1212}, \objectname{HD 271832}, and \objectname{W60 D29}) were classified in \citet{Groenewegen2009} as likely foreground objects based solely on their blue colours/low mass-loss rates or a small MACHO amplitude. We confirm that three of these are indeed foreground sources. The fourth \objectname{HD 269788} has a radial velocity of 248 km/s  \citep{Kunder2017}, we retain a LMC association for this source.

\section{Summary and conclusions}
\label{sec:conclusion}

In this paper, we present the {\em Spitzer} IRS spectra of 789 point sources in the LMC. The spectra were analyzed primarily using their spectral features to determine the nature of their gas and dust chemistries and to gain knowledge of their evolutionary state. This was done using a decision-tree method of spectral classification, as previously implemented in Papers I and II. We find that 789 unique point sources within the SAGE footprint were observed with the IRS, within a total of 1\,080 staring-mode observations. 
We find that the {\em Spitzer}-IRS LMC sample contains
            225 AGB stars,
            42 post-AGB objects,
            71 RSGs,
            41 PNe,          
            203 YSOs,
            134 H\,{\sc ii} regions,
            30 stars,
            7 WR stars,
            8 B[e] stars,
            8 galaxies,            
            and 24 other objects, including 6 RCrB stars, 3 LBVs, 2 BSGs, 1 Nova, 1 YSG and 2 SNR.
We identify one mixed-chemistry (carbon/oxygen-rich) object, we correct an erroneous classification of an ERO from \citet{Gruendl2009} and we identify a new background galaxy.  
            
After verifying our spectral classifications against others in the LMC \citep[e.g.,][]{Buchanan2006, Seale2009,Stanghellini2007}, we then compared them to a number of seminal photometric classification schemes.
In terms of Young Stellar Objects, we find that the selection methods which target YSOs in star-forming regions are very successful. However when these methods are applied globally the samples become contaminated by evolved stars.
For AGB stars, the photometric selection methods for carbon-rich AGB stars were very reliable,  whilst the clean selection of oxygen-rich AGB stars is more arduous. The dustiest O-rich evolved stars can enter carbon star and YSO selection boxes, which means that the total dust production budget of the LMC has been underestimated. Once these stars have been correctly accounted for, we would  expect the O-rich AGB dust production rate to be revised upwards  with respect to the \citet{Riebel2012} and \citet{Matsuura2009} results.
It is notoriously difficult to select massive stars and post-AGB stars photometrically \citep[c.f.][]{Kastner2008, vanAarle2011}. Using our large IRS sample we can confirm that the colour--magnitude cut we proposed in Paper I based on J-band and 3.6 $\mu$m photometry, for the selection of red supergiants, creates a very clean sample. These colours may also be useful for identifying B[e] stars. 

Finally this large IRS sample and point-source classifications allows us to obtain samples of similar objects for future studies.

\vspace{0.4cm}

We would like to thank the referee for comments and careful review of the data tables that improved this paper.

Jones and Meixner acknowledge support from NASA grant, NNX14AN06G, for this work. The authors also acknowledge financial support from the Ministry of Science and Technology (MoST) Taiwan under grant MOST104-2628-M-001-004-MY3.
McDonald and Zijlstra were funded by the Science and Technology Facilities Council under grant ST/L000768/1.
The NASA Astrophysics and Data Analysis Program (Grant NNX13AE66G) supported G.C.S. and the addition of high-resolution IRS spectra to CASSIS, the Combined Atlas of Sources with {\it Spitzer} IRS Spectra. Readers are referred to ApJS, 196, 8 (2011).
Sargent acknowledges support from NASA grants NNX13AD54G and NNX15AF15G.
The work of M. Sewi{\l}o was supported by the appointment to the NASA Postdoctoral Program at Goddard Space Flight Center. 
This research has made use of the VizieR catalogue access tool, CDS, Strasbourg, France. The original description of the VizieR service was published in A\&AS, 143, 23 (2000).

{\it Facilities:} \facility{{\em Spitzer} (IRS)}, \facility{{\em Spitzer} (MIPS)}, \facility{{\em Spitzer} (IRAC)}.

\appendix

\section{Corrections to Paper I.}

Source OBJID 126 was previously labeled as \textquotedblleft KDM 4554". This was erroneous, since KDM 4554 is nearly 2\arcmin\ away from the position of OBJID 126. Instead, we use the 2MASS designation for this source, \objectname{2MASS J05304499-6821289}.
We re-classify OBJID 24, 112 and 160 from {\tt UNK} to {\tt No Detection}  as no infrared point source was located within 3\arcsec\ of the position where the IRS spectrum was extracted.
OBJID 78 \objectname{SSTISAGE1C J051618.70$-$\allowbreak715358.8} has been identified as \objectname{IRAS 05170$-$7156} consequently its classification has been changed from {\tt UNK} to {\tt GAL}.
We also re-classify OBJID 185 from {\tt O-AGB} to {\tt UNK}. 

\section{The decision tree}

\begin{figure}
\centering
\includegraphics[scale=0.70]{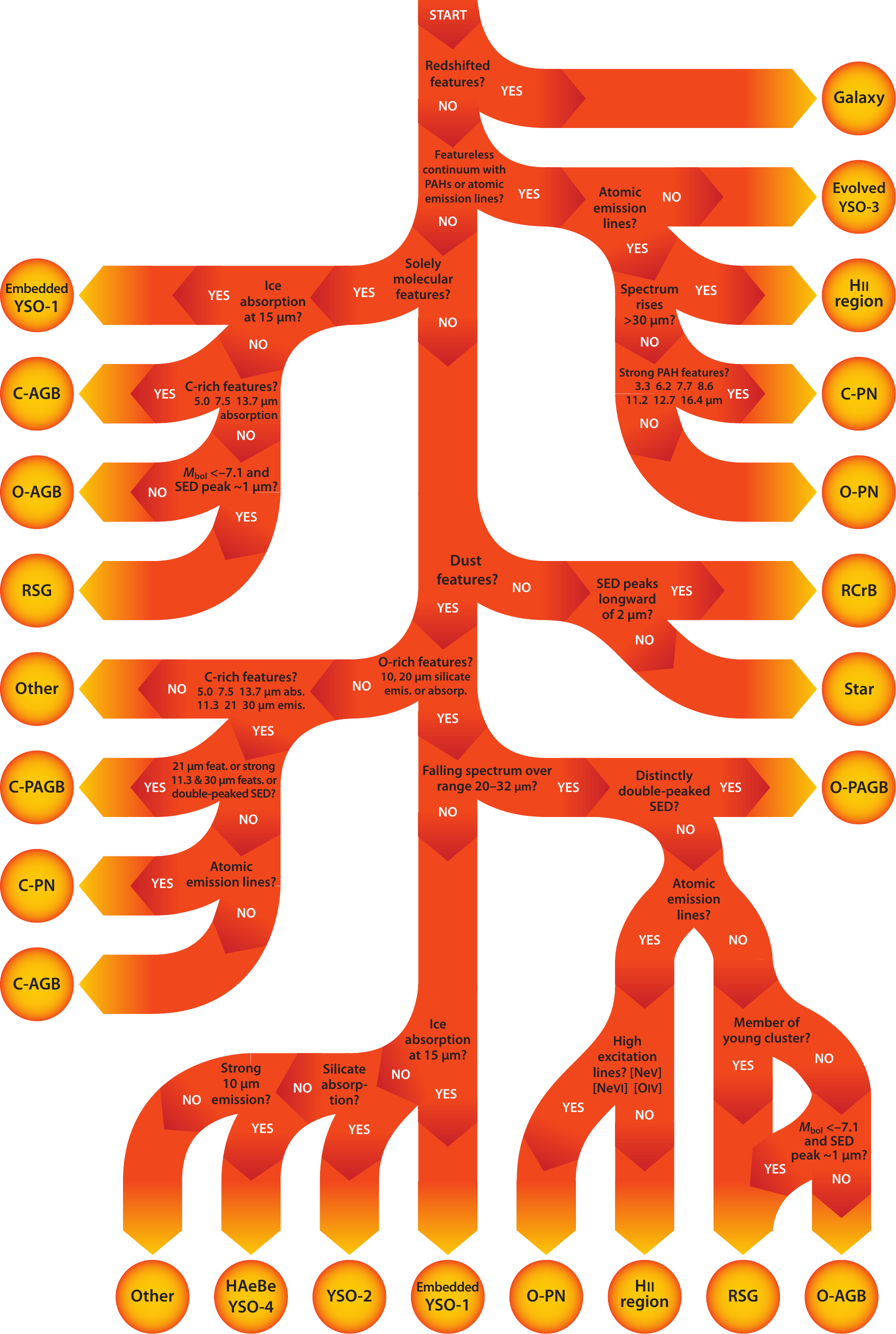}
\label{fig:classtree}
\caption{The classification process used as a basis in this work. Starting from the top of the diagram, {\sc yes}/{\sc no} decisions are made until a final classification is arrived at (shown by the circles).  Figure reproduced from Paper II.}
\end{figure}

To classify each spectra in our IRS sample we use a classification tree, as presented in Papers I and II and reproduced in Fig.~\ref{fig:classtree}. One proceeds through the tree, responding to the {\sc yes} or {\sc no} questions in order to reach a terminal classification. We have added an additional classification to our paper ({\tt H\,{\sc ii}}/{\tt YSO-3}) as it is unclear if the atomic emission lines present in some spectra are the result of poor background subtraction when using cluster mode in complex ISM regions. 

\def\aj{AJ}					
\def\actaa{Acta Astron.}                        
\def\araa{ARA\&A}				
\def\apj{ApJ}					
\def\apjl{ApJL}					
\def\apjs{ApJS}					
\def\ao{Appl.~Opt.}				
\def\apss{Ap\&SS}				
\def\aap{A\&A}					
\def\aapr{A\&A~Rev.}				
\def\aaps{A\&AS}				
\def\azh{AZh}					
\def\baas{BAAS}					
\def\jrasc{JRASC}				
\def\memras{MmRAS}				
\def\mnras{MNRAS}				
\def\pra{Phys.~Rev.~A}				
\def\prb{Phys.~Rev.~B}				
\def\prc{Phys.~Rev.~C}				
\def\prd{Phys.~Rev.~D}				
\def\pre{Phys.~Rev.~E}				
\def\prl{Phys.~Rev.~Lett.}			
\def\pasp{PASP}					
\def\pasj{PASJ}					
\def\qjras{QJRAS}				
\def\skytel{S\&T}				
\def\solphys{Sol.~Phys.}			
\def\sovast{Soviet~Ast.}			
\def\ssr{Space~Sci.~Rev.}			
\def\zap{ZAp}					
\def\nat{Nature}				
\def\iaucirc{IAU~Circ.}				
\def\aplett{Astrophys.~Lett.}			
\def\apspr{Astrophys.~Space~Phys.~Res.}		
\def\bain{Bull.~Astron.~Inst.~Netherlands}	
\def\fcp{Fund.~Cosmic~Phys.}			
\def\gca{Geochim.~Cosmochim.~Acta}		
\def\grl{Geophys.~Res.~Lett.}			
\def\jcp{J.~Chem.~Phys.}			
\def\jgr{J.~Geophys.~Res.}			
\def\jqsrt{J.~Quant.~Spec.~Radiat.~Transf.}	
\def\memsai{Mem.~Soc.~Astron.~Italiana}		
\def\nphysa{Nucl.~Phys.~A}			
\def\physrep{Phys.~Rep.}			
\def\physscr{Phys.~Scr}				
\def\planss{Planet.~Space~Sci.}			
\def\procspie{Proc.~SPIE}			
\def\icarus{Icarus}
\let\astap=\aap
\let\apjlett=\apjl
\let\apjsupp=\apjs
\let\applopt=\ao








\end{document}